\documentclass{jfm}
\usepackage{graphicx}
\usepackage{url}
\usepackage{epstopdf}
\usepackage{xcolor}
\usepackage{placeins}
\usepackage{pdfpages}
\usepackage{amsmath}

\usepackage{lineno}

\newcommand\N{{\bf n}}
\renewcommand\P{{\bf p}}

\newcommand\F{{\bf f}}

\newcommand\U{{\bf u}}

\newcommand\X{{\bf x}}

\newcommand\be{\begin{equation}}
\newcommand\nd{\end{equation}}
\newcommand\bed{\begin{displaymath}}
\newcommand\ndd{\end{displaymath}}

\newcommand\ba{\begin{array}}
\newcommand\ea{\end{array}}
\newcommand\bea{\begin{eqnarray}}
\newcommand\nda{\end{eqnarray}}

\newcommand\remove[1]{}

\renewcommand\Re{{\rm Re}}
\newcommand\We{{\rm We}}

\newcommand\Oh{{\rm Oh}}

\shorttitle{Pulsed Jet}
\shortauthor{Y. Kulkarni, C. Pairetti, R. Villiers, S. Popinet and S. Zaleski}

\title{The atomizing pulsed jet}

\author{Yash Kulkarni\aff{1}\corresp{\email{kulkarniyash2398@gmail.com}},
  Cesar Pairetti\aff{1}\corresp{\email{paire.cesar@gmail.com}},
  Rapha\"el Villiers\aff{1},
  St\'ephane Popinet\aff{1}
 \and St\'ephane Zaleski\aff{1,}\aff{2}\corresp{\email{stephane.zaleski@sorbonne-universite.fr}}}

\affiliation{\aff{1}Sorbonne Université and CNRS, UMR 7190, Institut Jean Le Rond $\partial$’Alembert, 75005 Paris, France
\aff{2}Institut Universitaire de France, UMR 7190, Institut Jean Le Rond $\partial$’Alembert, 75005 Paris, France}

\newcommand\blue[1]{\textcolor{red}{#1}}
\newcommand\red[1]{\textcolor{red}{#1}}
\newcommand\purp[1]{\textcolor{purple}{#1}}

\renewcommand\blue[1]{{#1}}
\renewcommand\red[1]{{#1}}
\renewcommand\purp[1]{{#1}}

\norule
\begin{document}

\newpage
\maketitle

\begin{abstract}

Direct Numerical Simulations of the injection of a pulsed round liquid jet in a stagnant gas are performed in a series of runs of geometrically progressing resolution. 
The  Reynolds and Weber numbers and the density ratio are sufficiently large for reaching a complex high-speed atomization regime but not so large so that the small length scales of the flow are impossible to resolve, except for very small liquid-sheet thickness. 
The Weber number based on grid size is then small, an indication that the simulations are very well resolved. Computations are performed using octree adaptive mesh refinement with a finite volume method and height-function computation of curvature, down to a specified minimum grid size $\Delta$. Qualitative analysis of the flow and its topology reveals a complex structure of ligaments, sheets, droplets and bubbles that evolve and interact through impacts, ligament breakup, sheet rupture and engulfment of air bubbles in the liquid. A rich gallery of images of entangled structures is produced.  Most processes occurring in this type of atomization are reproduced in detail, except at the instant of thin sheet perforation or breakup.  We analyze droplet statistics, showing that as the grid resolution is increased, the small-scale part of the distribution does not converge, and contains a large number of droplets close in order of magnitude to the minimum grid size with a significant peak at $d = 3\Delta$. 

This non-convergence arises from the {\em numerical sheet breakup} effect, in which the interface becomes rough  just before it breaks.
The rough appearance of the interface is associated to a high-wavenumber oscillation of the curvature.
To recover convergence, we apply the controlled ``manifold death'' numerical procedure, in which thin sheets are detected, and then pierced by fiat before they reach a set critical thickness $h_c$ that is always larger than $6 \Delta$. 
This  allows convergence of the droplet frequency above a certain critical diameter $d_c$  above and close to $h_c$. 
A unimodal distribution is observed in the converged range. 
The number of holes pierced in the sheet is a free parameter in the manifold death procedure, however we use the Kibble-Zurek theory to predict the number of holes expected on heuristic physical grounds. 
\end{abstract}


\section{Introduction}

\label{sec:intro}

Atomization simulations have progressed at an amazing rate, however
the topic is still far from mature. As we shall show in this paper the
prediction of the breakup of liquid masses in a typical large-speed
flow is marred by vexing numerical effects and profound physical
uncertainties. New developments, such as better codes, rapidly
increasing processing power and new numerical methods are however
poised to mitigate the difficulties. In this paper, we investigate the
pulsed jet, a paradigmatic case of atomizing flow inspired by diesel
engine jets, although we do not aim at solving a particular applied
problem or even any experimental configuration, but rather to investigate
the potential and limitations of Direct Numerical Simulation (DNS) of
atomizing flow, and in particular the ability of such DNS to reveal
physically relevant properties of the flow through {\em statistically converged}
numerical approaches. The emphasis here is on statistical
convergence, rather than ``trajectory" convergence, since only the
former is conceivable in very complex, irregular flows.

The study of such convergence already has a long history, despite
rapid progress.  We review specifically the history of round jet atomizing simulations and we refer the reader to reviews on the general topic of atomization and its simulation  
\cite[]{gorokhovski08,villermaux2020fragmentation}. The first attempts at testing the convergence of the
Probability Distribution Function (PDF) of droplet sizes were those of
\cite{Herrmann2011}. His simulations of a round jet used a variant of the Level-Set
method and showed that the number of small droplets in the PDF was
underestimated by coarse-grid simulations. This can be understood by
the fact the the Level-Set methods tend to eliminate small droplets by
``evaporating" them. On the other hand Volume-Of-Fluid (VOF)
methods keep too many droplets in an intermediate range around the grid
size $\Delta$, and in that range the PDF of
droplet sizes is overestimated. This can be seen for example in a study of 
the convergence of the droplet size distribution in the round jet by
\cite{pairetti2020mesh} where coarse grids overestimate the number of large droplets. 

\begin{figure}
    \centering
    \includegraphics[width=14cm]{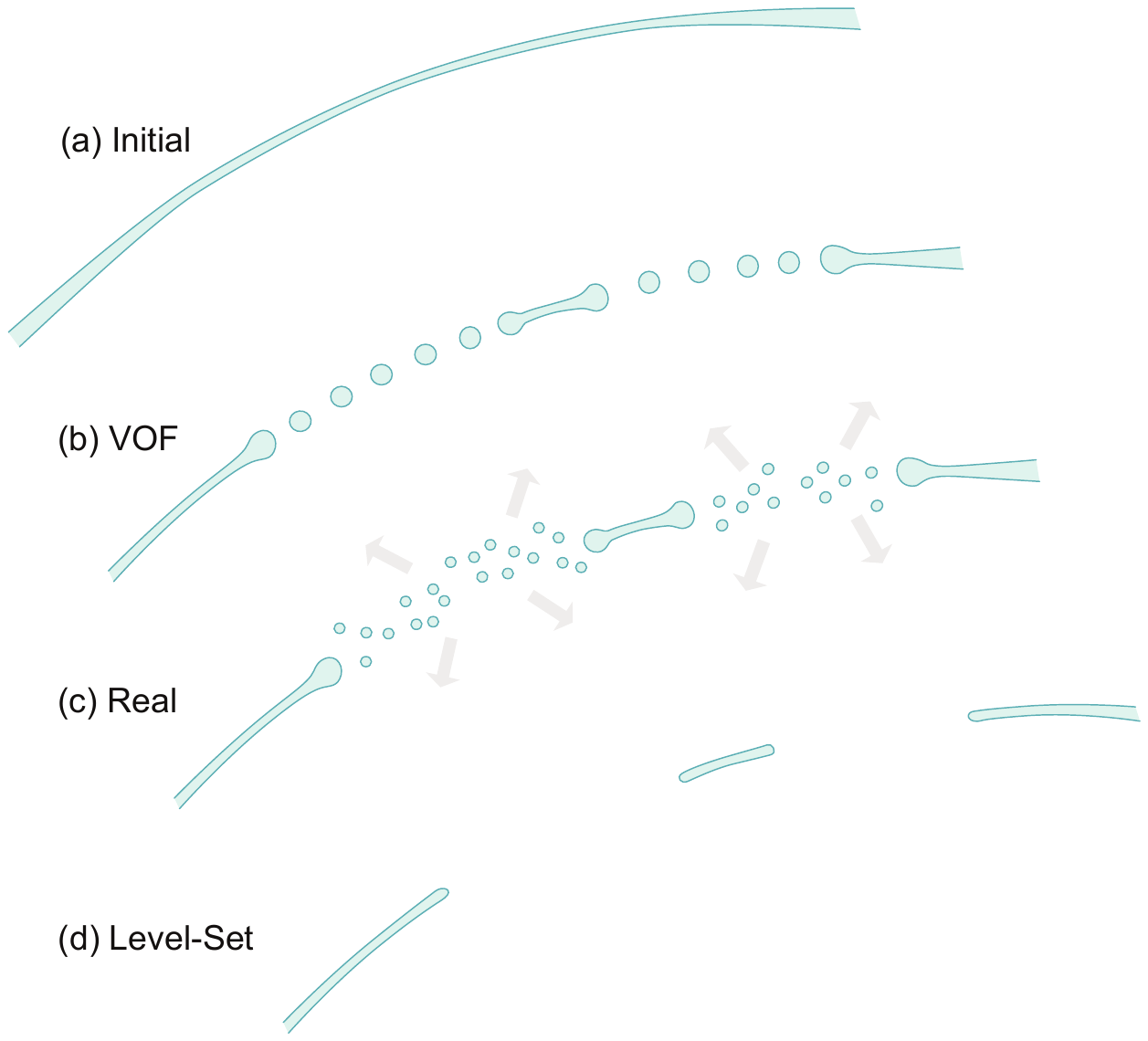}
    \caption{\red{An illustration of the outcomes for the numerical simulation of a thinning liquid sheet; (a) shows the initial configuration, before breakup. The other three images (b-d) show a schematic view of the outcome either in reality or in various types of numerical simulation. It is arbitrarily assumed that there are two breakup locations. Both the Volume-of-Fluid (VOF) and the Level-Set methods yield topology changes when the sheet thickness reaches the grid size. In (b) fragments larger than the grid size are obtained because of mass conservation in the VOF method. (c)
 In reality, the sheet thinning continues until much later than in the numerics, unless extremely fine grids are used. The final size of some of the droplets is then be much smaller than in the VOF simulation. (d) The Level-Set or Diffuse-Interface methods on the other hand evaporate the thin parts of the sheet and loses much more mass.}}
    \label{nicefig}
\end{figure}

A graphical illustration of the contrasting effects of
Level-Set and VOF methods on the droplet sizes in atomization is shown on Figure
\ref{nicefig}. Despite the inaccuracy of the distribution for small droplet diameters, both methods may converge as the grid is refined progressively.  There is thus some hope that very large simulations will eventually produce converged distributions, a hope we want to explore in this paper. 

\begin{figure}
    \centering
    \includegraphics[width=12cm]{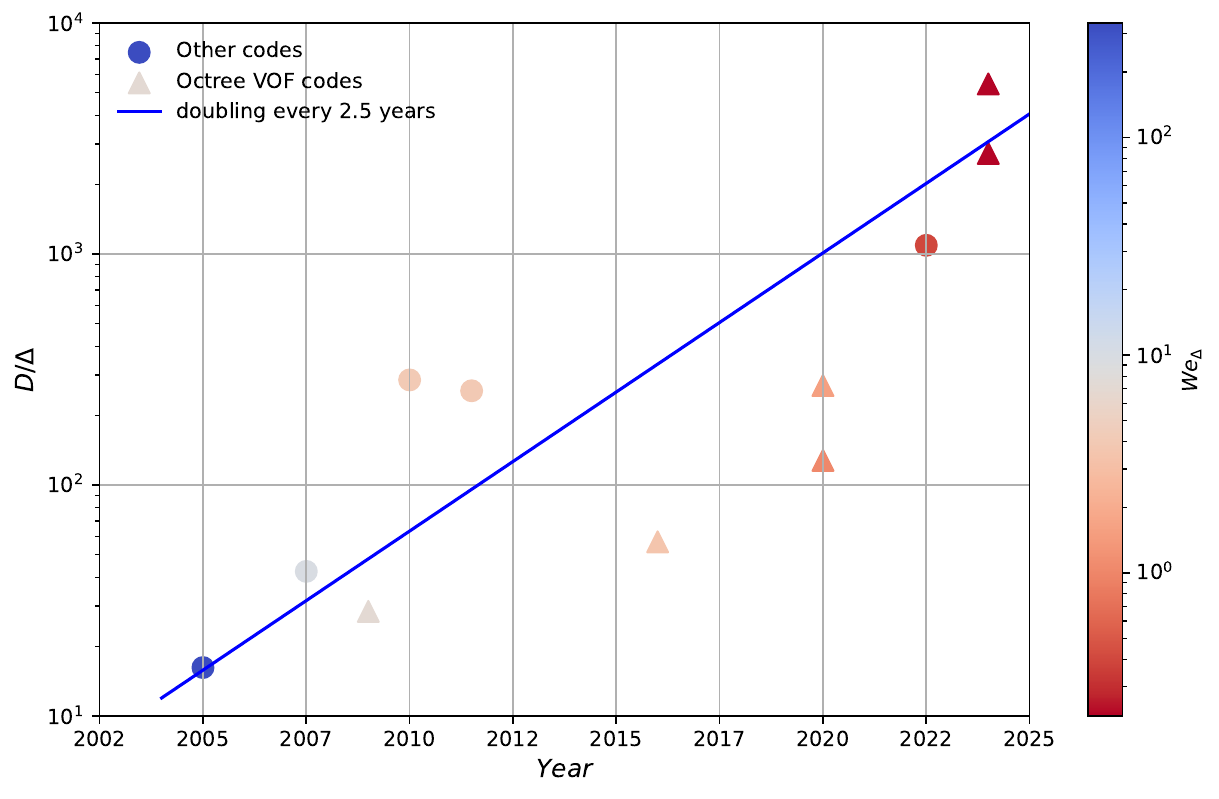}
    \caption{The increase in two-phase round-jet grid resolution in time. The graph includes two simulations published only on the Gerris and Basilisk websites  (and in other channels outside of academic journals) before 2017. The 2024 simulations are those reported in this paper.}
    \label{review}
\end{figure}

Aside from the two investigations cited above, very few studies address the issue of PDF convergence, despite detailed analyses of aspects of the flow. Perhaps the first round, single-jet atomization simulation in the conditions of a diesel jet was that of \cite{Bianchi05} using the VOF method, followed by more detailed simulations by 
\cite{menard07} using the Combined Level-Set VOF mod (CLSVOF). This was followed by other simulations using CLSVOF by  \cite{lebas2009}, \cite{chesnel2011subgrid} and \cite{Anez2018} and by simulations 
using the VOF method \cite[]{fuster2009a}. The latter studied both the coaxial jet cases in the so-called ``assisted atomization" setup and the conical jet.  For the conical jet  \cite{fuster2009a} showed that the PDF changed drastically with grid resolution. The PDF never peaked as the droplet size was decreased towards the grid size. 
Until 2022, the most detailed (or highest-fidelity) published simulations of a round jet were performed by \cite{shinjo2010simulation} using a combination of Level-Set and VOF techniques with a ratio of jet diameter to grid size of 286. These simulations, together with those of \cite{Herrmann2011} are the two outliers above the trend line in Figure \ref{review}.
In addition to revealing a wealth of details about ligament and sheet formation, and the perforation of sheets,  \citeauthor{shinjo2010simulation} have also shown distributions of droplet sizes, but without investigating explicitly their convergence.  As in earlier simulations of the round jet, the droplet size distribution has a single maximum close to the small droplet end of the spectrum, almost at the grid size.
Studies by \cite{Jarrahbashi2014,Jarrahbashi2016} focused on a round, spatially-periodic jet and  analyzed the effect of vortex dynamics on the development of instabilities along with the jet core. 
Studies by \cite{zhang2020modeling}  and \cite{pairetti2020mesh} applied the VOF method with octree Adaptive Mesh Refinement. As already mentioned the later paper investigated the  distribution of droplet sizes showing some difficulties in reaching statistical convergence. 
Several authors including \cite{torregrosa2020study} and  \cite{salvador2018analysis} have focused on the effect of injection conditions and turbulence. \cite{khanwale2022breakup} and \cite{saurabh2023scalable}, using a diffusive interface method with octree Adaptive Mesh Refinement, obtained the most refined simulations published at time of writing with a special treatment of the refinement in thin regions.
It should be noted that all of the round-jet studies cited above involve a variety of physical parameters and grid resolutions, despite the fact that they share similarities, such as moderate density ratios and liquid Reynolds numbers (defined below) in a range $5000 \le Re_l \le 13400$ that are attainable by Direct Numerical Simulation for single-phase flow. 
For example the two octree studies of \cite{zhang2020modeling}  and \cite{pairetti2020mesh} have gas-based Weber numbers (defined below)  of respectively  177 and 417. As a general rule, it is better to select relatively small density ratios, Reynolds and Weber numbers to increase the likelihood of reaching convergence in computation. In this work, we thus chose the parameters in the lower range of Re, We and density ratio for this type of flow.
Similar attempts at obtaining convergence were realized in the simulations of \cite{ling17}, investigating assisted quasi-planar atomization after the experiment of Grenoble \cite[]{Benrayana06,Fuster2013}. \cite{ling17} performed computations on four grids of increasing resolution. Despite the huge computational effort involved in that latter simulation, convergence was clearly not reached, both quantitatively as droplet size distributions had a very narrow region of overlap, and qualitatively as clearly under-resolved structures in thin sheet-like regions were observed. One of our objectives in this work is to see what happens if ever finer grids are used, until very thin sheets are clearly resolved. In experiments as well as in some simulations it is clear that thin sheets are the site of weak spots 
\cite[]{lohse2020double}. 
The presence of these weak spots as one of the mechanisms leading to atomization forces the following somewhat sobering conclusion. 
Numerical simulations of  Diffuse-Interface, Level-Set or VOF type can never be converged if thin sheets break or perforate only when they reach the grid size. Instead a physical mechanism for weak spots or perforations must be present to nucleate holes in a manner independent of the numerics. 
Although such a mechanism may not be known yet, we suggest as a backstop procedure to define a critical sheet thickness $h_c$ beyond which perforations occur with relatively high frequency, that is at the rate of one or more perforations per connected thin-sheet region. 
\cite{Chirco22} have called {\em manifold death} (MD) this procedure\footnote{This choice of words partially echoes that of \cite{debregeas1998life}.}. 
It should be noted that bags or membranes are observed at relatively low velocities in many atomization configurations (droplets, jets etc...) and that at higher velocities, bags are not visible and perhaps non existent. 
\red{We note that the literature \cite[]{Lasheras00,tolfts2023statistics} distinguishes two kinds of atomization, a bag or membrane-type atomization at low Weber numbers
and a fiber-type atomization at high Weber number. The transition between the two behaviors 
was found by \cite{tolfts2023statistics} to lie between $\We_g=70$ and $\We_g=400$.}
From the connection between thin sheets and the lack of statistical convergence one may conjecture 
that in the fiber-type regime statistical convergence could be achieved. This is however contradicted by the lack of convergence observed by \cite{Herrmann2011} at $\We_g = 500$. A definite answer to this argument may have to wait for a study similar to the one in this paper in the fiber type regime, a study which would probably be even more expensive. 

In what follows we first describe the characteristics of our test case. 
We then describe our numerical method, including the manifold death (MD) procedure. Then we continue with our results which are both qualitative and quantitative. Qualitatively our results reveal new processes of numerical rupture. Quantitatively they allow an analysis of the droplet size distributions or PDF.
\red{A discussion of the phenomenology of hole expansion in a thin sheet is given in Appendix \ref{app:pheno_hole_expansion}}.  We end with a conclusion including perspectives and discussion. 

\section{Mathematical model and numerical method}

Our mathematical model is based on the mass and momentum-conservation equations for incompressible and isothermal flow
\begin{equation}\label{eq:mass}
\nabla\cdot\U = 0,
\end{equation}
\begin{equation}\label{eq:NS-continuum}
\frac{\partial \rho \U}{\partial t} + \nabla\cdot(\rho\U\U)
=
- \nabla p
+ \nabla\cdot \left(2\mu\mathbf{D}\right) + \F_\sigma,
\end{equation}
where $\U(\X,t)$ is the velocity field and $p(\X,t)$ is the pressure field. The tensor $\mathbf{D}$ is defined as $\frac{1}{2}\left[\nabla \U + \left(\nabla \U\right)^T\right]$. The density and viscosity of the flow are noted as $\rho$ and $\mu$ respectively. The last term on the right-hand side of the Navier--Stokes Equation (\ref{eq:NS-continuum}) represents the surface tension force
\begin{equation}\label{eq:STF}
\F_\sigma= \sigma \kappa \N \delta_s,
\end{equation}
which depends on the surface tension $\sigma$ and the interface shape, particularly on its curvature $\kappa$ and unit normal vector $\N$. The Dirac distribution $\delta_s$ indicates that the force only acts at the free surface.
We consider the phase distribution function $c(\X,t)$ that takes value  unity in the reference phase and  null outside of it. The transport equation for this function is:
\begin{equation}\label{eq:c-advect}
\frac{\partial c}{\partial t} + \nabla\cdot(c\U)= c\,\nabla\cdot \U.
\end{equation}
In the context of incompressible flow, the right-hand side above is equal to zero.
The $c$ function implicitly defines the interface at its discontinuity surface, defining also the $\delta_s$, $\N$, and $\kappa$ fields, e.g. $\nabla c = \N \delta_s$. 

The Volume of Fluid method represents the evolution of $c$ using the Piecewise Linear Interface Capturing (PLIC) method of \cite{Hirt1981} and \cite{DeBar74}. In this context, the mean value of the color function on a cell is
\begin{equation}\label{eq:VOFDef}
f  = \frac{1}{| \Omega | }\int_\Omega c(\X,t)\,dV,
\end{equation}
where $| \Omega |$ is the volume of the cell $\Omega$. Then $f$ is the volume fraction of the reference phase in the cell.
The mixture properties of the cell may then be computed by arithmetic averages
\begin{equation}\label{eq:properties}
\rho_\Omega  = f \rho_l + (1 - f)\rho_g,
\qquad
\mu_\Omega  = f\mu_l + (1 - f)\mu_g.
\end{equation}
Spatial discretization of the above Equation is realized on a network of cubic cells obtained by a tree-like subdivision of an initial cubic cell of size $L_0$. The subdivision is realized using a wavelet based error estimate and is adapted dynamically as the simulation progresses. At all times the maximum subdivision level is a fixed number $\ell$ and the smallest grid size is 
\begin{equation}
    \Delta_\ell = 2^{-\ell} \textcolor{red}{L_0}. \label{eq:ell}
\end{equation}
From now on, we consider the algebraic equations derived from applying the Finite Volume Method on each cell. In this context, the approximate projection method by \cite{chorin68} can be used to solve the coupling between equations (\ref{eq:mass}) and (\ref{eq:NS-continuum}), considering that the velocity $\U$ is staggered in time with respect to the volume fraction $f$ and the pressure $p$. The discrete set of equations can then be expressed as
\begin{equation}\label{eq:NS-FVM}
\frac{\rho\U^{*} - \rho\U^{n}}{\Delta t} 
+ \nabla\cdot\left(\rho^{n + \frac{1}{2}}\,\U^{n+1/2}\U^{n+1/2}\right) 
=
\nabla\cdot [\mu^{n + \frac{1}{2}}(\mathbf{D}^n + \mathbf{D}^*)] 
+ \F_{\sigma}^{n+1/2}
\end{equation}
\begin{equation}\label{eq:Poisson}
\nabla \cdot \left(\frac{\Delta t}{\rho^{n+\frac{1}{2}}} \nabla p^{n + \frac{1}{2}}\right) = \nabla \cdot \U^*
\end{equation}
\begin{equation}\label{eq:correction}
\U^{n+1} = \U^* - \frac{\Delta t}{\rho^{n+\frac{1}{2}}} \nabla p^{n + \frac{1}{2}}
\end{equation}
In the above the expression $\nabla\cdot\left(\rho^{n + \frac{1}{2}}\,\U^{n+1/2}\U^{n+1/2}\right)$ must be interpreted as the use of  a predictor-corrector scheme for advection of the velocity field.  Both the predictor and the corrector use the Bell-Collela-Glaz scheme  \cite[]{popinet2003,popinet2009}, and involve two projections. A full  description of the predictor corrector scheme with projection is best found in the ``literate'' source code \url{http://basilisk.fr/src/navier-stokes/centered.h}.
In addition to these equations, $f$ is advanced in time using the split-volume-fraction advection scheme of \cite{Weymouth:2010hy}. 
We use the semi-implicit Crank-Nicholson time-stepping to compute the diffusive flux due to the viscous term in Equation (\ref{eq:NS-FVM}). An important aspect of the method is the numerical approximation of the surface tension force $\F_{\sigma}$. We use the Continuous Surface Force (CSF)  well-balanced formulation $\F_{\sigma} = \sigma \kappa \nabla c$ for Eq. \ref{eq:STF}  where we discretize $\nabla c$  at cell faces with the same scheme employed to compute the pressure gradient $\nabla p$. This is useful to reduce spurious currents provided curvature is accurately computed, as explained by \cite{Popinet2018,popinet2009}. \red{The interface curvature is computed using a complex method described in \cite{popinet2009}. When the grid resolution is sufficient 
second-order stencils based on height functions are used to compute curvature. In some large-curvature cases the height functions are not defined and fits by quadratic functions are used instead.} Equations (\ref{eq:Poisson}) and (\ref{eq:correction}) are the projection steps that will ensure continuity for the velocity field at time step $n + 1$. The mesh is adapted dynamically, by splitting the grid cells to eight smaller cells whenever the estimated local discretization error exceeds a set threshold (see Appendix \ref{app:adap-threshold}).
Details of the procedure can be found on the web site \url{basilisk.fr}, and in particular in the documented free code
\url{http://basilisk.fr/src/examples/atomisation.c}.

\red{In all the data reported in this paper, the velocities are expressed in
multiples of the jet velocity $U$.  All the numerical values for time are made dimensionless by $\tau_0 = L_0/(3 U)$ and all the reported lengths are made dimensionless by $L_0/3$. However in the theoretical discussions the dimensional values are used unless otherwise indicated. To avoid confusion we note with a $^*$ exponent the dimensionless values, so that
\begin{equation}
t^* = t/\tau_0 \qquad x^* = 3 x / L_0 \qquad \Delta^* = 3 (2^{-\ell}).
\end{equation} }
In addition to the aforementioned model and procedure, we implement the {\em manifold-death} procedure as described by \cite{Chirco22}. This procedure involves two primary steps: (1) the identification of thin liquid sheets with a thickness of $h_c$ or less, achieved through a local integration of the characteristic function $c$. This integration results in the calculation of the {\em signature} of a quadratic form, based on the local main inertia moments; (2) the creation of holes within the so identified thin sheets, using a probabilistic approach. This procedure is applied at a given frequency, defined by the user within this method. From a physical standpoint, this technique aligns with the weak spot model, wherein the likelihood of hole formation is exceedingly low for $h > h_c$ and substantially increases,  rapidly approaching certainty, for $h < h_c$.
\purp{Note that this procedure distinguishes thin sheets from other shapes such as ligaments, only the thin sheets are detected and perforated.  The specific implementation of the Manifold Death method we used is called the {\em Signature Method}. 
The complete manifold death code with the signature method can be accessed at \url{http://basilisk.fr/sandbox/lchirco/signature.h}.  }
\red{In the present case hole punching is attempted every time interval of $\tau_m=0.01 \tau_0$, that is the {\em raw dimensionless breakup frequency}
is $\tilde f_b^* = \tau_0/\tau_m = 100$. At these instants $N_b=200$ holes are attempted. We loosely term the pair of parameters $(\tilde f_b^* ,N_b)$ the {\em breakup frequency}. The breakup frequency chosen in this study is probably too high as will appear in the 
discussion below. However it is chosen to maintain the same parameters as in 
other studies, and possible improvements are discussed in the conclusion section.}
As we show in Appendix \ref{app:pheno_hole_expansion} the characteristic time for sheet thinning  is $\tau_c$ given by 
\begin{equation}
   \tau_c = \frac  D U \left( \frac {\rho_l }{\rho_g}  \right)^{1/2},
\end{equation}
\red{where $D$ is the jet diameter. 
From the definition 
of these quantities we can obtain the {\em breakup attempt frequency in units of the sheet thinning characteristic time}
$f_b = \tau_c/\tau_m$ 
\begin{equation}
   f_b = \frac{3 \tilde f_b^* D}{L_0}  \left( \frac {\rho_l }{\rho_g} \right)^{1/2} ,
\end{equation}
that is with the parameters of our simulation
\begin{equation}
   f_b = \frac{50}{3}  \left( \frac {\rho_l }{\rho_g} \right)^{1/2} .
\end{equation}
The numerical value is $f_b \simeq 87.9$ ensuring that punching attempts are sufficiently frequent as not to ``miss" transient thinning sheets. It is large enough to eliminate numerical breakup events but is likely to be too high as discussed below.}

Using this numerical method, we analyze the atomization of a circular jet injected at average velocity $U$ in a gas-filled cubical chamber of edge length $L_0$. We consider incompressible, isothermal flow. The dimensionless groups most relevant for this problem are the gas-based Weber number, the liquid-based Reynolds number and the  ratios
\begin{equation}\label{eq:dimensionlessPar}
\We_g =  \frac{\rho_g U^2 D}{\sigma},
\qquad
\Re_l =  \frac{\rho_l U D}{\mu_l},
\qquad
\rho^* =  \frac{\rho_l}{\rho_g},
\qquad
\mu^* =  \frac{\mu_l}{\mu_g},
\qquad
\frac{L_0}D,
\end{equation}
where $\rho_l$ and $\rho_g$ are the densities of liquid and gas respectively.
The viscosity coefficients are noted by $\mu_l$ and
$\mu_g$, and $\sigma$ is the surface tension coefficient. 
The boundary condition on the injection plane $x = 0$ imposes the no-slip condition everywhere except on the liquid section (${y^2 + z^2} < D^2 / 4 $) where the injection velocity is:
\begin{equation}\label{eq:BC}
u_x(t) = U\,\left[ 1 + A_p\sin\left(\omega_{p} U t/D \right)\right].
\end{equation}
On the remaining cube sides, we allow free outflow: $\partial_n \U_\Gamma = 0$ and $p_\Gamma = 0$.

The dimensionless parameters of our simulation are given in Table \ref{tab:simSetup}.
\begin{table}
\centering
\begin{tabular}{c c c c c c c}
\hline
\textbf{$\Re_{l}$}& $\We_g$ & \textbf{$\rho^*$} & \textbf{$\mu^*$} & \textbf{$A_{p}$} & \textbf{$\omega_{p}$} &  $L_0/D$  \\
\hline
5800 & 200  & 27.84 & 27.84 & 0.05 & $\pi$/5 & 18 \\
\hline
\end{tabular}
\caption{The dimensionless numbers characterizing our simulation. The Reynolds number  $\Re_l$ based on the liquid is rather moderate. The density and viscosity ratios are identical, which implies that  $\Re_g  = \Re_l$.}
\label{tab:simSetup}
\end{table}
These numbers are in the low range of the dimensionless numbers 
used in the above cited work. 
They are much smaller than the numbers of the spray G case of the Engine
Combustion Network \cite[see][]{duke2017internal}
used by \cite{zhang2020modeling}. 
The Reynolds number and density ratio are identical to those of \cite{menard07} and of the same order as those of \cite{Herrmann2011}. 
\red{The equality of the gas and liquid Reynolds numbers stems from the 
fact that the viscosity and density ratios are equal, so the ratio of 
kinematic viscosities is one.
However while the moderate density ratio is realistic in the context of Diesel jets, the viscosity ratio is not and would be much larger in a realistic setup, which would lead to much larger values of Re$_g$ that are expensive to resolve by DNS. This motivates the choice of viscosity ratio. We note that the same choice of equal viscosity and density ratios and moderate We$_{g}$ was made by \cite{Ling2017} who also studied atomization and PDF convergence in a simplified setup.
}
The Reynolds number is somewhat larger than that of \cite{shinjo2010simulation} while the gas Weber number $\We_g$ is  smaller than that of all the other papers in the literature. The rationale for selecting such a small Weber number is twofold. First the simplest prediction for the droplet size, based on the ratio of Bernoulli pressure $\rho_g U^2$ to surface tension, is $d = D \We_g^{-1}$.  Indeed if a purely inviscid flow with a velocity jump is considered, the cutoff wavelength for the Kelvin-Helmholtz instability scales as $D \We_g^{-1}$. The wavelength thus obtained is a natural scale for the subsequent formation of sheets and ligaments. 
The number of grid points in the diameter of a droplet of  this diameter $d$ is then  $\We_\Delta^{-1}$ 
where 
\begin{equation}
\We_l = {\rho_g U^2 l }/{\sigma} \label{wedef}
\end{equation}
is the Weber number based on length scale $l$ 
and $\Delta$ is the grid size. 
Similarly
$$
\Oh_l = \mu_l (\sigma \rho_l l)^{-1/2}
$$
is the Ohnesorge number based on scale $l$.
The values of $\We_\Delta$ for two of the grids we use are given in Table \ref{tab:delta}. We also give the value of the Ohnesorge numbers $\Oh_{\Delta}$  and  $\Oh_{h_c}$ which characterize the interplay of viscous and surface tension effects. In particular they
typify the regime of the Taylor-Culick rims \cite[]{song99,savva2009viscous} occurring at the edge of a sheet of thickness $\Delta$ or $h_c$. The $\Oh$ number of the thin sheets at the moment of hole formation is relatively large, implying a viscous Taylor-Culick flow and the absence of unsteady fragmentation as in \cite{wang2018unsteady} and \cite{kant2023bag}.

The second important consequence of selecting the relatively small value $\We_g =200$ is that it makes the simulation sit on the boundary between membrane type and fiber type defined in \cite{Lasheras00}. Membrane-type in that later paper refers to the formation of bags and sheets. Thus the prevalence of thin sheets discussed in this paper may be less marked at higher Weber numbers, the sheets being replaced by fibers, and also less important at smaller Weber numbers where the flow may be less unstable. 
\begin{table}
\centering
\begin{tabular}{c c c c }
\hline
Level $\ell$ & $D/\Delta$ & $\We_\Delta$ & $\Oh_{\Delta}$  \\
\hline
11 & 114  & 1.75 & 0.137 \\
15 & 1820 & 0.11 & 0.55 \\
\hline
\end{tabular}
\caption{Characteristic numbers related to the grid size. The level $\ell$ is defined in Eq. (\ref{eq:ell}).} 
\label{tab:delta}
\end{table}

\begin{table}
\centering
\begin{tabular}{c c c c}
\hline
          MD level $m$& $D/h_c$   & $\We_{h_c}$   & $\Oh_{h_c}$    \\   \hline
 12        & 227       & 0.88          & 0.194         \\
  13       & 455       & 0.44          & 0.27          \\
\hline
\end{tabular}
\caption{Characteristic numbers based on the critical sheet thickness in our simulations. The manifold-death level $m$ is defined in the text.} 
\label{tab:hc}
\end{table}

\section{Results}
\label{sec:results}
\subsection{Uncontrolled perforation: numerical sheet breakup}
\begin{figure}
    \centering
    \includegraphics[width=\textwidth]{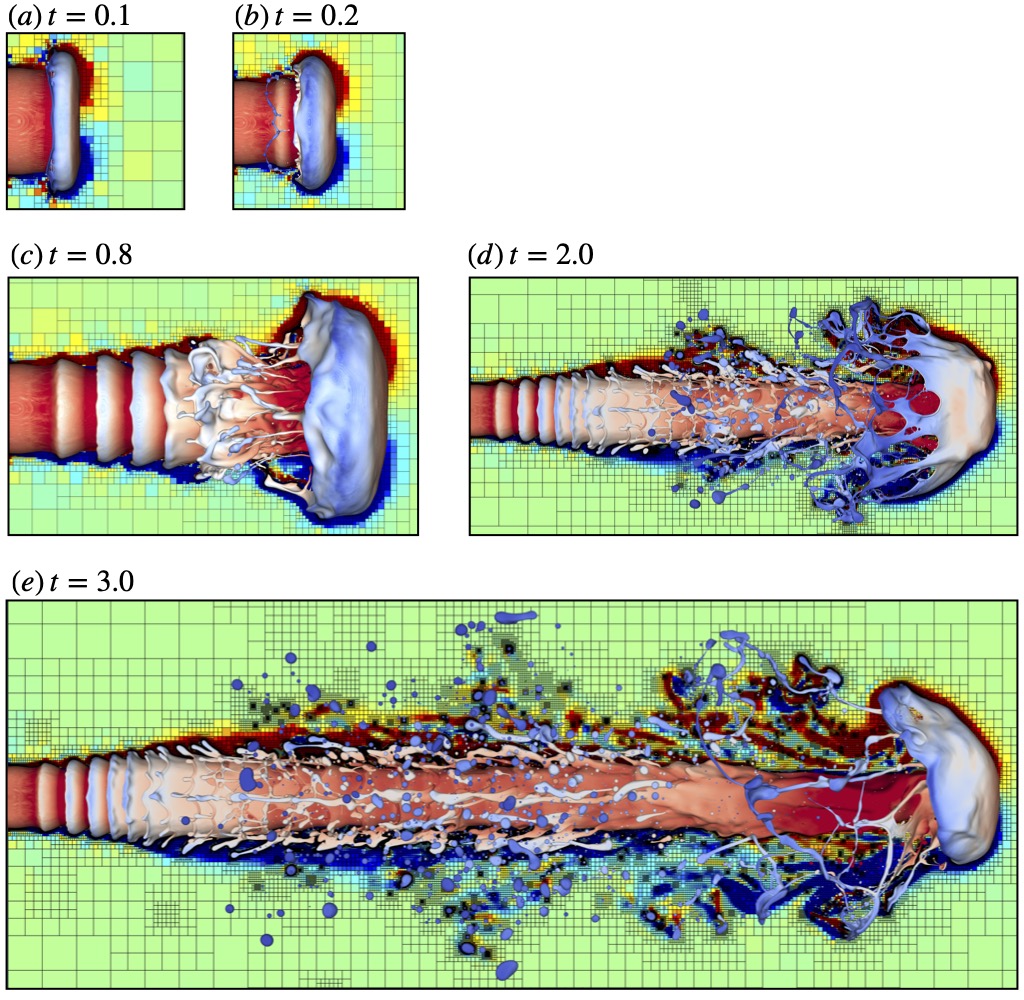}
    \caption{The advancing pulsed jet at various time instants $t$ and level $\ell=14$.  The fluid interface is colored by the axial velocity and the background is colored by the vorticity. The background also shows the mesh refinement. (a) The pulsed jet develops a mushroom head and a rim. In (b) the rim detaches. (c) Development of flaps coming from the sinusoidal pulsation. (d) Jet entering in a regime where effect of pulsation is lost at the mushroom head. (e) A fully developed jet and a rich spectrum of droplets and ligaments. Since $\Re_g$ is rather low at $5800$ there is relatively little vorticity away from the interface unlike in the case of \cite{kant2023bag}.}
    \label{fig:Illustration_basic_No_MD}
\end{figure}

Figure \ref{fig:Illustration_basic_No_MD} shows a global view of the jet evolution (see also Movie1, Movie2, Movie3, Movie4, Movie5 and Supplementary Material. (For the prepublication, see
\url{https://zenodo.org/doi/10.5281/zenodo.11110044}
). The dynamics is initialized with a slightly penetrated jet at $t=0$.  At $t=0.1$, a mushroom-like structure starts rolling up. The annular structure stretches and extends in the form of a ``corona flap" that develops behind the mushroom head. A rim forms at the edge of this flap then detaches. It is visible at $t=0.2$ as a corrugated and partially fragmented annular ligament. As the jet evolves pulsation results in a periodic series of corona flaps seen at $t=0.8$. These flaps interact with the ligaments coming from the mushroom head. At $t=2$, the jet has evolved further and we can identify an interval in the core structure  where the effect of pulsation is lost. Several mechanisms of droplet formation like sheet rupture, ligament breakup and their interactions with the jet core and corona flaps are seen. At $t=3$, the jet is fully developed and we see long axial ligaments with a marked velocity gradient made visible by the color change along the axis,  circular ligaments, corona fingers merging and sheet rupture. 
A large number of droplets and ligaments are produced in the mushroom tip region. 
Most of the phenomena illustrated above are also observed with controlled perforation, except the initiation of sheet rupture and annular ligament detachment. 

\begin{figure}
    \centering
    \includegraphics[width=\textwidth]{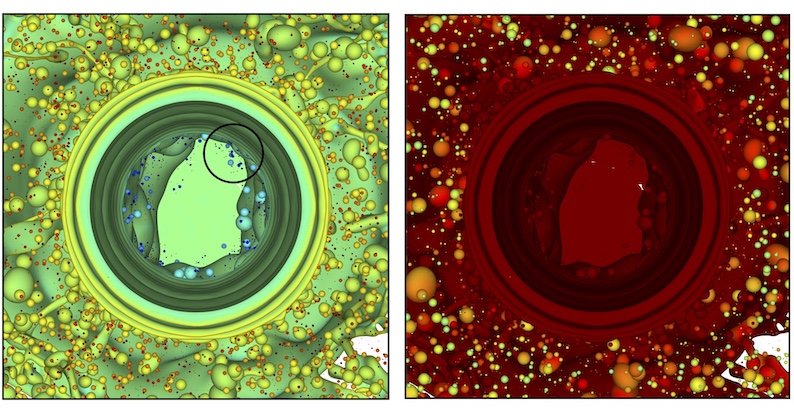}
    \caption{View from the inlet at time $t=3.04$ showing the inner region of the central core liquid jet. Left image is colored by the curvature showing the encapsulation of gas bubbles identified by the negative curvature (blue) in the liquid core encircled in the black circle. The droplets have positive curvature (red). The entrained bubbles travel with the core jet velocity and could also result in the formation of a few compound droplets during atomization or provide a physical breakup mechanism for thin sheets. The right image is the same as left one but colored by the axial velocity. The simulation corresponds to level $\ell=14$ with manifold death method applied at level $m=13$.}
    \label{fig:bubble_encapsulation}
\end{figure}

\begin{figure}
    \centering
    \includegraphics[width=\linewidth]{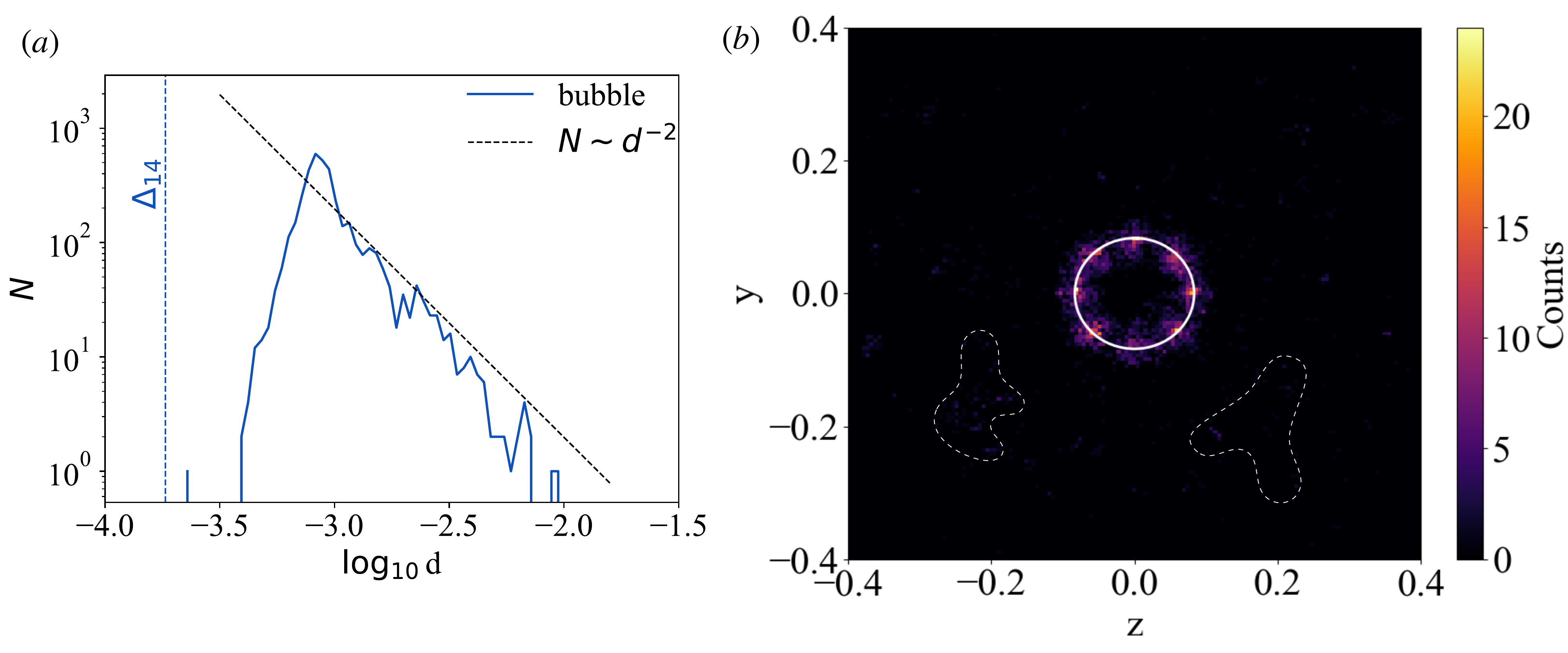}
    \red{\caption{(a)The bubble size distribution for the image shown in figure \ref{fig:bubble_encapsulation}. The number $N$ is defined in Eq. (\ref{definition:N}). (b) A 2D histogram for the same image showing bubble size distribution on the transverse coordinates. The jet is advancing in the x direction which is the axial direction. The solid white circle is the inlet region. The curly dotted white lines are drawn to highlight pale patches to indicate that some bubbles also exist outside the jet core indicating possible compound drops. \label{fig:bubble_enclosed_stats}}}
\end{figure}
Figure \ref{fig:bubble_encapsulation} shows a contrasting view of the jet with the camera direction aligned with the jet axis and the view positioned at the inlet. In Figure \ref{fig:bubble_encapsulation}a the objects are colored by curvature, so small droplets are red and small bubbles, with negative curvature, are blue. The bubbles have been trapped in the bulk or core of the jet or inside other liquid masses. The likely mechanism for the formation of a bubble is liquid mass collision, for example droplets and \red{wavelets of fluid mass} impacting on the jet core. Many such droplet or ligament impacts are seen in Movie1. A surprising aspect of the display is the relatively large number of droplets \red{and bubbles} seen.  In Figure \ref{fig:bubble_encapsulation}b the objects are colored by axial velocity, as with Movie1 and Movie3. A further interesting aspect of this view is to show the large dispersion in the axial velocities of the droplets. 

\red{At this point in the exposition it may be useful to define the number distribution and the probability density function (PDF) and to recall the connection between the two. We consider a fixed instant of time $t_0$. 
We partition the interval of droplet sizes
$(0,d_M)$ into a discrete set of $M$ intervals $[d_i, d_{i+1} = d_i + \delta_i]$. 
The index $i$ varies from 0 to $i_m$. The intervals are geometrically distributed, with 
$d_{i+1} = 1.047 d_i$, so 
$log(d_i)$ is evenly distributed in log space. 
The number of droplets 
with diameter in the interval $[d_i, d_i + \delta_i]$ is $n_{d,i} \delta_i$. The probability density function $p(d)$ is proportional to the expected value $\langle n_{d,i} \rangle$ as follows
\begin{equation}
    p(d_i) = \frac 1 {\cal N} \langle n_{d,i} \rangle ,  \label{definition:PDF}
\end{equation}
where the normalizing factor is
$$
{\cal N} = \sum_{i=0}^{M} \langle  n_{d,i} \rangle \delta_i.
$$
We show two kinds of plots, one with the number $N(d_i)$ of droplets in each bin $i$, that is 
\begin{equation}
    N(d_i) =  n(d_i) \delta_i , \label{definition:N}
\end{equation}and the other with the PDF $p(d)$.
Because of the difference between the definitions of $N$ (which is bin-size dependent) and $n$ and $p$ (which are both bin-size independent), there is a scaling relation 
\begin{equation}
    N(d) \sim d p(d) \sim d n(d) 
\end{equation}
between the two distributions. (We note that in the literature both $n$ and $N$ are reported with the name "number distribution"). 
The pulsed jet process is both deterministic \red{(except for the random character of sheet perforations)} and transient, so the value of  $n_{d,i}$ 
obtained at $t=t_0$ in a simulation is well defined. The expected value
$\langle n_{d,i} \rangle$ differs very little from the numerically obtained value  $n_{d,i}$ .  
Moreover in such a deterministic process no difference is expected between two realizations, that is two identical simulations should yield identical droplet counts. 
On the other hand in a jet with a random upstream injection condition instead of a pulsed injection condition, the simulation is not deterministic and the value of  $n_{d,i}$  obtained in a simulation at some time $t$  is fluctuating. The transient nature of the jet is also important. 
In a steady state simulation (which should be obtained some time after the jet starts flowing out of the simulation domain) initial perturbations are amplified chaotically and the system is again probabilistic. 
 Thus our pulsating initial condition simplifies somewhat the analysis by yielding a deterministic  instead of a probabilistic number density. }

\red{Figure \ref{fig:bubble_enclosed_stats} shows the number distribution of bubbles. The ordinate $N$ is the number $n_i$ of droplets defined in the text. We see that above the critical thickness, a distribution $N \sim r^{-\gamma}$ is seen. This distribution may have consequences in the sheet rupture and hence the drop size distribution. We have discussed the physical implications of bubble size distribution in Appendix \ref{app:pheno_hole_expansion}. In figure \ref{fig:bubble_enclosed_stats}b, we plot a 2D distribution of the bubble  size on a y-z plane. The white circle represents the inlet. The plot shows that most of the bubbles are near the outer boundary of the jet core where we see wavelet interactions and drop impacts the most (Figure \ref{fig:Illustration_basic_No_MD}c and Movie1). The distribution also shows eight characteristic high bubble-density zones. The occurrence of these zones and what decides their number is beyond the scope of current paper. We also see some bubbles present outside the jet core encircled in dotted white lines. This indicates that we do have a relatively small number of compound droplets.}
\label{subsec:numerical_rupture_VoF_No_MD}
\begin{figure}
    \centering
    \includegraphics[width=\textwidth]{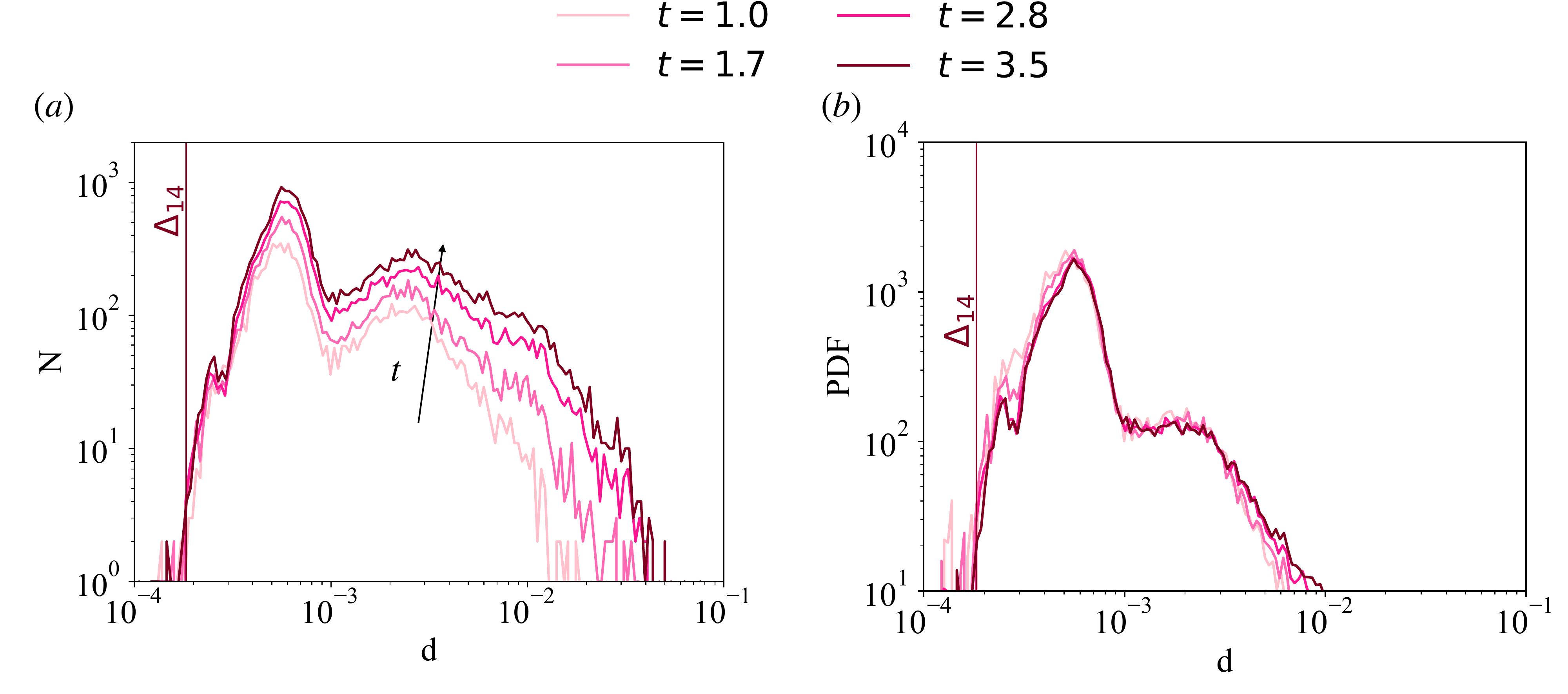}
    \caption{(a) The droplet-size number distribution ($N$ is defined in Eq. (\ref{definition:N})) and (b) the probability density function of the droplet diameter 
    (defined in (\ref{definition:PDF})) at various time instants. The plot shows that the PDF has converged in time $t \sim 2.5$.   This time convergence determines the choice of of the end time of the simulation at t = 3.5. Both (a) and (b) have 200 bins. The simulation corresponds to level 14 and the vertical dashed line represents the grid size.}
    \label{fig:PDF_Hosto_time_L14}
\end{figure}

We now discuss the statistical properties of the jet, in particular the droplet-size distribution in the case of uncontrolled sheet perforation or ``no manifold death" case (no-MD case). As the jet advances, the droplet-size distribution evolves. This is shown in Figure \ref{fig:PDF_Hosto_time_L14}. One can see that the droplet-size distribution has started to converge at $t=1.8$ and is approximately converged at $t=2.8$. The histogram thus only shifts in the vertical direction (the number of droplets), while maintaining the same qualitative nature. This histogram is bimodal with two peaks or local maxima. There is a large amplitude peak corresponding to a smaller droplet size (referred to as peak 1) and a smaller-amplitude one corresponding to a larger droplet size (referred to as peak 2). 
\begin{figure}
    \centering
    \includegraphics[width=\textwidth]{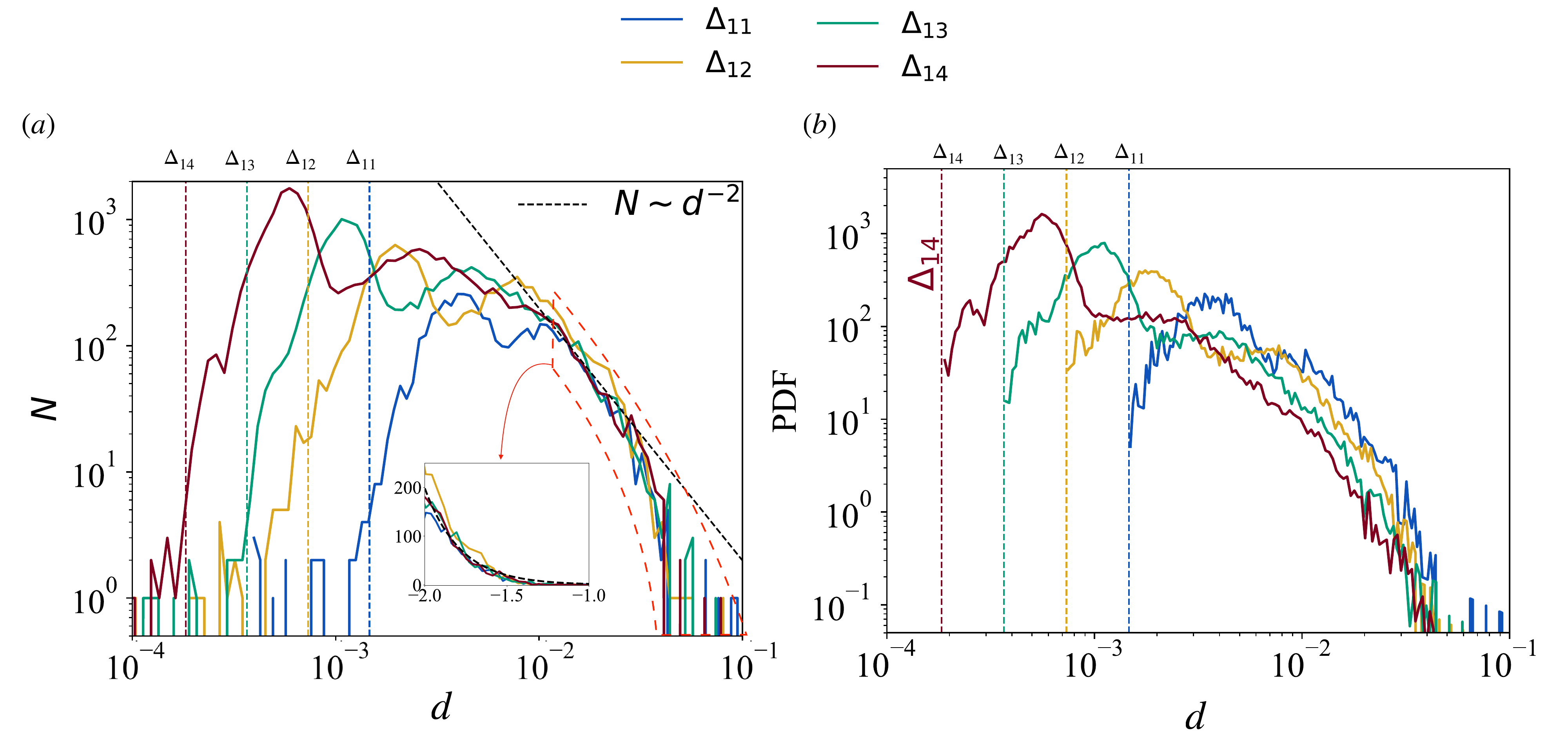}
    \caption{(a) The droplet-size distribution and (b) the probability density function of the droplet diameter $d$ at the final time $t=3.5$ for various grid resolutions. The grid sizes are shown with vertical dashed lines and $\Delta_{\ell}$ is defined as in (\ref{eq:ell}). The inset in (a) shows the converged tail region. All the plots are done with 200 bins. A Pareto distribution $N(d) \sim d^{-2}$ is added to compare with the converged region of the distributions.
    \label{fig:PDF_Hosto_lvl_t3p5}}
\end{figure}
We investigate the statistical convergence of the droplet size distribution by plotting in Figure \ref{fig:PDF_Hosto_lvl_t3p5} the histogram at $t=3.5$ when the jet is fully developed. We plot various grid resolutions or levels on that same Figure. The highest resolution 
in the no-MD case corresponds to the $\Delta_{14}$ line. This implies $D/\Delta = 920$ grid points per initial diameter. Figure \ref{fig:PDF_Hosto_lvl_t3p5} has two important characteristics:  (i) the histograms at various resolutions are qualitatively similar, that is bimodal with a major peak 1 at small scales and a minor peak 2 at large scales,  and (ii) despite such high resolution, it is clear that there is no convergence for the droplet size distribution. However the tail region of the distribution,  to the right of peak 2 shows some degree of convergence. In that tail region a Pareto distribution $n(d) \sim d^{-2}$ is seen reminiscent of such distributions found in other contexts \cite[]{bala2020,pairetti2021numerical} . 
\begin{figure}
    \centering
    \includegraphics[width=\textwidth]{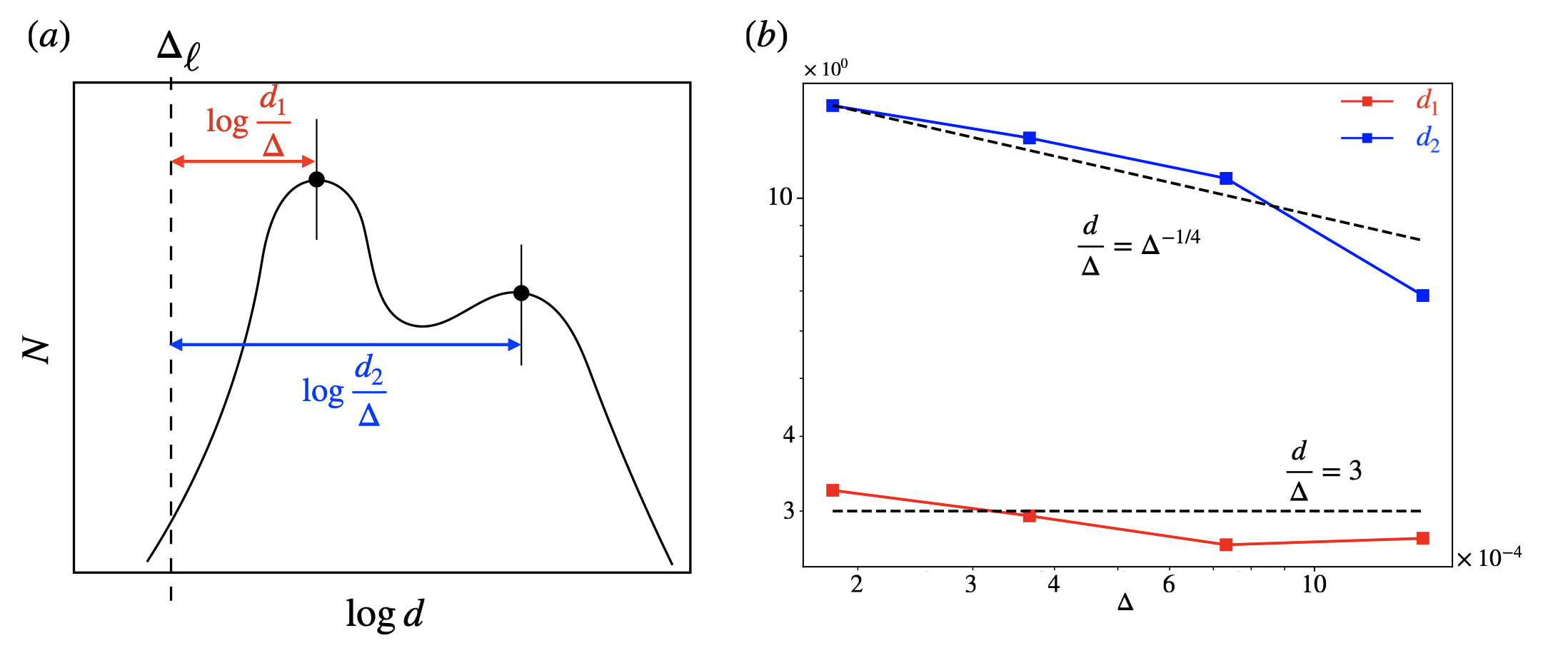}
    \caption{(a) Schematic of the bimodal droplet size distribution. The arrows indicate the logarithmic distance from the grid size $\log d_i/\Delta = \log d_i - \log \Delta$. (b) The values of $d_i/\Delta$ as a function of the maximum grid refinement level $\ell$ for the distribution of Figure \ref{fig:PDF_Hosto_lvl_t3p5}. The dashed line is at $3\Delta$ implying the droplets at peak 1 have a grid dependent diameter of $d_1 \sim 3\Delta$.} 
    \label{fig:peak_distances_No_MD}
\end{figure}
The dependence of the two peaks with the grid size is shown explicitly in Figure \ref{fig:peak_distances_No_MD}b where we plot the ratios  $d_i/\Delta_\ell$ corresponding to the peaks as a function of the maximum level $\ell$. We observe that the ratio for peak 1, $d_1/\Delta_\ell$ remains constant around a value of 3,  while for peak 2, $d_2/\Delta_\ell$
weakly increases with $\ell$. The grid dependence of these two peaks is synonymous with the absence of statistical convergence.  We will discuss possible explanations for the behavior of these two peaks below, and in particular the relation between peak 1 and  the curvature ripples appearing just before sheet rupture. We also discuss a mechanism and a possible scaling for peak 2 in Appendix \ref{app:pheno_hole_expansion}.
Since the grid dependence behavior is maintained at all resolutions,  the constant scaling of the diameter $d_1$ of peak 1 with the grid size implies that no matter how much we refine the grid the distribution of droplet sizes is grid-dependent. 

\begin{figure}
    \centering
    \includegraphics[width=0.9\textwidth]{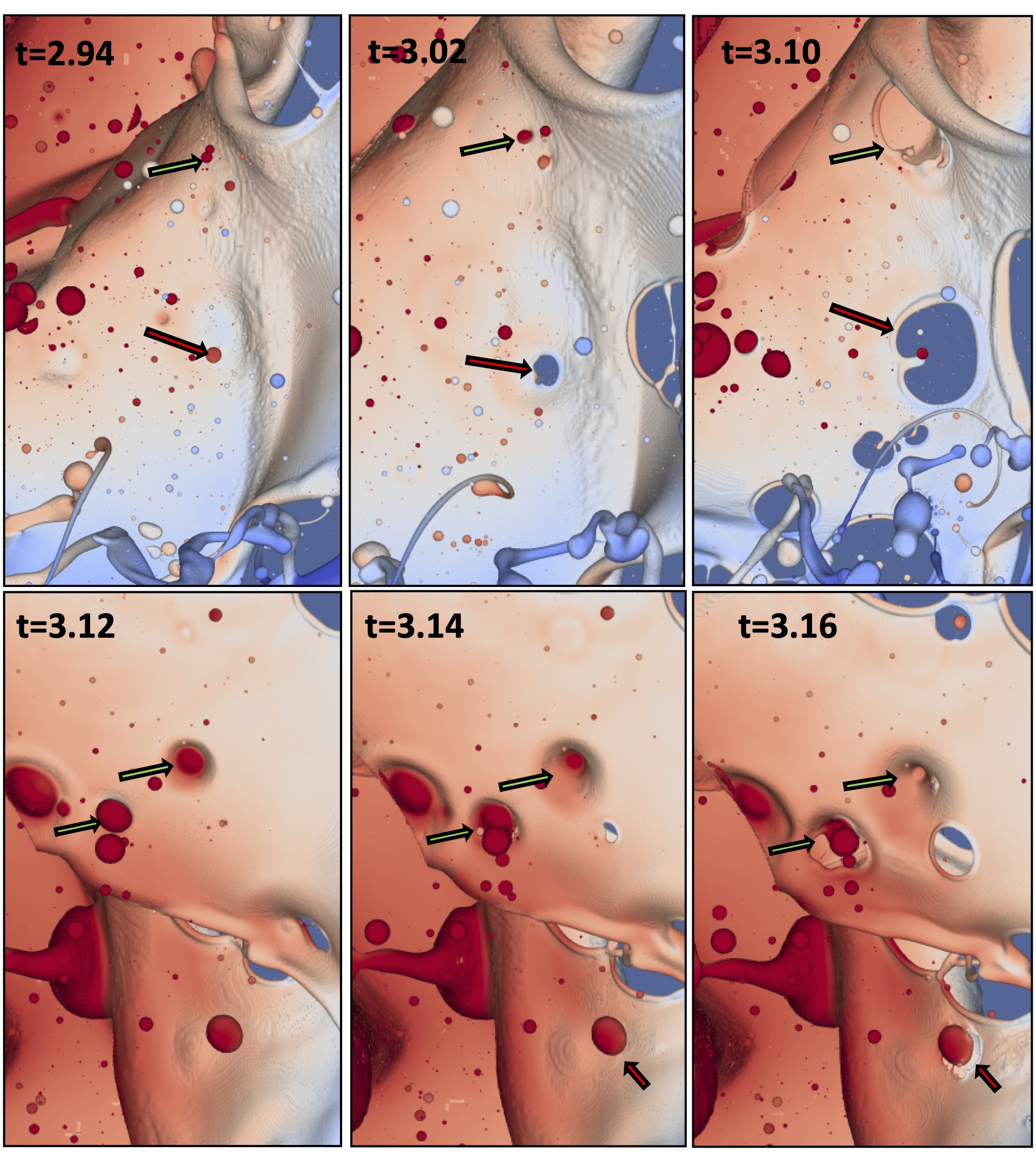}
    \caption{Droplet impacts on  the frontal liquid sheet. The points of interest are indicated by the arrows. Droplet impacts can result in holes with characteristic ligaments as seen at $t=3.10$.  In some cases droplets coalesce into the sheet seen at $t=3.12$ and $t=3.14$. At $t=3.16$, the sheet is ruptured  but the droplet is still identifiable. All simulations are for maximum level $\ell = 13$. The manifold death method is not applied. The interface is colored by axial velocity.  The artefacts or corrugated surfaces particularly visible at t=3.02 and t=3.10 at the top right are not aliasing artefacts (due to an approximate isosurface interpolation) but are representative of the curvature oscillations seen in the next figures. }
    \label{fig:morphology_drop_impact} \label{fig:sheet_droplet_impact}
\end{figure}

\begin{figure}
    \centering
    \includegraphics[width=\textwidth]{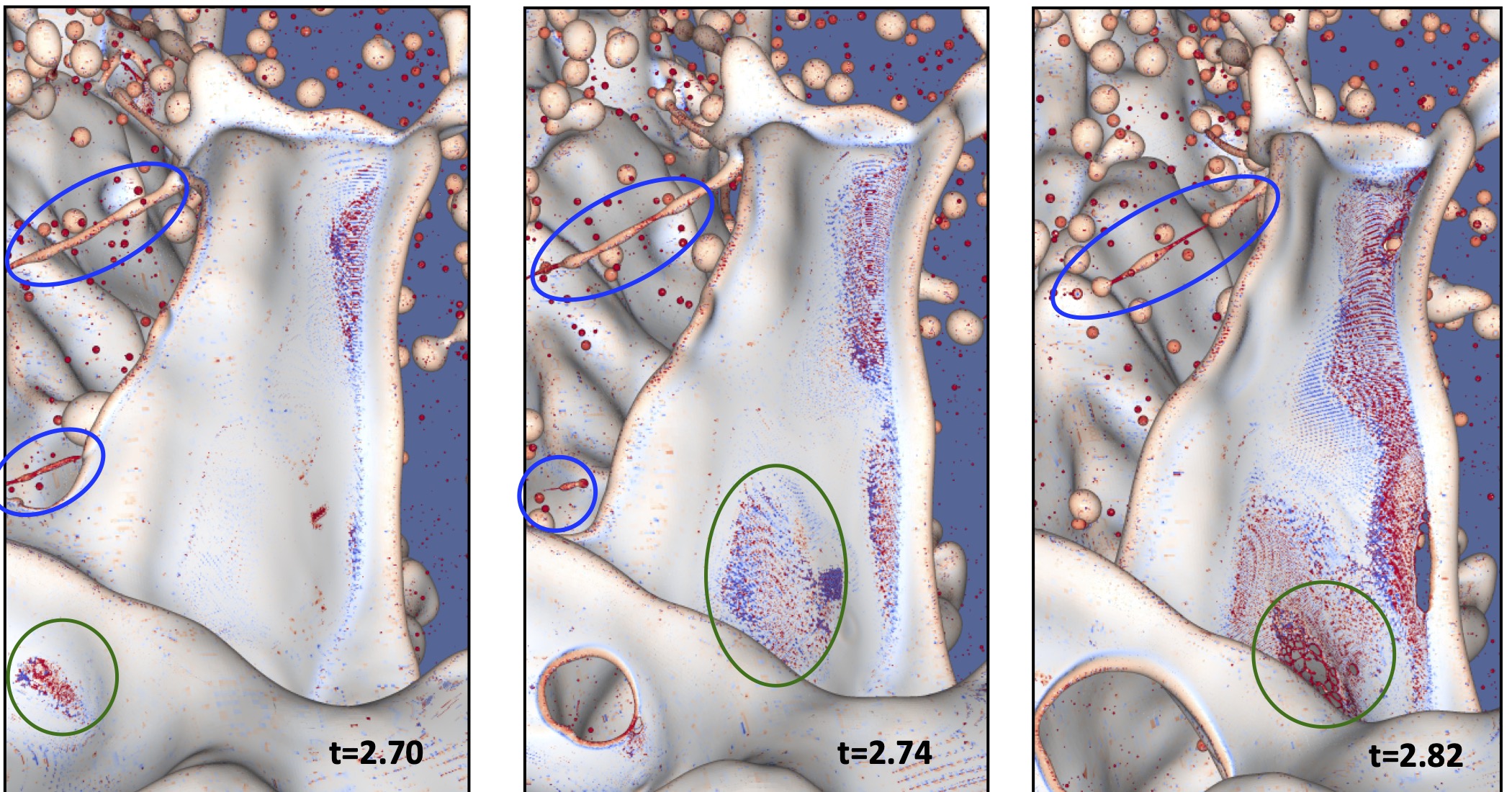}
    \caption{Appearance and evolution of the curvature ripples on the interface. The interface is colored by the interface curvature. These ligaments are colored a darker red and the interfaces on the thin sheets are closer to white. Ligament rupture is encircled in blue. The rupture or ligament pinch off appears in a darker red color. Sheet rupture, encircled in green, displays curvature ripples in the form of red-blue oscillations. At time $t=2.82$, the green-circled rupture region displays a grid-dependent ligament network that has evolved from these ripples. These images correspond to a level $\ell = 14$ simulation. (See also Movie5.)}
    \label{fig:kappa_ligament_sheet_flat_No_MD}
\end{figure}

The mechanism leading to peak 1 is related to the perforation of thin sheets. This perforation is observed to occur in two contrasted ways.  
A simple mechanism is the impact of small droplets on the liquid sheet. The droplets are formed by breakup previously occurring elsewhere in the simulation. The impacts yield in some cases the formation of an expanding hole, as shown on Figure \ref{fig:sheet_droplet_impact}. 
We note that such impacts have already been observed in previous studies such as those of \cite{menard07} and \cite{shinjo2010simulation}.
Not all droplet impacts create perforations. In Figure \ref{fig:morphology_drop_impact}, at time $t=2.94$, we track the droplets indicated by arrows. We see that droplet impacts often create holes with a characteristic finger-like ligament inside the hole. Impacts are shown by arrows at time $t=3.10$ in Figure \ref{fig:morphology_drop_impact}. In the lower panel at $t=3.12$ and $t=3.14$ some droplet impacts do not result in holes but that instead droplets merge with the sheet. 
This hypothesis was confirmed by repeated observations of the sheets at various angles and by visioning Movie5. The red arrow from $t=3.14$ and $t=3.16$ captures an interesting moment, where the droplet impact and sheet rupture happen simultaneously and it is unclear if the hole was created by the numerical rupture process discussed below or by droplet impact.

Another mode of thin sheet perforation, much more complex and purely numerical, is illustrated on Figure \ref{fig:kappa_ligament_sheet_flat_No_MD} as well as Movie5.
The curvature coloring of the interface helps us identify the perforation spots.
We see high frequency oscillations of the curvature field in form of alternating red and blue colors. These indicate the ripples originating at the fluid interface before the sheet rupture happens. Finally, at $t=2.82$, we see a dense network of small-diameter ligaments, scaling with grid size.
Another mechanism that may contribute to peaks 1 and 2 is the rupture of ligaments, also shown on Figure \ref{fig:kappa_ligament_sheet_flat_No_MD}.
Small diameter ligaments have large curvature and are thus easily identified by their color. They break physically and not numerically by the Rayleigh-Plateau instability. The origin of these ligaments themselves if often an expanding hole whose rim collides into other rims. 

We note that the high-frequency pressure oscillations are due in part to the use of the Height-Function method to compute curvature. The Height-Function method ceases to work when the sheet is too thin. However, other methods will also fail as represented in Figure \ref{nicefig}.
\begin{figure}
    \centering
    \includegraphics[width=\textwidth]{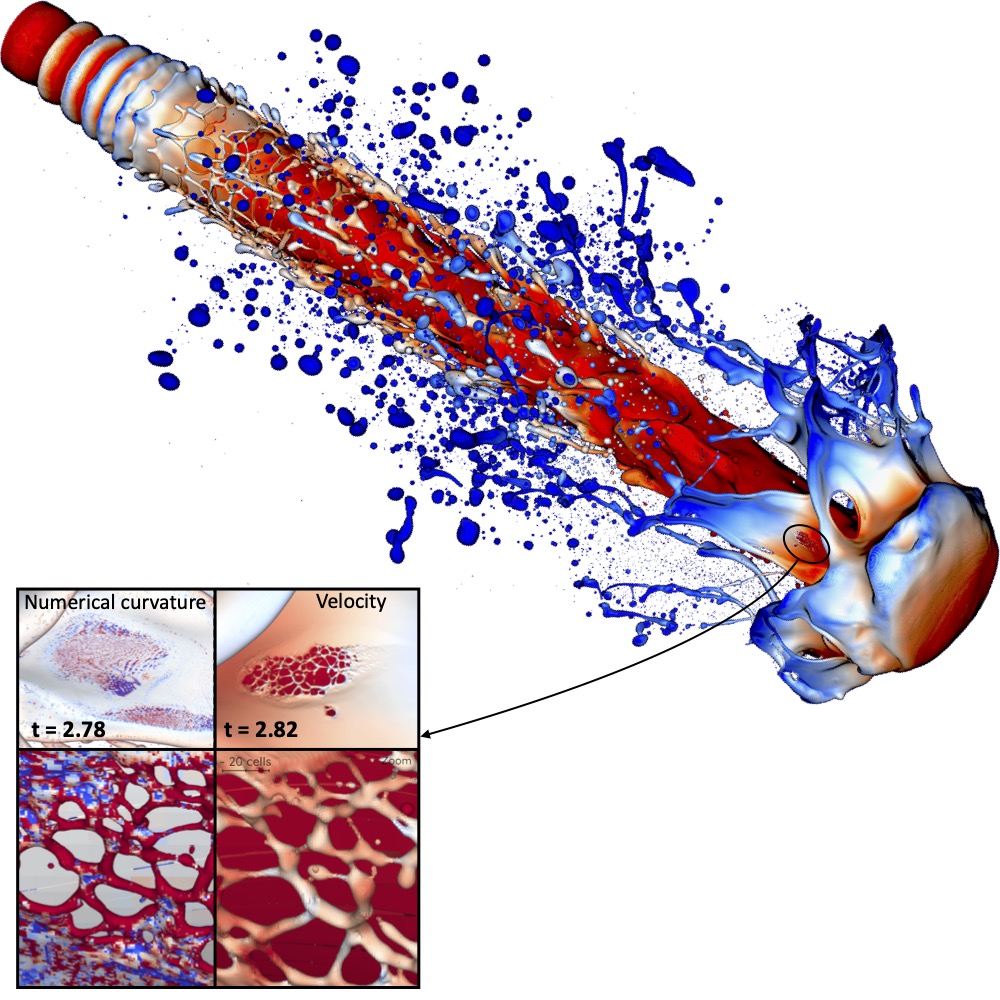}
    \caption{The fully-developed jet at $t= 2.8$. The inset shows a zoom-in at the sheet rupture spot showing the curvature ripples in the weak spot about to be punctured and the resulting ligament network. This ligament network eventually produces grid-dependent droplets. The simulation shown here is at level $\ell=14$.}
    \label{fig:sheet_fully_dev_No_MD}
\end{figure}
\begin{figure}
    \centering
    \includegraphics[width=\linewidth]{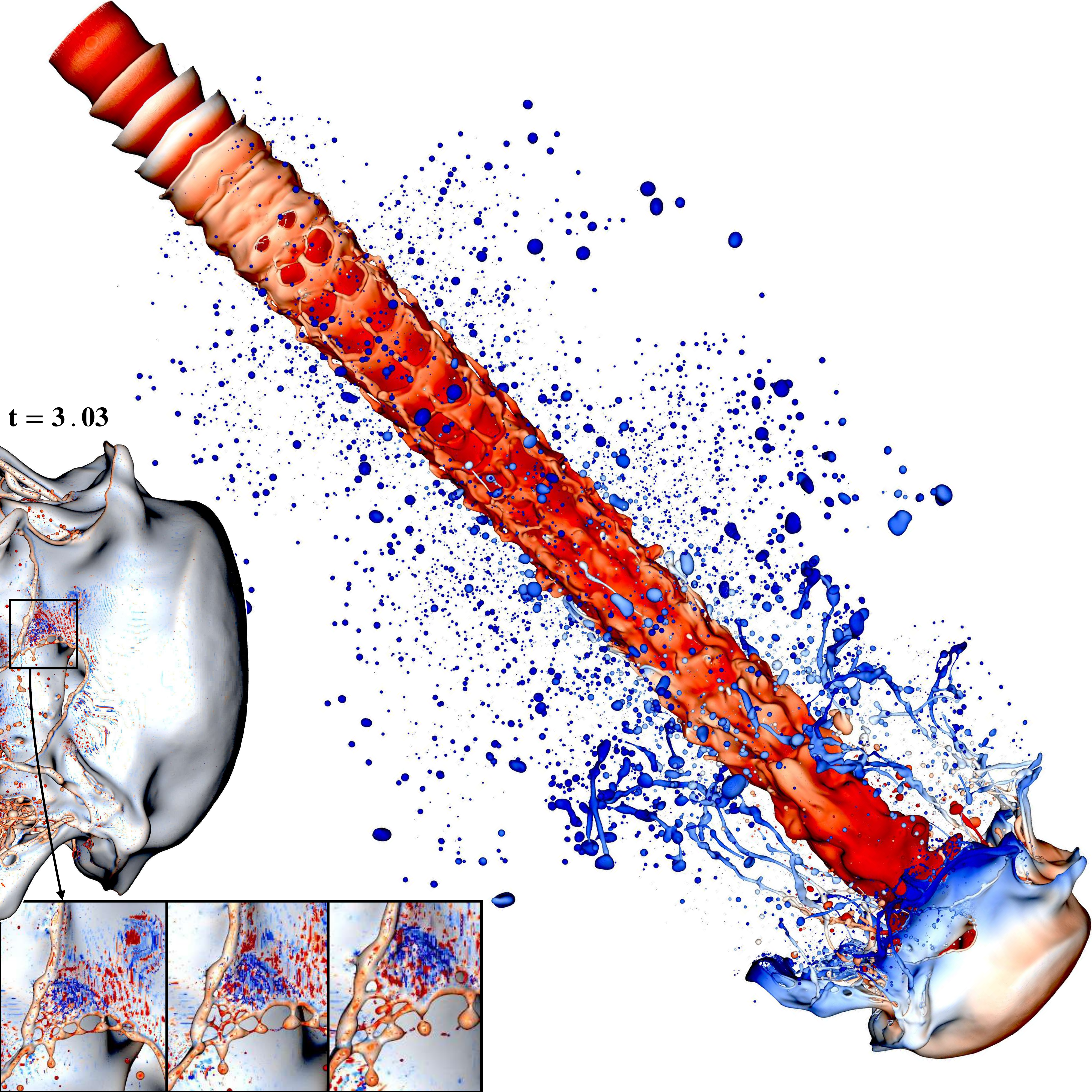}
    \caption{The fully-developed jet at $t= 3.03$. The inset shows a zoom-in at the sheet rupture spot showing the curvature ripples in the weak spot about to be punctured. The simulation shown here is at level $\ell=13$.      \label{fig:fully_dev_No_MD_FRONT_LVL13}}
\end{figure}
\begin{figure}
    \centering
    \includegraphics[width=\linewidth]{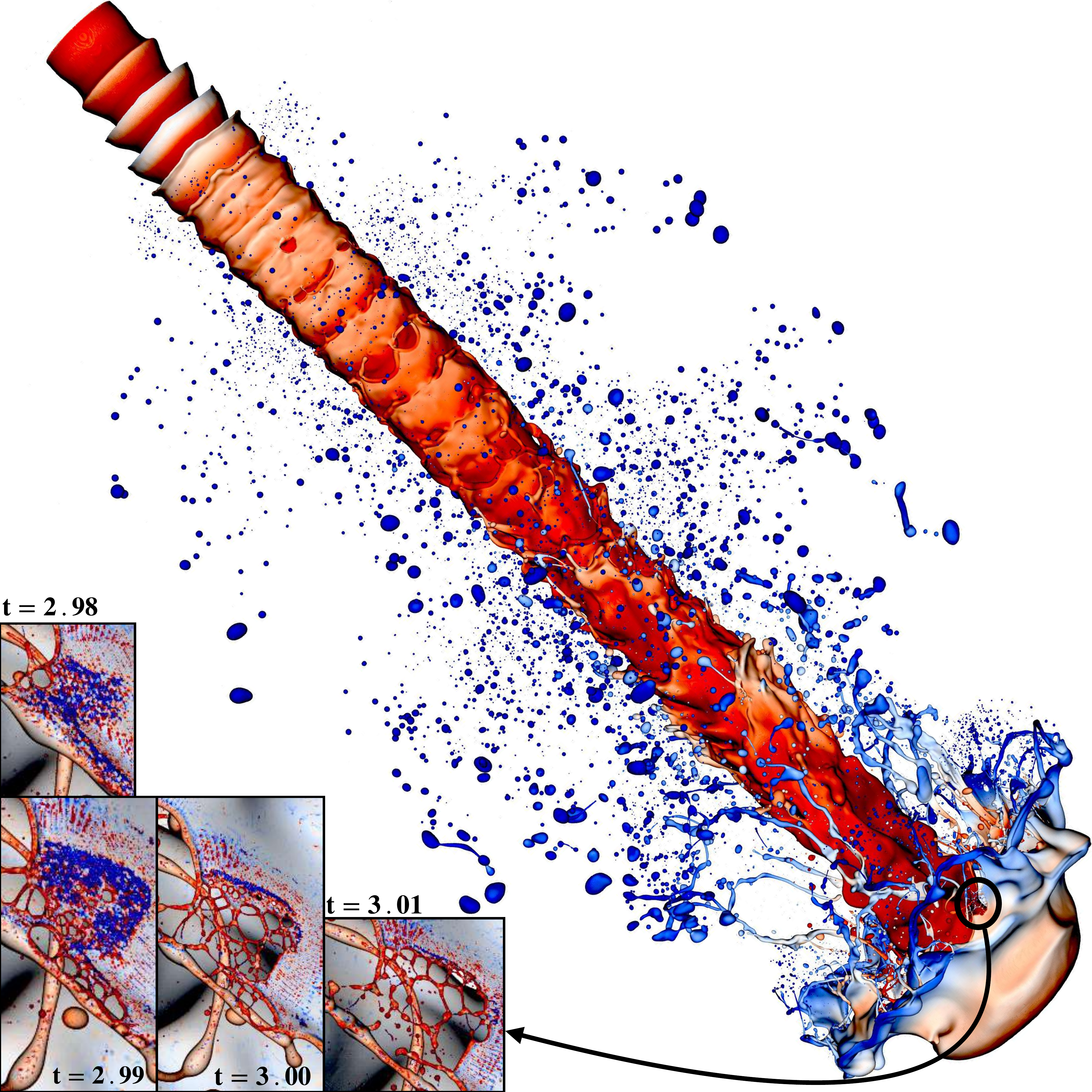}
    \caption{The fully-developed jet at $t= 3.01$. The inset shows a zoom-in at the sheet rupture spot showing the curvature ripples in the weak spot about to be punctured. The simulation shown here is at level $\ell=14$.}
    \label{fig:fully_dev_No_MD_FRONT_LVL14}
\end{figure}
\begin{figure}
    \centering
    \includegraphics[width=\textwidth]{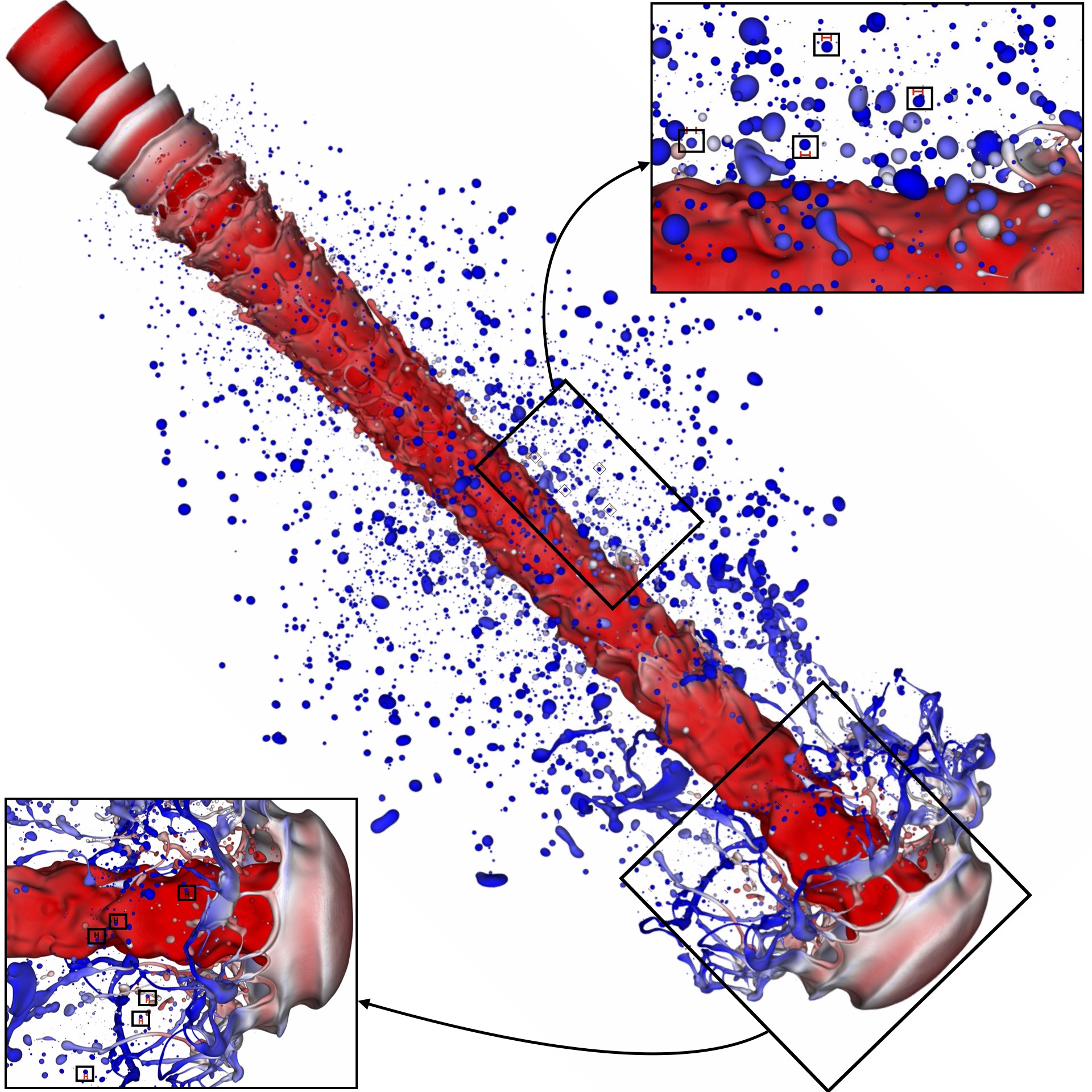}
    \caption{The $(14,13)$ jet at $t= 3.1$ with Manifold Death (MD) applied. The basilisk level $\ell =14$ and MD level is $13$. 
    The insets and the boxes indicate droplets that are near the maximum size of the distribution in the converged region.}
    \label{fig:sheet_fully_dev_with_MD}
\end{figure}
To investigate the small-scale ligament networks, we zoom around such a breakup event in Figure \ref{fig:sheet_fully_dev_No_MD}.  We see that a weak spot develops and curvature ripples appear at $t=2.78$ as soon as the sheet reaches a thickness a few times the cell size as shown in the inset. These curvature ripples then give rise to ligament networks inside the expanded hole eventually leading to the grid-dependent droplets. To summarize, as the sheet reaches a thickness of order $ 3 \Delta$, the rupture mechanism is as follows: \textit{thin sheet regions} $>$ \textit{curvature ripples} $>$ 
\textit{ligament networks} and \textit{tiny droplets} $(d \sim 3\Delta)$. 
Since this sequence of events is inherent to the numerical breakup of thin sheets, 
it is impossible to escape its occurrence by increasing the grid refinement, although refinement can delay it.
The numerical sheet breakup identified here is the direct cause of the lack of convergence of peak 1.  Indirectly, the formation of droplets and ligaments of a given size in a non-converged manner leads to all the droplet size counts to be unconverged as these other size droplets are formed through coalescence and breakup events from the peak 1 droplets and similarly scaled ligaments. 

\red{We also note that the point-of-view and region of Fig. \ref{fig:morphology_drop_impact} were selected to show a large number of drop impact situations, which are however scarcer than the numerical breakup of sheets.}

\purp{To give an overall idea of the evolution of the jet, we show the fully evolved jet at $\ell = 13$ in Figure \ref{fig:fully_dev_No_MD_FRONT_LVL13} and the same fully evolved jet but at $\ell = 14$ in Figure \ref{fig:fully_dev_No_MD_FRONT_LVL14}. 
The mist of tiny droplets near the mushroom head and near the inlet flaps is more pronounced in the $\ell = 14$ case than in the $\ell = 13$ case. 
The $\ell=14$ case retains more large drops and deformed ligaments at mid-length compared to the $\ell=13$ case. The inset zoom shows the numerical sheet rupture. The ligament network structure of is more clearly seen at  $\ell = 14$ than at $\ell=13$.}

\subsection{Controlled perforation by the manifold death method}
\label{subsec:Manifold_Death_rupture_results_MD}
We now present the results of atomization when we apply the manifold death (MD) method. Here we adopt the following conventions.
\begin{enumerate}
    \item The octree maximum refinement level $\ell$ defines the finest grid size as given in Equation (\ref{eq:ell}).
    \item The manifold death level $m$ implies that the critical thickness for punching holes is given by $h_c = 3 \Delta_{m}$.
\end{enumerate}
As an example, the notation $(\ell,m)=(13,12)$ would mean that the finest cell size in the simulation is $\Delta_{13}$ and the critical thickness at which holes are being punched is $h_c = 3\Delta_{12} = 3\times (2\Delta_{13}) = 6 \Delta_{13}$.  This convention for the definition of  the ``manifold death level $m$" maintains consistency with the comments in the code \url{http://basilisk.fr/sandbox/lchirco/signature.h}.
The extent of the integrated region in the quadratic form computation leads to the correspondence $h_c = 3 \Delta_m$. 
\begin{figure}
    \centering
    \includegraphics[width=\textwidth]{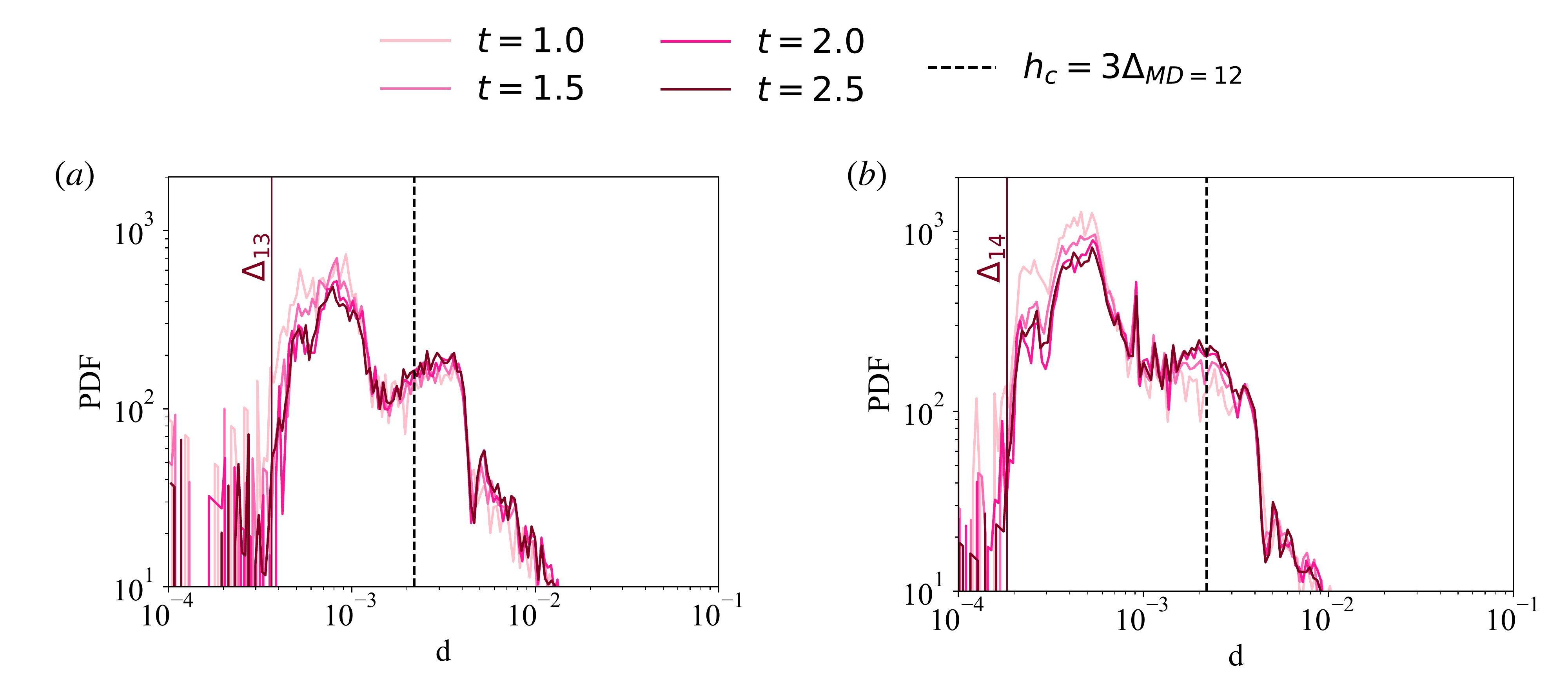}
    \caption{The probability density function at various times $t$ for the droplet size distribution when manifold death is applied. The PDF has converged  in time  at $t=2.5$. (a) PDF at maximum level $\ell=13$ and MD level $m=12$ and (b)  the PDF at maximum level $\ell=14$ and MD level $m=12$. The critical hole punching thickness $h_c = 3 \Delta_{m=12}$ is shown as the black dashed line and is the same for both plots.}
\label{fig:MD_stats_PDF_all_time_conv}
\end{figure}
\purp{The overall evolution of the jet shows important variations with the no-MD case. Comparing Figure \ref{fig:sheet_fully_dev_with_MD} MD $(14,13)$ to Figure \ref{fig:fully_dev_No_MD_FRONT_LVL13} ($\ell=13$) and Figure \ref{fig:fully_dev_No_MD_FRONT_LVL14} ($\ell=14$), 
i) the mist of tiny droplets near the mushroom head has disappeared in Figure \ref{fig:sheet_fully_dev_with_MD},
ii) the mushroom-head corona-flap sheet is more affected by holes and breakups and is closer to Figure \ref{fig:fully_dev_No_MD_FRONT_LVL14} $\ell=14$ than to Figure \ref{fig:fully_dev_No_MD_FRONT_LVL13} $\ell=13$,
iii) the cloud of droplets at mid-length is similar to both no-MD cases but is more uniform and closer to the Figure \ref{fig:fully_dev_No_MD_FRONT_LVL13}  $\ell=13$ case.} 

Figure \ref{fig:MD_stats_PDF_all_time_conv} shows the evolution of the probability density function of the droplet size distribution with time. Similar to the no-MD plots, we see that here also the statistics are converged in time as soon as $t=2.5$. When compared to Figure \ref{fig:PDF_Hosto_lvl_t3p5}, we see that the peak corresponding to the smaller droplet size has a smaller amplitude. 

\begin{figure}
    \centering
    \includegraphics[width=0.9\textwidth]{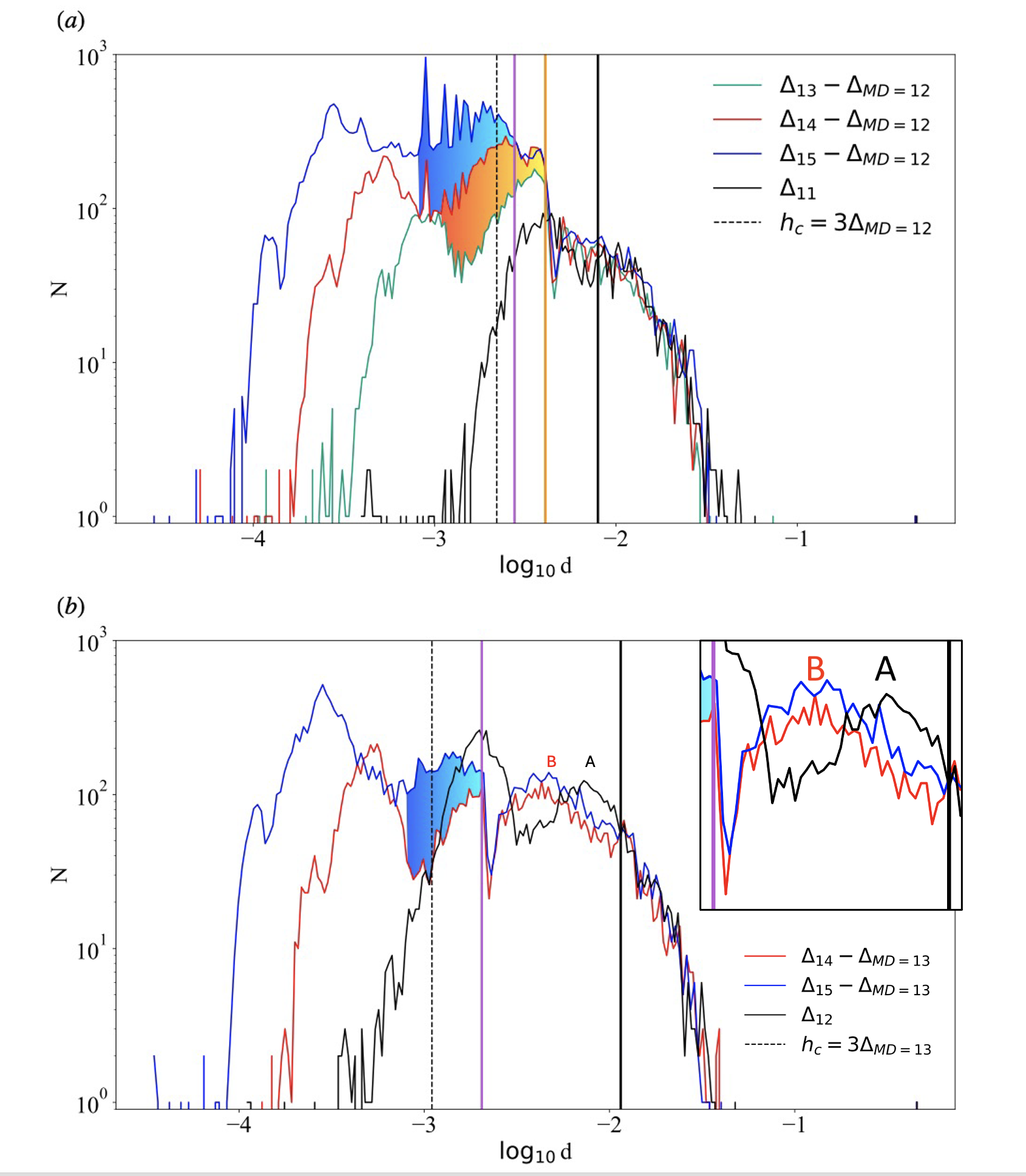}
    \caption{The droplet size distribution from simulations using the manifold death method. Each curve is labeled  $\Delta_{\ell}-\Delta_{MD=m}$, where $\ell$ is the grid level and $m$ is the the MD level. The dashed black vertical line represents $h_c = 3 \Delta_{m}$, that is the critical thickness of punching holes. The solid vertical lines indicate the values of the converged diameter $d_c$, corresponding to the smallest diameter above which the difference between subsequent distributions is not significant. All curves in (a) have  fixed $h_c = 3\Delta_{12}$ and all curves in (b) have fixed $h_c = 3 \Delta_{13}$. Note that the $\Delta_{11}$ curve in (a) and the $\Delta_{12}$ curve in (b) indicate the no-MD method and roughly correspond to respectively $h_c = 3\Delta_{MD=12}$ and  $h_c = 3 \Delta_{MD=13}$, given the breakup due to lack of sheet resolution in VOF simulations. The shading highlights the departure from the converged region. We see that $d_c$ approaches $h_c$ as the grid size is reduced. The inset shows the important change in the typical sizes of droplets in the first peak above $h_c$ as MD is applied. Peak A corresponds to peak 2 in the no-MD distributions of Figure \ref{fig:PDF_Hosto_lvl_t3p5} while peak B, with much smaller droplets, to droplets seen in the MD Figure \ref{fig:sheet_fully_dev_with_MD}.}
\label{fig:MD_stats_histo_all_LVL_conv}
\end{figure}

We  show the number density $n(d)$ in Figure \ref{fig:MD_stats_histo_all_LVL_conv}.
Figure \ref{fig:MD_stats_histo_all_LVL_conv}a shows the histograms at various grid resolutions for a fixed manifold death threshold $h_c = 3 \Delta_{12}$. The distribution is again bimodal, but the peak corresponding to the smaller droplets is weaker. We also observe a clear converged region in the interval
$[d_c, d_{i_m+1})$.  where $d_c$ is the minimum converged equivalent droplet diameter or  ``converged diameter" for simplicity. The extent of the converged region increases, that is $d_c$ decreases as the grid size $\Delta_l$ decreases. In Figure \ref{fig:MD_stats_histo_all_LVL_conv}a the vertical line marks $d_c$ for two consecutive grid sizes. We have shaded the onset of the non-converged region between two consecutive grid refinements for clarity. 

In practice, the ``converged diameter" $d_c$ is determined from the comparison of a pair
of distributions from simulations of two different grid sizes $\Delta_l$ . We note this pair
\{$\Delta_{l_1} - \Delta_{l_2}$\}. 
In  Figure \ref{fig:MD_stats_histo_all_LVL_conv}a  the orange vertical line marks $d_c$ for the pair \{$\Delta_{14} - \Delta_{13}$\}. 
Similarly, the purple vertical line marks  $d_c$ for the pair \{$\Delta_{15} - \Delta_{14}$\}. 
Some additional information may be obtained by
investigating a no-MD case. In such a case, there is a typical 
sheet thickness at breakup $h_e$ that is comparable to the controlled breakup for $h_c \simeq h_e$. 
Indeed the critical breakup film thickness $h_e$ for no-MD 
is $h_e \simeq 1.5 \Delta$, 
Then the above case $h_c = 3\Delta_{12}$ with manifold death
has a critical sheet thickness equal to, per expression
(\ref{eq:ell})  to  $h_e = 1.5 \Delta_{11}$ 
in the no-MD case at level 11. 
Thus the no-MD simulation at  $\ell=11$ is a rough equivalent
to the three simulations with MD level $m=12$.
The vertical black line is the converged thickness $d_c$
for  the pair \{$\Delta_{13}, \Delta_{11}$\} where the first simulation is MD and 
the second is no-MD. One sees 
on Figure \ref{fig:MD_stats_histo_all_LVL_conv}a that the converged diameters, indicated by the solid  vertical lines 
decrease and tend to approach 
the critical sheet thickness $h_c$ for MD breakup, 
represented by the dashed vertical line. 
We now show the droplet size distribution and converged region for a smaller critical sheet thickness $h_c = 3 \Delta_{13}$.
In Figure \ref{fig:MD_stats_histo_all_LVL_conv}b we see similar results 
but with only two instead of three levels $l=14$ and 15. 
Since a simulation at $\Delta_{15}$ is already quite expensive (it implies an equivalent of $35$ trillion cells for a uniform grid with $\Delta_{15}$), we did not perform a third simulation at $l=16$ in that case. A zoomed-in view for the converged region $d_c$ for each consecutive pair is shown in Appendix \ref{appex:d_c_ZOOM_IN}. 

\red{The closely packed peaks around $log(d)=-3$ in Fig. \ref{fig:MD_stats_histo_all_LVL_conv}a, mostly visible for $(\ell,m)=(15,12)$ are related to a process that produces a narrow spectrum of small droplets of size $\bar d$ with 
$\Delta_{15} < \bar d < h_c$. These small droplets coalesce at least two or three times to produce additional small droplets of size  $2\bar d, 3\bar d$ etc. We are unsure of the mechanism producing these initial droplets of well-defined diameter $\bar d$ and why it appears most clearly for the pair ($15,12$).}

\begin{figure}
    \centering
    \includegraphics[width=\textwidth]{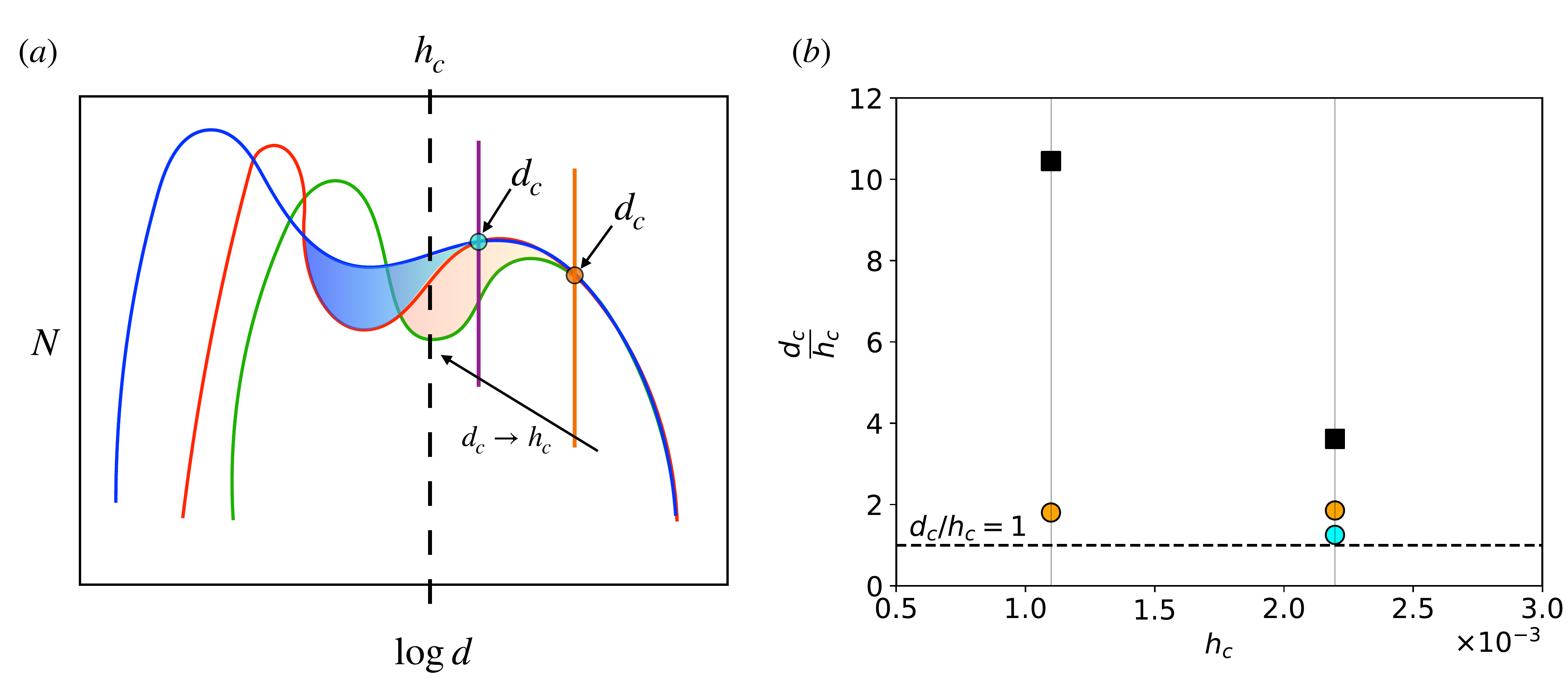}
    \caption{(a) Schematic of the comparison of two simulations and how it allows to define the ``converged diameter" $d_c$.  It is seen that $d_c$ approaches $h_c$ as the simulations are refined. (b) Plot of $d_c/h_c$ versus $h_{c}$ corresponding to the $d_c$ values of Figure \ref{fig:MD_stats_histo_all_LVL_conv}. Along a fixed $h_c$, the $d_c/h_c$ points move closer to unity as the maximum level is increased. The black squares corresponds to the No MD case while all circles are for the cases with MD. 
    }
    \label{fig:h_c_d_c_plane_movement}
\end{figure}

The variation of the converged diameter $d_c$ as 
a function of $h_c$ and $\Delta_\ell$ is shown explicitly in Figure \ref{fig:h_c_d_c_plane_movement}. Figure \ref{fig:h_c_d_c_plane_movement}a illustrates schematically how the converged diameter approaches $h_c$ as
$\Delta_\ell$ is reduced. 
Figure \ref{fig:h_c_d_c_plane_movement}b shows the data in the $d_c/h_c, h_c$ plane. Each data point in that plane corresponds to a simulation pair discussed above such as \{$\Delta_{14} - \Delta_{13}$\}, and to a  vertical line in Figure \ref{fig:MD_stats_histo_all_LVL_conv}.
Following a vertical line in Figure \ref{fig:h_c_d_c_plane_movement}b, one can see 
$d_c$ tending to $h_c$  as the grid is refined for a fixed $h_c$.

\begin{figure}
    \centering
    \includegraphics[width=\textwidth]{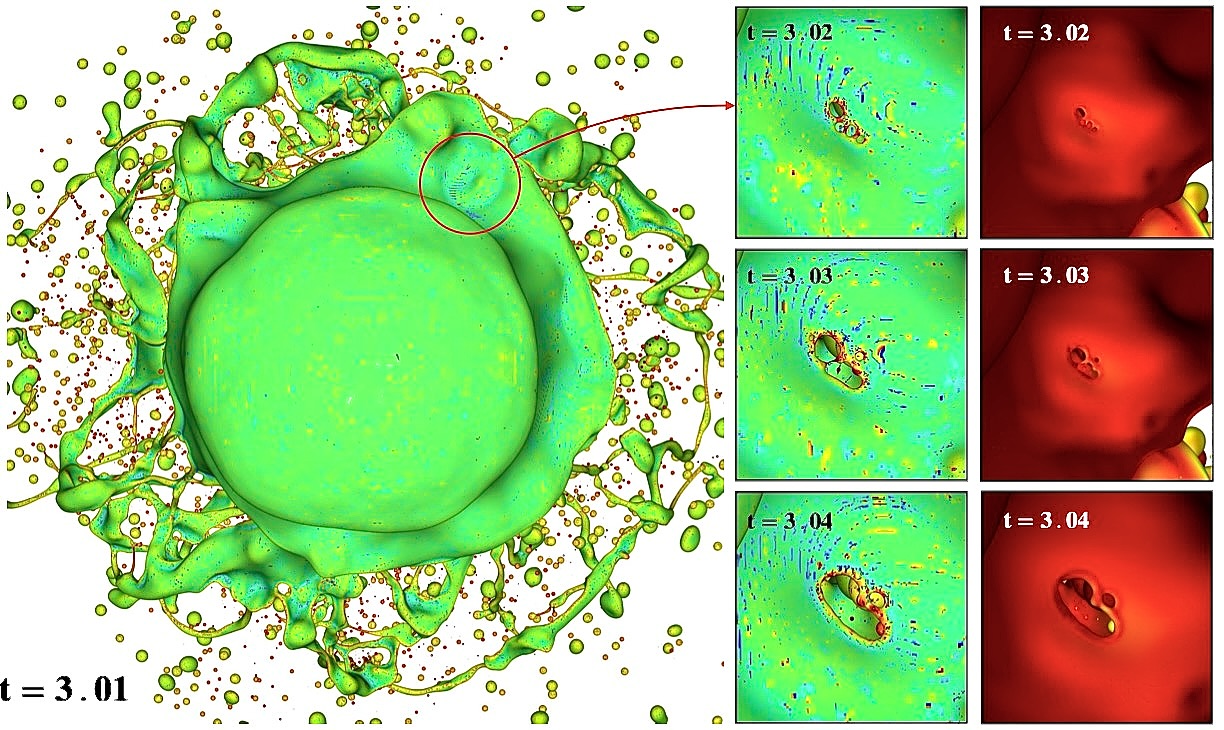}
    \caption{Sheet perforation when the manifold death method is applied. Unlike Figure \ref{fig:sheet_fully_dev_No_MD}, the holes are not preceded by large regions containing curvature ripples. Holes are punched at $t=3.01$ and expand with time as shown. The interface in the left part is colored by curvature and on the right is colored by the axial velocity. The simulation shown here corresponds to  $(\ell,m)=(14,13)$ }
    \label{fig:MD_holes_GOOD}
\end{figure}

Figure \ref{fig:MD_holes_GOOD} shows a zoom-in at a hole expansion that was initiated by the manifold death method. The phase modification region is a cube but in less than 5-10 timesteps the hole takes a rounded shape due to surface tension and starts expanding.  
When these holes expand, we do not see the noisy curvature ripples that resulted in grid-dependent droplets in the no-MD simulation of Figure \ref{fig:sheet_fully_dev_No_MD}. This qualitative improvement in sheet collapse helps to obtain grid convergence. 
\begin{figure}
    \centering
    \includegraphics[width=\textwidth]{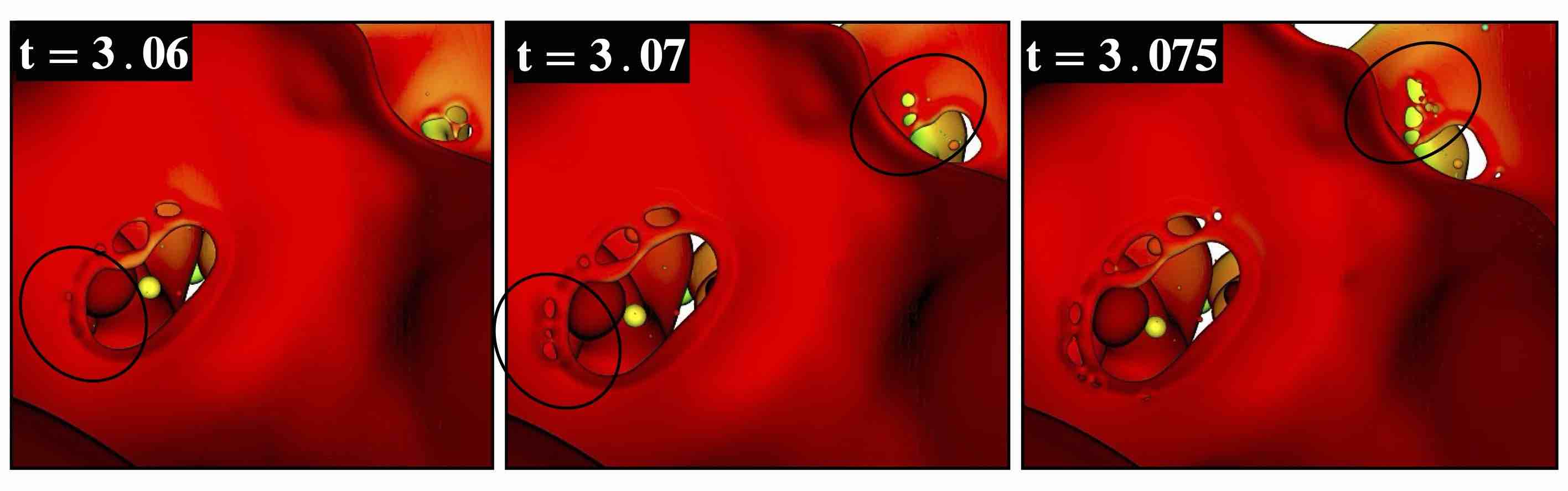}
    \caption{Further evolution of the holes of Figure \ref{fig:MD_holes_GOOD} . The image is colored by axial velocity. We see that near the rims of the main expanding holes, additional holes are punched, encircled in black. (Manifold death is applied.)} 
    \label{fig:MD_holes_BAD}
\end{figure}

The manifold death procedure we use is tuned to a relatively large frequency of hole formation. This results in features such as those shown in Figure \ref{fig:MD_holes_BAD}. We see that once we detect a thin sheet and the MD process generates holes, these expand.  However, it often happens that before the holes have expanded over the entire 
sheet more holes are punched just outside the rim. These kind of holes are identified in black circles in Figure \ref{fig:MD_holes_BAD}. \red{This effect is also seen in Movie4 that shows the evolution of thin sheets with MD.} Experimental photographs and the reasoning in \red{Appendix \ref{sec:mystery_stuff_outlook} point to a smaller number of holes}. This smaller number could be achieved by an improved MD algorithm that would involve a more realistic model of the physical process of hole formation. 

\begin{figure}
    \centering
    \includegraphics[width=0.49\textwidth]{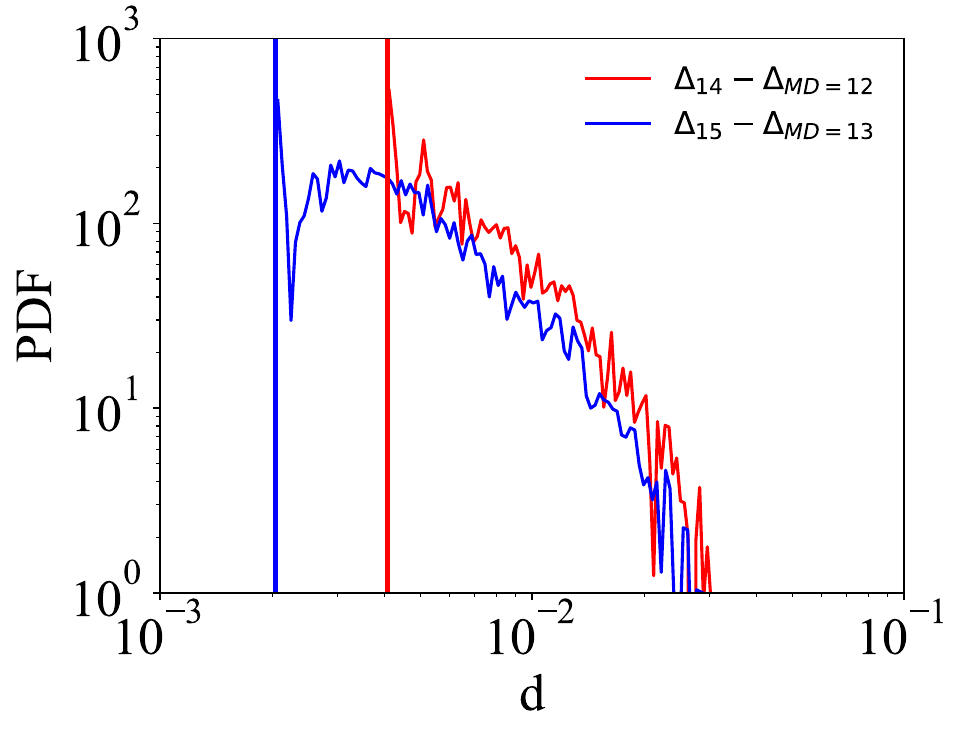}  \includegraphics[width=0.49\textwidth]{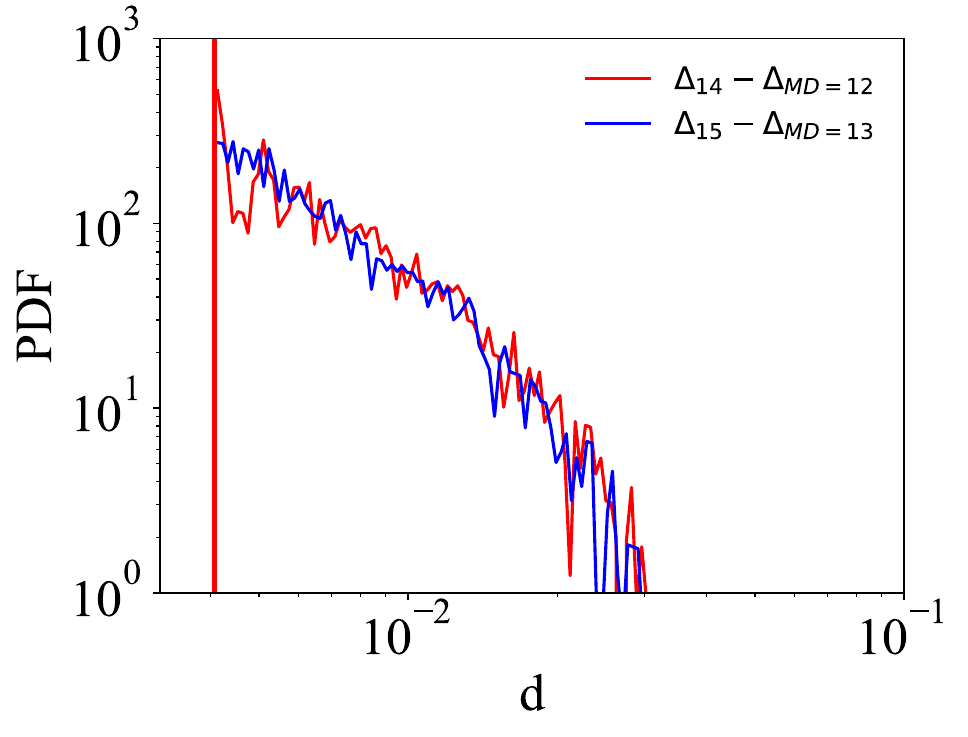}
    \caption{Two distributions of droplet sizes at two MD levels, $m=12$ and $m=13$, with the same ratio of the grid size to the MD threshold $h_c/\Delta_\ell$. We only plot the converged region, that is $d>d_c(h_c)$.
    The most refined simulation $(l,m)=(15,13)$ yields a distribution somewhat but not exactly similar to the less refined. Both distributions show a sudden increase in droplet number (a ``ridge") as the droplet size is decreased at the edge of the converged region. The vertical line shows the limit of convergence $d_c$. The left plot shows the un-rescaled number distribution, while the right plot shows the PDFs. The right plot is clipped to show the PDFs $p(d)$  for $d> d_c(m=12)$,  where $d_c$ is the edge of the converged region for $m=12$. The Pareto distribution $n(d) \sim d^{-2}$ is also plotted.}
    \label{highest}
\end{figure}

The final result of our manifold-death-enhanced study of droplet sizes is summarized on Figure \ref{highest}. In this Figure we select two distributions of droplet sizes at two MD levels, $m=12$ and $m=13$  and we keep the ratio of the grid size to the MD threshold $h_c/\Delta_\ell$ constant so the level difference is
always $\ell - m= 2$. The two obtained distributions are similar to each other,  with a sudden increase in droplet number at the edge of the converged region.
\begin{figure}
    \centering
    \includegraphics[width=0.9\textwidth]{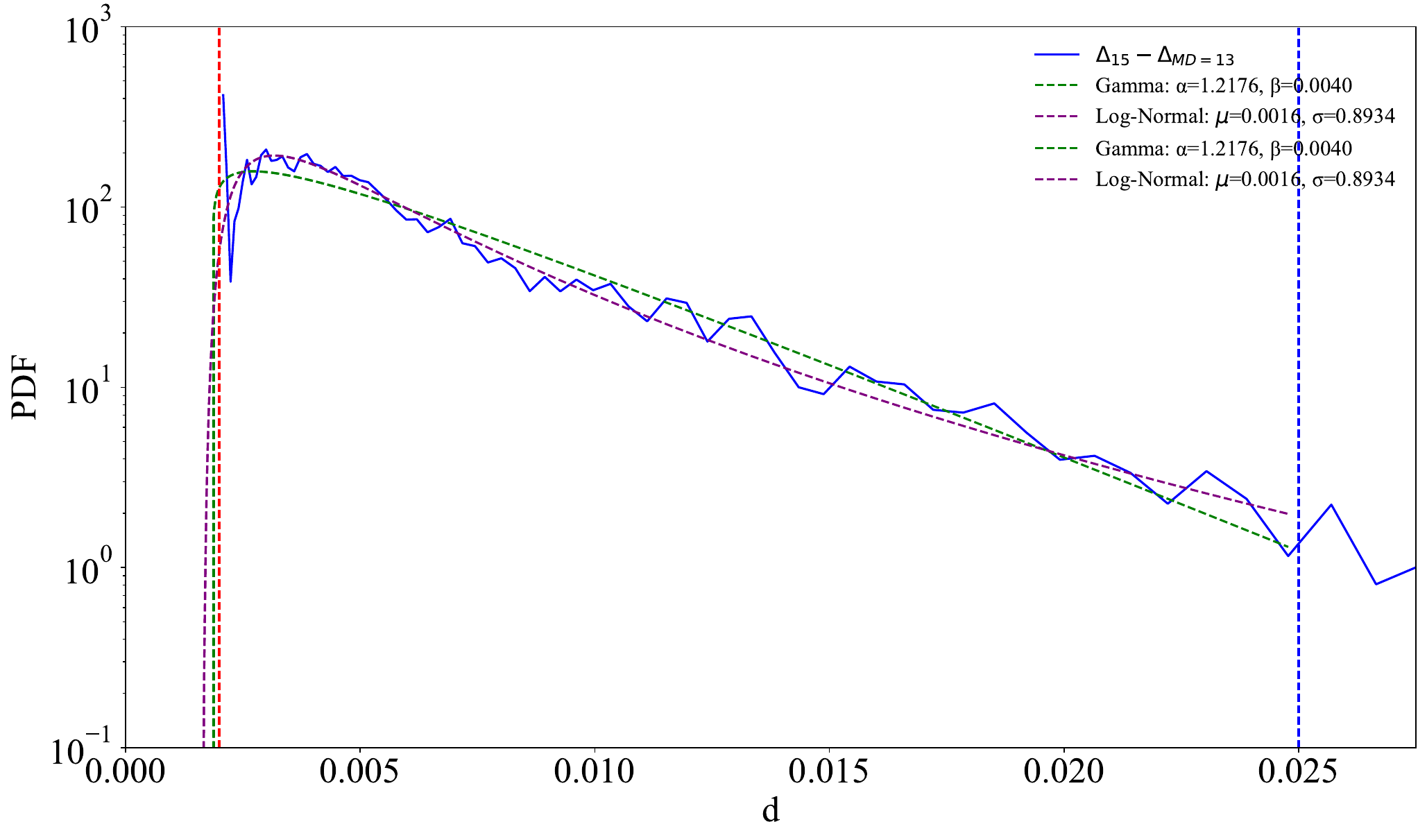} 
    \caption{Same as Figure \ref{highest} but with a linear abcissa and a fit to Gamma and Log-Normal distributions.}
    \label{converged-fitted}
\end{figure}
Moreover, these plots show that a sufficiently thin grid produces a bump-shaped distribution with a clear maximum (this is the case at $m=13$ but not at $m=12$). This maximum is located at $d_{\rm max}/D = 4.3 \, 10^{-2}$ in our case. It is interesting to try to understand the origin of this bump. By dimensional analysis the maximum should be expressed as
\begin{equation}
    d_{max} = D f(h_c/D,N_1, \cdots, N_n),
\end{equation}
where the $N_i$ are the dimensionless numbers of the problem other than $h_c/D$. It is quite tempting to assume that in some range of $h_c/D$  and of the dimensionless numbers a proportionality relation holds. From the existing data it would be 
\begin{equation}
    d_{max} = 4.3 \, h_c.  \label{dmax43}
\end{equation}
We note that this is to our knowledge the only example of a {\em converged} bump-shaped (unimodal) distribution in simulations of jet atomization  at large Weber number. We attempt to fit this bump shape to standard distributions in the literature, namely the Gamma and the Log-Normal. Both distributions are acceptable in the large $d$ range, above 0.005, as seen in Figure \ref{converged-fitted}. The fitting distributions are defined as follow:

\begin{align*}
	\text{Log-Normal : } \, P\left( x ; \mu , \sigma \right) &=
	\frac{1}{x \sigma \sqrt{2\pi}} \,\textrm{exp}\left[-\frac{1}{2}\left(\frac{\text{ln}( x) - \mu}{\sigma}\right)^2\right] \,,\\
	\text{Gamma : } \, P\left( x ; \alpha, \beta \right) &=
	\frac{\beta^{\alpha}}{\Gamma(\alpha)} x^{\alpha - 1} \textrm{exp}\left(-\beta x\right)  \,.
\end{align*}

\section{Conclusions}
\label{sec:conclusions}

We have studied an atomizing, pulsed round jet at a succession of grid sizes with unprecedented resolution. Visual inspection of the resulting interface topology reveals a new numerical phenomenon preceding hole formation in the VOF method: high frequency curvature oscillations followed by the appearance of a small-scale network of ligaments. 
\purp{While the experimental observation and theoretical analysis of droplet formation by expansion of holes is not new, this is the
first time that these high frequency oscillations of curvature are observed in VOF simulations.}
The inspection of droplet size frequency distributions reveals the presence of two large peaks tied to the grid size. It is clear that the small-$d$ peak and probably the second
one are related to the unphysical breakup.

{To mitigate this effect we apply the manifold-death procedure of \cite{Chirco22}. The procedure forces hole formation when sheets reach a set thickness $h_c$. Number frequency plots show convergence in a range of droplet sizes starting at a critical diameter $d_c$. The extent of this converged range increases when either the grid size $\Delta_\ell$ or the critical thickness $h_c$ are decreased. 
For a fixed $h_c$, when we reduce $\Delta_\ell$, the converged diameter $d_c$ approaches the critical thickness $h_c$. 
We thus recover the convergence properties observed by \cite{Chirco22} and by \cite{Tang_Adcock_Mostert_2023}. We characterize them in the $h_c$-$d_c$ plot 
of  Figure \ref{fig:h_c_d_c_plane_movement}b.
We note that the statistical accuracy is much stronger in the present case since 
our pulsed jet produces three orders of magnitude more droplets than the phase inversion case of \cite{Chirco22}.}
We do not have a definite explanation for the lack of convergence below the diameter $h_c$, that is we are not sure why $d_c$ does not decrease below $h_c$ as $\Delta_\ell$ is decreased. We can however offer several avenues for improvement of this feature and for future exploration. One is the smoothness of the manifold death procedure itself and the choice of its parameters such as the number of holes punched or the punching frequency $1/\tau_m$. 
For the first time in VOF simulations of atomization, a bump-shaped, converged droplet size distribution is obtained. Even without convergence, bump shaped distributions are harder to obtain in VOF simulations than in Level-Set or Diffuse-Interface simulations where they are common \cite[]{Herrmann10,khanwale2022breakup,saurabh2023scalable} because these two other methods tend to ``evaporate'' small droplets, as shown on Figure \ref{nicefig}. A scaling relation for the position of the bump is suggested. The bump only appears at the highest level. Whether the  proposed  scaling  (\ref {dmax43}) for the bump would persist if yet more refined simulations are performed is an interesting open question. 

Beyond the physics of hole formation and expansion, the simulations reveal a rich spectrum of elementary phenomena that are the building blocks of atomization. These include networks of well resolved ligaments in and around holes, ligaments detached from the main liquid core and either aligned or at an angle with it,  bubbles trapped in liquid regions, droplets impacting on sheets etc.
This rich physics is seen in increasing detail as grid resolution is increased. For example, the sharpness and apparent realism of the various expanding holes, with attached ligaments or networks of ligaments inside, was not seen in previous simulations, for example in \cite{ling17}.  Moreover a large number of each type of object or ``building block" are visible at each instant of time, leading to more significant and converged statistics. 

A tantalizing question is that of the physical, as opposed to numerical,  origin of the perforations. We attempt some discussion of this issue in Appendix \ref{sec:mystery_stuff_outlook}. Two types of rate processes for heterogeneous nucleation are postulated, one based on in-sheet germs or nuclei, such as small bubbles, the other on external perturbations such as floating droplets or particles. Our simulations do not allow to decide at present between these two processes for hole formation. However they may in the future help understand how to discriminate between them. 

Unlike most of the other round jet simulations in the literature, ours is bimodal and even trimodal after the application of manifold death. It seems however that a single mode emerges in the {\em converged range} of the best resolved MD simulations, indicating that the appearance of a maximum of droplet size well in the converged region is possible only if the physics is correctly captured down to scales much smaller than the integral scales such as the jet diameter $D$. 

Another important characteristic of simulations and experiments are the state of the fluids at the inlet. The primary mixing-layer instability in our simulations is initiated not by upstream turbulence as in most round jet simulations in the literature but by a pulsating flow. \red{In future investigations, it would be interesting to compare the visual aspects of the 
fragmenting interface as well as number distributions and PDFs, between simulations with or without noisy/turbulent inflow conditions.}

The other perspectives opened by this work are numerous. 
Clearly, simulations of comparable quality at higher Weber number may allow investigation of the ``fiber-type" regime observed in many experiments. 
Currently our simulations sit on the boundary between membrane type and fiber type. Simulations of comparable resolution should also be performed in the planar mixing-layer case investigated experimentally by the Grenoble group \cite[]{Benrayana06,Fuster2013} and numerically
by \cite{jiang2021impact}. The latter authors performed very high resolution simulations with hole formation, and interestingly obtained a bimodal distribution, but they did not apply a controlled perforation scheme as in this paper. Simulations of secondary atomization, that is the atomization of a droplet suddenly plunged in a high speed flow has already been performed together with controlled perforation by  \cite{Tang_Adcock_Mostert_2023} achieving convergence of the PDF. The secondary atomization case is probably the most promising for an eventual comparison of the distribution of droplet sizes obtained by simulation with the rich experimental data reviewed and modelled for example by \cite{jackiw2022prediction}.

More distant perspectives, or more difficult problems remaining to solve are in our opinion twofold. One is clearly the improvement of the manifold death procedure, to have a better control of the number of holes punched in each thin sheet region, and a smoother manner to ``cut-out" the initial hole. 
\red{The breakup frequency and number of attempted breakups were chosen in a rather arbitrary way. These values are too high as shown above.
An adequate breakup frequency would correspond to smaller values of  $f_b$ and $N_b$.  A caveat is however that if the breakup frequency ($f_b$, $N_b$)  is too low, some bags may escape MD and breakup numerically.
An expensive trial-and-error procedure could be attempted 
to tune the breakup frequency.
Another avenue of improvement would be to modify the MD method to perform an analysis of the bag region to identify  the connected thin sheet regions. 
Then a controlled number of perforations could be performed in any of these regions. This controlled number of perforations per thin region could be inspired by experimental observations or by a theory such as the one in Appendix  \ref{app:pheno_hole_expansion}.
A final, intriguing possibility is that if a Level-Set or Diffuse-Interface method is used in conjunction with Manifold Death, and the frequency is on the low side, numerical breakup will be avoided in the regions that would not be sufficiently perforated because of the low frequency. There would still be mass loss because of the evaporation of thin sheets, but no curvature oscillations nor an excessive number of small droplets scaling with grid size.}

Another challenging future task is to use this type of simulation to infer, by careful matching of simulation results and experiments, the physical mechanisms for weak spots and sheet perforations. Given the wide variety of such mechanisms that have been advanced, this is a formidable enterprise for which we hope this paper could provide partial guidance. 

\section*{Acknowledgements}  We thank Marco Crialesi and Daniel Fuster for fruitful discussions. S.Z. would like to thank Ousmane Kodio for mentionning the Kibble-Zurek theory in discussions on pattern formation. This project has received funding from the European Research Council (ERC) under the European Union's Horizon 2020 research and innovation programme (grant agreement nr 883849). 
We thank the European PRACE group, the Swiss supercomputing agency CSCS, the French national GENCI supercomputing agency and the relevant supercomputer centers for their grants of CPU time on massively parallel machines, and their teams for assistance and the use of Irene-Rome at TGCC and ADASTRA at CINES. This project was provided in particular with HPC computing and storage resources by GENCI at TGCC thanks to the grants 2023-A0152B14629 and
2022 SS012B15405 on the supercomputer Joliot Curie's SKL partition .

\section*{Declaration of Interests}

The authors report no conflict of interest.
\appendix 

\section{A phenomenological theory for hole nucleation and expansion}
\label{app:pheno_hole_expansion}
\label{sec:mystery_stuff_outlook}
\red{
In what follows we describe hole nucleation and expansion as a random process. 
Its consequences on ligament size are given based on the hole nucleation rate in a manner independent on the exact physical mechanism of nucleation. An analogy between this description and the Kibble-Zurek theory is made.}

\red{In atomization processes, sheet rupture may have several causes. Without deciding between these causes, we distinguish two mathematical forms of
the hole forming process \cite[]{villermaux2020fragmentation}. In one it is in-sheet germs or nuclei, such as small bubbles or small droplets of an immiscible liquid \cite[]{Vernay:2017bz} that create the holes. In the other one, external nuclei such as floating droplets, particles or hot puffs are the cause of perforation. 
The probability of punching a hole per unit area of thin sheet can be written to model both processes as
\begin{equation}
    {\mathrm d}\mathcal{P} = \mathcal{A}(h) {\mathrm d}t + \mathcal{B}(h) {\mathrm d}h. \label{ABmodel}
\end{equation}
There $\mathcal{A}(h) {\mathrm d}t$ is the probability that an external object that would be able to punch a hole in a sheet of thickness $h$ actually hits the sheet during the time interval ${\mathrm d}t$, while $\mathcal{B}(h) {\mathrm d}h$ is the probability that an in-sheet heterogeneity is present that will nucleate a hole while the sheet thins from $h$ to $h - {\mathrm d}h$. In our present study we do not have external objects such as solid particles, but we still have holes appearing due to the drop impact. Thus, in our simulations, the atomization driven by drop impact is mostly contained in $\mathcal{A}(h)$, while those appearing due to numerical perforation of sheets are contained in $\mathcal{B}(h)$. Our simulation captures an overall picture and it is difficult to isolate the effect of any one of the two. In the manifold death method, we focus on tuning the hole appearance due to in-sheet heterogeneity part having a simplistic model for $\mathcal{B}(h)$ as a step function with critical size being $h_c$.}  
\red{However, in fluids, heterogeneities such as bubbles do not have a single size but a distribution of sizes, and correlations obtained experimentally may be obtained that yield $n_b(r)$ \cite[]{brujan2010cavitation} such that $n_b(r)$ d$r$ is the number of
bubbles with size between $r$ and $r+$d$r$ per unit volume. The number of bubbles between $r$ and $2r$ then scales like $n_b(r) r$ and the average distance between such bubbles is $l_1 = [r n_b(r)]^{-1/3}$.
A natural requirement is that this {\em  distance between  bubbles is much larger than their radius}. 
Then one must have
$[r n_b(r)]^{-1/3}\gg r$. Since we probe the distribution at length scales $r \sim h$ 
that are relatively small it makes sense to require that 
$n_b(r) \ll r^{-4}$ in the limit of small  $r$. 
A simple model for the distribution of nuclei or number of nuclei per unit volume is
\begin{equation}
    n_b(r) \sim c_n r^{-\gamma} \label{distribn} 
\end{equation}
where $\gamma$ is an exponent to be determined by measurement and $c_n$ is a constant. By the requirement above the exponent $\gamma < 4$ and indeed  \cite{gavrilov1969size} found $\gamma \simeq 3.5$ while measurements by \cite{shima1987equation}  give $\gamma$ between $2$ and $4$.  As a result the 
volume fraction $r^4 n_b(r)$ of bubbles between  $r$ and $2r$  vanishes at small $r$.
Now suppose that bubbles of radius $r$ are the cause of weak spots in a sheet of thickness $r$.
Then from  (\ref{ABmodel})
the probability of ``in-sheet" nucleation in a region of extent $\lambda^2$ while the sheet thickness decreases
from $h$ to $h/2$ scales like $\mathcal{B}(h) h \lambda^2$. This nucleation is caused by small bubbles
in the volume $\lambda^2 h$
then  $\mathcal{B}(h) h \lambda^2= n_b(h) \lambda^2 h^2$, and the ``in-sheet" nucleation  probability
per unit area is 
just 
\begin{equation}
    \mathcal{B}(h)= n_b(h) h.  \label{Bnb}
\end{equation}
The typical distance $\lambda$ between weak spots in a sheet of size $h$ thinning to $h/2$ is given by the condition
that the expected number of weak spots in area $\lambda^2$ is one, which reads
$$
n_b(h) \lambda^2 h^2 \sim 1.
$$
Then
$$
\frac \lambda  h \sim ( n_b(h) h^4 )^{-1/2} \sim h^{-2 + \gamma/2}
$$
so for $2 <\gamma < 4$ one has $\lambda \gg h$, while $\lambda$ still tends to vanish at small $h$. The distance between holes $\lambda$ is much larger than the thickness of the sheet but is still small compared to the total size of the sheet. We shall call $L$ the total size of the sheet so the above reads  $L \gg \lambda$. In other words the holes are rather numerous in each sheet. \\
This is not what is observed in experiments, for example by \cite{kant2023bag}, 
and one instead seems to see  one or two holes per sheet. To resolve this contradiction one must take into account the speed of expansion of the holes that form during sheet thinning. 
Indeed expansion of the holes brings the process of hole formation to a halt after a short time $\tau_1$ that we determine below. 
%
We now can give a physical meaning to the critical thickness $h_c$ in the manifold death algorithm. The expected value of the number of nucleations in a thin sheet of area $L^2$ while the sheet retracts from $h_c$ to $h_c/2$ is 
\begin{equation}
    N_0(h_c) = L^2 \int_{h_c/2}^{h_c} \mathcal{B}(h)  {\mathrm d}h \sim h_c^{2 - \gamma}.
\end{equation}
The critical thickness $h_c$ is then that for which 
\begin{equation}
    N_0(h_c) \sim 1 \label{N01}
\end{equation} is of order unity. 
Independently of the above prediction for $\mathcal{B}(h)$, and only on the basis of a model where the nucleation probability increases as $h(t)$ decreases it is possible to  predict the effective number of holes that will be observed. First consider the typical size $L(t)$ of an expanding bag. Experimental observations \cite[]{kant2023bag,opfer14} show that bags inflate exponentially. We consider only large scale bags governed by the large scale properties of the flow that thus inflate at a rate $\omega = 1/(2\tau_c)$. Then 
$L(t) \sim L_0 \exp ( \omega t)$ and the bag thickness decreases as 
\begin{equation}
    h(t) \sim h_0 e^{ - t/\tau_c}. \label{hrate}
\end{equation}
Theory and experimental observations \cite[]{Villermaux:2009ccz,opfer14,marcotte2019density,kant2023bag} indicate the scaling 
\begin{equation}
    \tau_c \sim \frac D U \left( \frac {\rho_l }{\rho_g}  \right)^{1/2}.   \label{tc}
\end{equation}
Once weak spots are activated in such a bag  they expand at the Taylor-Culick velocity \cite[]{Taylor59c,Culick60} 
\begin{equation}
    V_{T} = \left( \frac{2 \sigma}{\rho_l h_c} \right)^{1/2}.
\end{equation}
If $N$ holes are activated in a typical bag of size $L$, the average distance between hole nuclei positions scales like $\lambda = L N^{-1/2}$. This distance is covered by the expanding hole edges in a time
$\tau_1 = \lambda/V_T$ and
\begin{equation}
    \tau_1(N,h) = L N^{-1/2} \left( \frac{2 \sigma}{\rho_l h} \right)^{-1/2}. \label{eq:tau1def}
\end{equation}
We now distinguish between two extreme cases. In the first case $\tau_1(1,h_c) \ll \tau_c$.
This means that  after the first hole
has formed, it will expand to the whole bag size before the sheet has significantly thinned. During the expansion of that first hole, $h$ will vary from the initial thickness noted $h_0$ to $h_0 e^{-\tau_1/\tau_c}
\simeq h_0 - h_0\tau_1/\tau_c$ and the expected number of other holes that could form during that time is 
\begin{equation}
    N_1(h_c) = L^2 \int_{h_0 - h_0\tau_1/\tau_c}^{h_0} \mathcal{B}(h) {\mathrm d}h 
    \simeq N_0(h_c) \tau_1/\tau_c
\end{equation}
and from (\ref{N01}) we obtain $N_1(h_c) \ll 1$, that is the expected number of additional holes is much less than one, which means that in practice typically a single hole will be observed. Another way of deriving this result is to say that the hole expansion time is much shorter than the hole nucleation time. 
In the {\em opposite extreme}, the hole expansion time for a single hole is much longer than the hole nucleation time. One then needs to consider the  expansion time $ \tau_1(N) $  for a large number of holes $N$.
At this point it is useful to borrow an argument from the theory of bifurcations or instabilities in time-varying environments. The Kibble-Zurek theory \cite[]{kibble1976topology,zurek1985cosmological} argues that the time scale of the environment variation is equal to the time scale of the growth of the instability. We thus argue that the multi-hole expansion time $\tau_1(N)$ scales like the characteristic bag expansion time $\tau_c$ that is 
\begin{equation}
    \tau_1(N) = \tau_c. \label{eq:times}
\end{equation}
To obtain this rule consider how nucleation occurs as time progresses from the first perforation at time $t_0$ to a time where most of the surface is covered by expanded holes. At first the nucleation rate is near the threshold given 
by (\ref{N01}). As time progresses, previously nucleated holes expand and the thickness decreases. The decrease in thickness implies a higher nucleation rate since ${\cal B}(h)$ increases as $h$ decreases. Hole density increases and the typical distance $\lambda(t)$ between holes decreases. When $\lambda(t)$ becomes of the same magnitude as the distance
covered by the expanding rims one has $\lambda(t) \sim V_T (t - t_0)$ and the the process stops with $t - t_0 \sim \tau_1(N)$. Since we are in the second extreme case $N \gg 1$ the hole density and nucleation rate have increased considerably from that at $t_0$ which means that the sheet thickness has decreased significantly.  A significant decrease of the sheet thickness requires a time of order $t-t_0 \sim \tau_c$, hence (\ref{eq:times}).
Then from the balance of time scales  (\ref{eq:times}) the number of holes in a bag of size $L$ can be obtained as
\begin{equation}
  N_1(h_c) \sim \frac{ h_c \rho_g L^2 U^2 }{D^2  \sigma } . \label{eq:N1_hc}
\end{equation}
There is a degree of uncertainty for the ratio $L/D$, however $L$ and $D$ are both integral scales of the turbulent flow and can be considered of the same order of magnitude. (We disregard the possible case where eddies of smaller scale than the integral scale of turbulence are sufficiently energetic to generate smaller scale bags.) The number of holes per bag in this mechanism is then the ratio of two typically large numbers, $D/h_c$ and $\We_g$. Equation (\ref{eq:N1_hc}) is valid in the second extreme case  when the right hand side is much larger than unity. A very rough extrapolation between the two extreme cases is then
\begin{equation}
  N(h_c) \sim \mathrm{Max} \left[1, N_1(h_c) \right]. \label{eq:N2}
\end{equation}
In most cases one expects $L$ to be of the same order of magnitude as $D$ which leads to the simplified prediction
\begin{equation}
    N_1(h_c) = \We_{h_c}, \label{eq:N1we} 
\end{equation}
where $\We_l = \rho_g l U^2/\sigma$. Table \ref{tab:delta} 
gives some typical values for  $\We_{h_c}$ and $\We_\Delta$, allowing a prediction of the expected number of holes in the corresponding simulations. 
We notice that the estimate above is only based on the assumption that ${\cal B}(h)$ increases as $h$ decreases and does not depend on a specific model such as the distribution of bubbles (\ref{distribn}). }\\

\blue{A more sophisticated model of the number of holes can be obtained as follows. Considering the time $\tau_2(N)$ taken by the rate of nucleation ${\cal B}(h)$  to increase from the initial rate, producing a single hole $N_0 \sim 1$, to the final nucleation rate, producing a large density of holes, the final number of holes $N(h_f)$  is given by 
\begin{equation}
    N(h_f)= L^2 \int_{h_f}^{h_c} \mathcal{B}(h) {\mathrm d}h ,
\end{equation} Then 
$$
N(h_f) \sim L^2 [ (h/h_c)^{2-\gamma}]_{h_f}^{h_c} ,
$$
and
$$
N(h_f)  \sim L^2 (h_f/h_c)^{2-\gamma} .
$$
On the other hand using (\ref{hrate}) 
$$
h_f = h_c e^{-\tau_2(N)/\tau_c}
$$
where for simplicity we write $N$ for $N(h_f)$.
hence
$$
N \sim L^2 e^{-\tau_2(N) (2-\gamma)/\tau_c} ,
$$
and thus
\begin{equation}
\tau_2(N) = \frac{\tau_c \ln N}{\gamma-2} \label{eq:logcorrection}
\end{equation}
Here we use again an equality of time scales assumption so that $\tau_1 \sim \tau_2$
and (\ref{eq:times}) becomes
\begin{equation}
    \tau_1(N) = \frac{\tau_c \ln N}{\gamma-2} \label{eq:times2}
\end{equation}
This  brings a $\gamma$-dependent and logarithmic correction to the prediction (\ref{eq:times}) and hence to (\ref{eq:N1_hc}).}\\

\red{Beyond the number of holes, another interesting prediction is that of the final size of ligaments obtained from hole expansion. There are at least two possible regimes for 
hole expansion, either unsteady fragmentation or smooth expansion. In the {\em unsteady fragmentation} regime the advancing rims bordering the expanding hole continuously shed small droplets \cite[]{wang2018unsteady}. 
In the smooth expansion regime, the rim advances without shedding droplets and volume is conserved. 
At the end of expansion a network of ligaments of diameter $\delta$ is formed as schematised in Figure \ref{fig:lig-net} and shown numerically by \cite{agbaglah2021breakup}.
\begin{figure}
    \centering
    \includegraphics[width=0.45\textwidth]{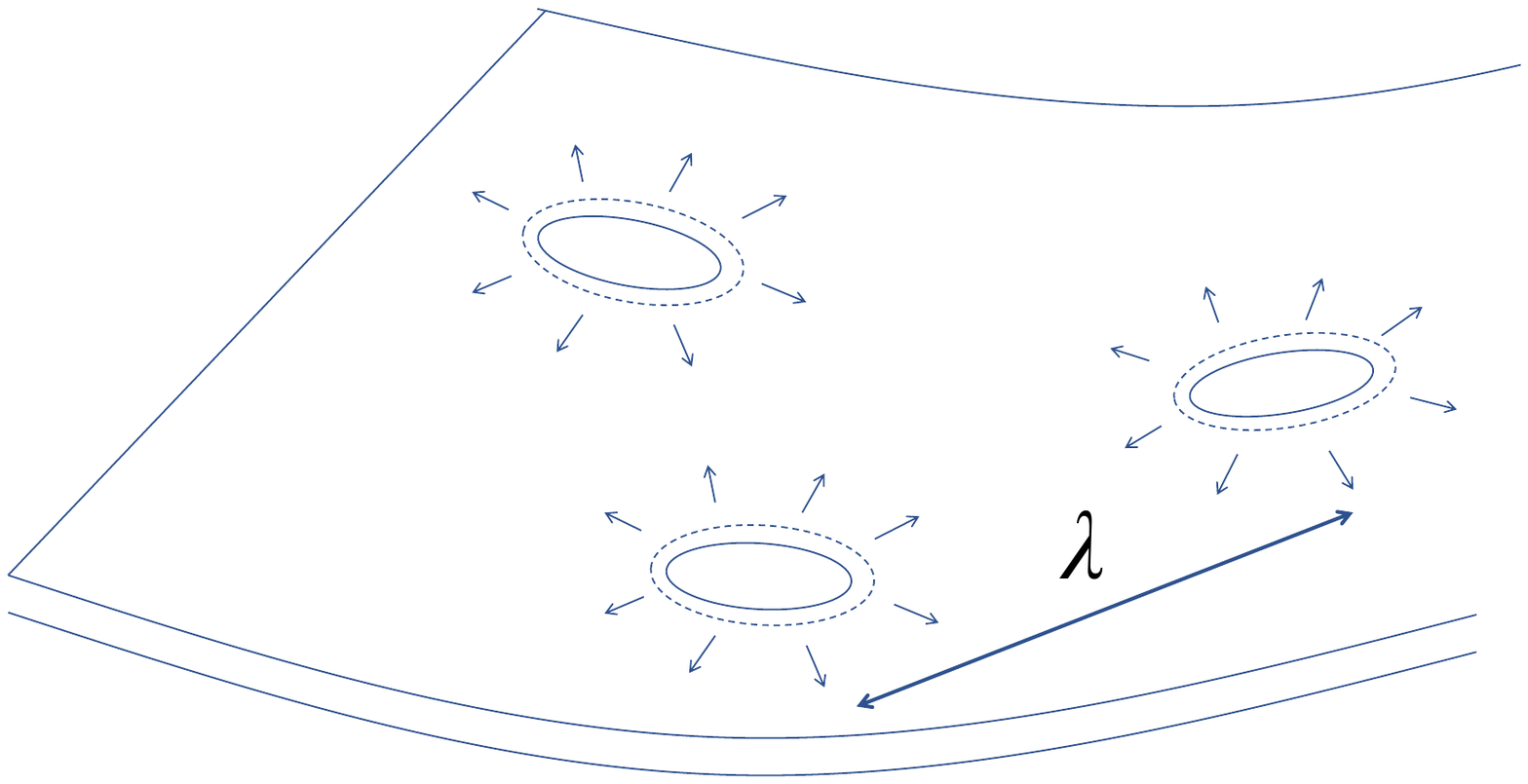}   \includegraphics[width=0.45\textwidth]{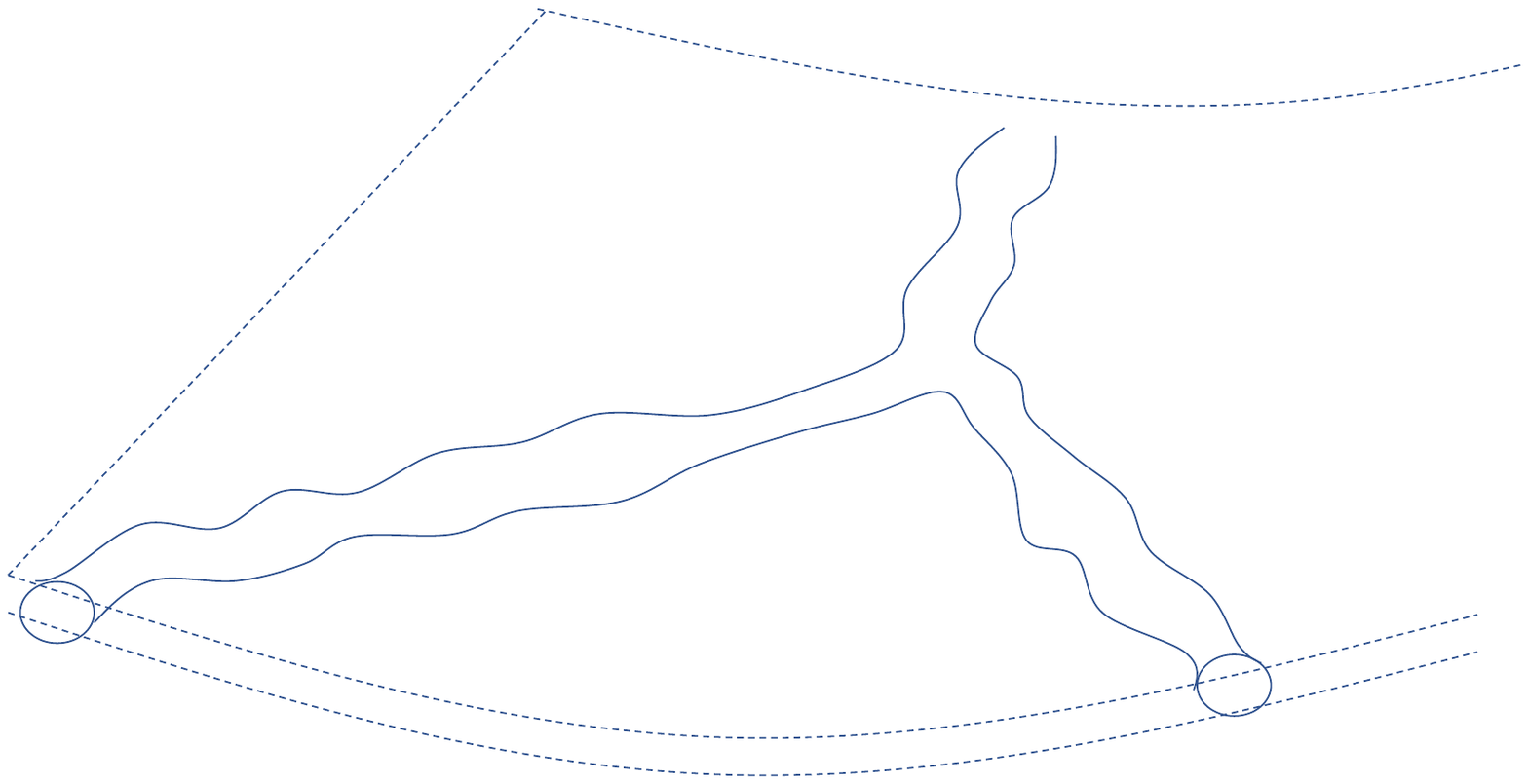}  
    \caption{A schematic view of how expanding holes in a thin sheet merge into ligaments.}
    \label{fig:lig-net}
\end{figure}
The number of ligaments is of the same order as the number of holes $N$ and their length is of order of the diameter of the hole at the end of expansion $\lambda$. The volume of ligaments is equal to the initial sheet volume so that equating the initial and final volumes
$$
 N \pi \lambda^2 h_c / 4 \sim   N \pi^2 \lambda \delta^2/4,
$$
from which we can deduce the ligament diameter 
$$
    \delta \sim \sqrt{h_c \lambda}.
$$
On the other hand the number of holes is related to $\lambda$ by
\begin{equation}
    N_1 = \frac{L^2}{\lambda^2}.  \label{N1lambda}
\end{equation}. Then Using (\ref{eq:N1_hc}) and (\ref{N1lambda}) one obtains 
\begin{equation}
    \delta \sim h_c^{3/4} L^{1/4} .
\end{equation}
It is possible, although this is a purely heuristic guess, that the second peak observed in the simulations with uncontrolled sheet perforation 
is caused by the break-up of this type of ligament into droplets. To obtain a simple scaling we consider $L \simeq D$ and $h_c \simeq \Delta$ when sheet perforation is
uncontrolled. The Rayleigh-Plateau instability of the ligament will yield $d_2 \sim \delta$ for the second peak.  One then obtains
$$
\frac {d_2} \Delta  \sim \Delta^{-1/4} .
$$
When plotted alongside our numerical data (Figure \ref{fig:peak_distances_No_MD}b) a fair agreement is obtained.}
\begin{figure}
     \centering
     \includegraphics[width=\textwidth]{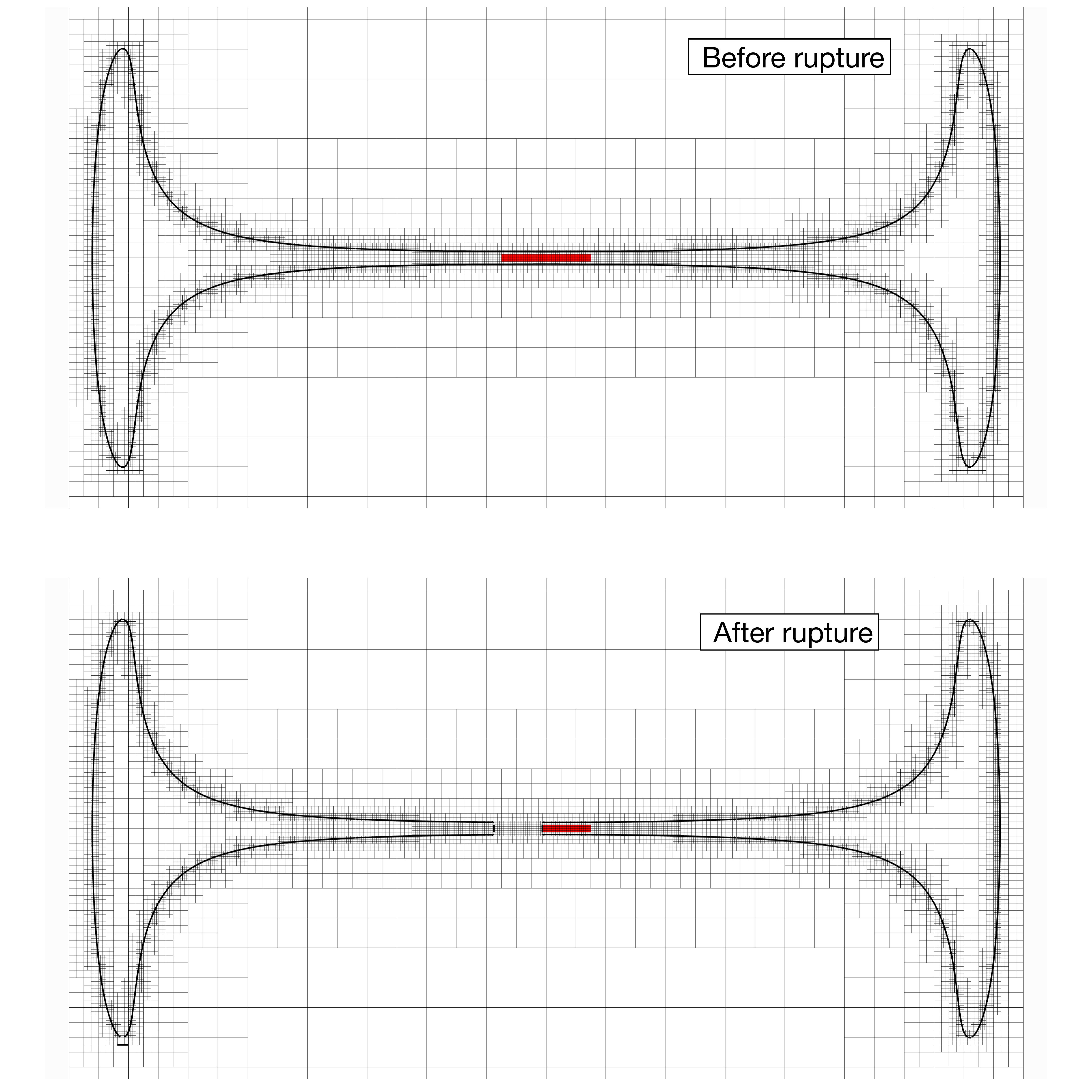}
     \caption{Thin structure detection and manifold death perforation in an example 2D case. The cells tagged as thin structure at the MD level are filled by red color. The upper image corresponds to the fluid structure right before perforation when a thin sheet is detected and the lower part shows the fluid structure after perforation. Note that the images are at same timestamp to explicitly show the perforation process. The image is done for case $(\ell,m) = (9,8)$. Adaptive mesh refinement can be seen in background.}
     \label{grid-ligaments}
\end{figure}
\section{Grid adaptation}
\label{app:adap-threshold}
\begin{figure}
    \centering
    \includegraphics[width=8cm]{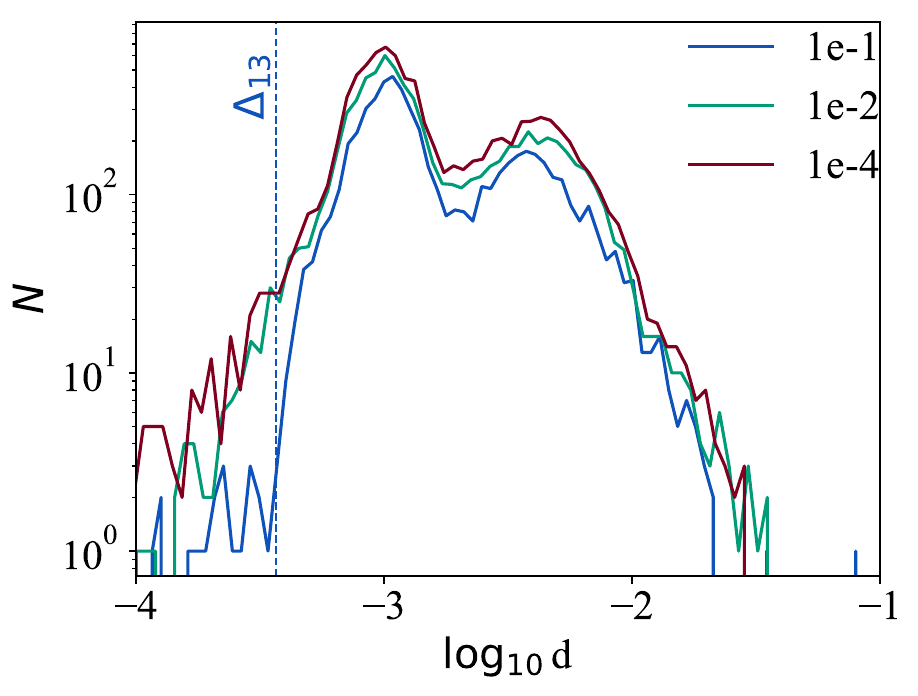}
    \caption{The droplet size frequency for maximum level $\ell=13$ and no manifold death at various error thresholds for the velocity.}
    \label{histo}
\end{figure}
As discussed in the main text, the mesh is adapted dynamically, by splitting the parent  cells whenever the local discretization error estimated by a wavelet approximation exceeds a threshold.
This threshold is set at $\epsilon_c = 0.01$ for the error on the volume fraction variable $c$ and to $\epsilon_u=0.1$ for the error on the velocity field. To investigate the influence of the relatively large error level on the velocity we plot the histograms of droplet sizes at various error thresholds in Figure \ref{histo}. On that Figure it is seen that although the total number of droplets changes significantly when $\epsilon_u$ decreases from 0.1 to 0.01, the position of the relative maxima and hence the typical droplet sizes vary relatively little. 
It is interesting to illustrate how grid adaptation interacts with ligament detection and perforation in the signature and manifold death methods. The signature method \cite[]{Chirco22} typifies the phase distribution with an index $i_s$. This index $i_s$ is determined by the signature  of the quadratic form of \cite{Chirco22}. In summary, $i_s$ can take four values. The index $i_s=-1$ designates gas phase, the index $i_s = 2$ designates a liquid phase while the index $i_s=0$ designates an interface. 
The fourth value that the index takes is $i_s = 1$ which designates a thin film region. Note that this detection is performed at the manifold death level and is later prolonged to the maximum level. 
Moreover the thin structure detection procedure acts without affecting the existing mesh refinement. Hence the mesh refinement is controlled only by the thresholds on the volume fraction and velocity field and in no way depends on the thin-sheet detection procedure. We illustrate the process of controlled sheet perforation using the manifold death method in Figure \ref{grid-ligaments}. We see that we have detected a thin region (red) in the upper image and a part of this region is eventually punctured in the lower image.

\section{Estimates of the mass loss due to the manifold death}

\newcommand{\pluseq}{\mathrel{+}=}

\begin{figure}
    \centering
    \includegraphics[width=\textwidth]{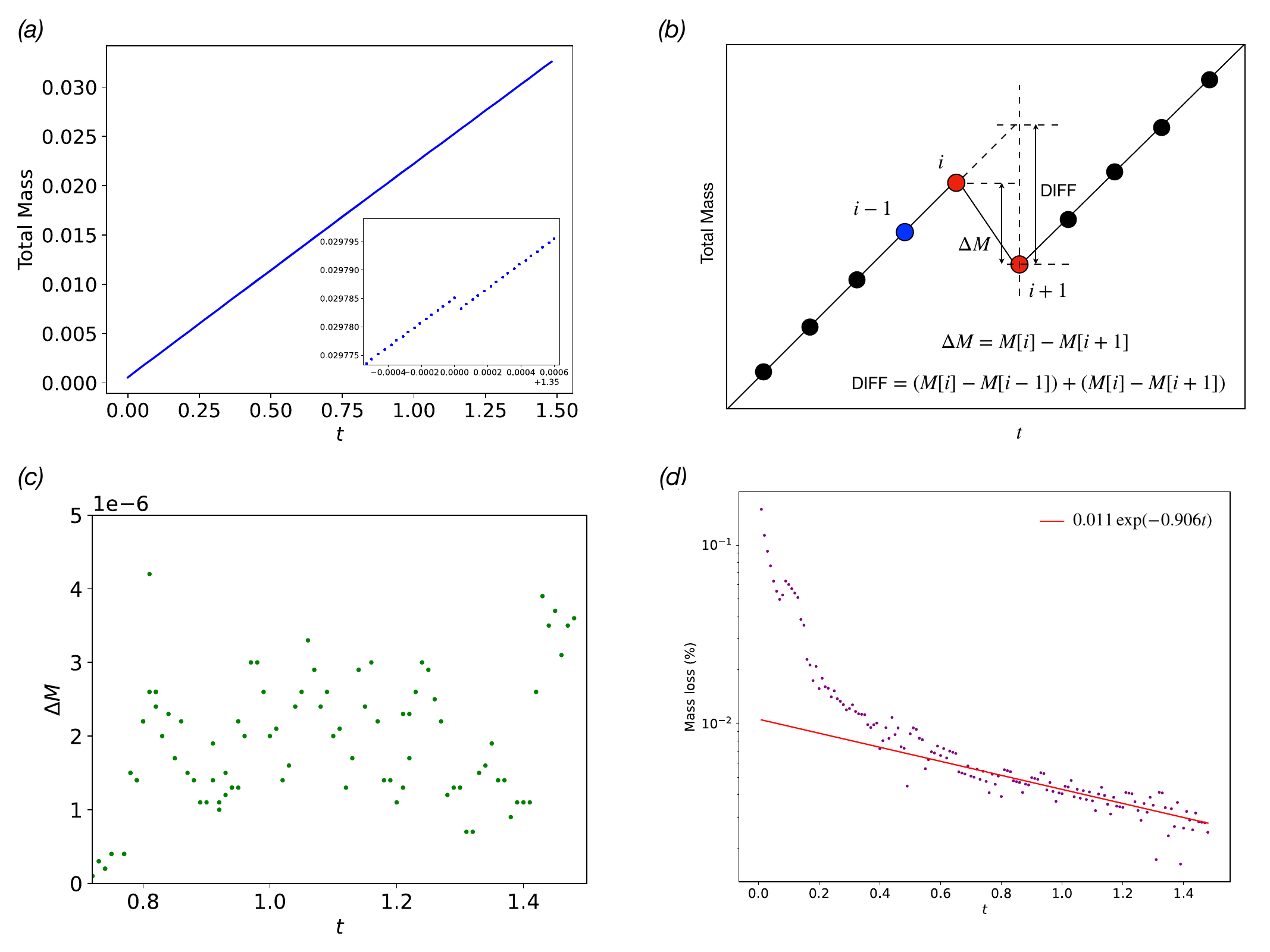}
    \caption{Estimates of the mass loss  with manifold death performed at levels $(\ell,m)  = (14,13)$. (a) Total mass as a function of time. Inset shows the tiny droplet at time $t=1.35$ due to artificial punching of thin structures every $t \pluseq 0.01$. (b) schematic of the data points and the procedure to estimate the mass loss. The value at $i$ represents the mass just before punching the holes and the value at $i+1$ represents the value just after punching the holes. $\Delta M$ and DIFF are used to do mass loss estimates. (c) $\Delta M$ at every $t \pluseq 0.01$. (d) Mass loss in percentage calculated as $\dfrac{\text{DIFF}}{M[i]} \times 100 $ and \red{the red line is the best fit line done for $t \in (0.8,1.4).$}}
    \label{fig:Mass_Loss_estimate_MD}
\end{figure}

In the manifold death method, we artificially punch holes in thin structures. This results in some fluid disappearance and hence mass loss. Here we give an estimate of the mass loss for the case $(l,m)=(14,13) $, that is the simulation corresponding to the red line in Figure \ref{fig:MD_stats_histo_all_LVL_conv}b. Figure \ref{fig:Mass_Loss_estimate_MD}a shows how the total mass behaves as a function of time. 
It is a linear rise in time with a tiny sinusoidal pulsation caused by the injection condition. Since we punch holes in these structures every $t \pluseq 0.01$, a tiny decrease from this linear rise is expected every $ \pluseq 0.01$ interval. The inset zoomed at $t=1.35$ shows this decrease in mass. The decrease is so small that it is not visible on the scale of the outer plot. To quantify this decrease, we calculate two quantities illustrated in the schematic of Figure \ref{fig:Mass_Loss_estimate_MD}b. The $\Delta M$ is the difference in the total mass at the timestamp just before punching a hole and the timestamp just after punching the hole. The plot as a function of time is shown in \ref{fig:Mass_Loss_estimate_MD}c, where the $\Delta M$ values are indicated every $t \pluseq 0.01$. Note that as mass is constantly being injected, this implies the number of thin structures shall increase with time and hence the $\Delta M$ is seen increasing with an oscillating behaviour. To get a non-dimensional estimate, we calculate the DIFF defined as 
$$
DIFF = 2 M(i) - M(i-1)  - M(i+1) 
$$
(see also the sketch in Fig.  \ref{fig:Mass_Loss_estimate_MD}b)  and calculate the percentage mass loss as

\begin{equation}
    \text{Mass loss } (\%) = \frac{DIFF}{M[i]} \times 100 .
\end{equation}

The rationale of using DIFF over $\Delta M$ is that since we are injecting some liquid constantly, we want to estimate how the mass differs not only from the value just before punching holes, but from the expected value that it was supposed to have. \red{The percentage loss is shown in Figure 22d, where we can see a decrease in time. We fit an exponential power law of mass loss percentage as $a\exp{(-bt)}$ and in the steady state time interval, we see that $b \sim \mathcal{O}(1)$. The mass loss relative to the total mass is going to zero as the jet advances.} \purp{The question of the effect of the mass loss on the PDF also arises. Since the mass loss occurs in very thin sheets, it affects the very small-scale droplets produced in these sheets and has a negligible effect on the part of the PDF above the converged radius $d_c$.}

\section{Evolving initial mushroom: manifold death in action}

\begin{figure}
    \centering
    \includegraphics[width=\textwidth]{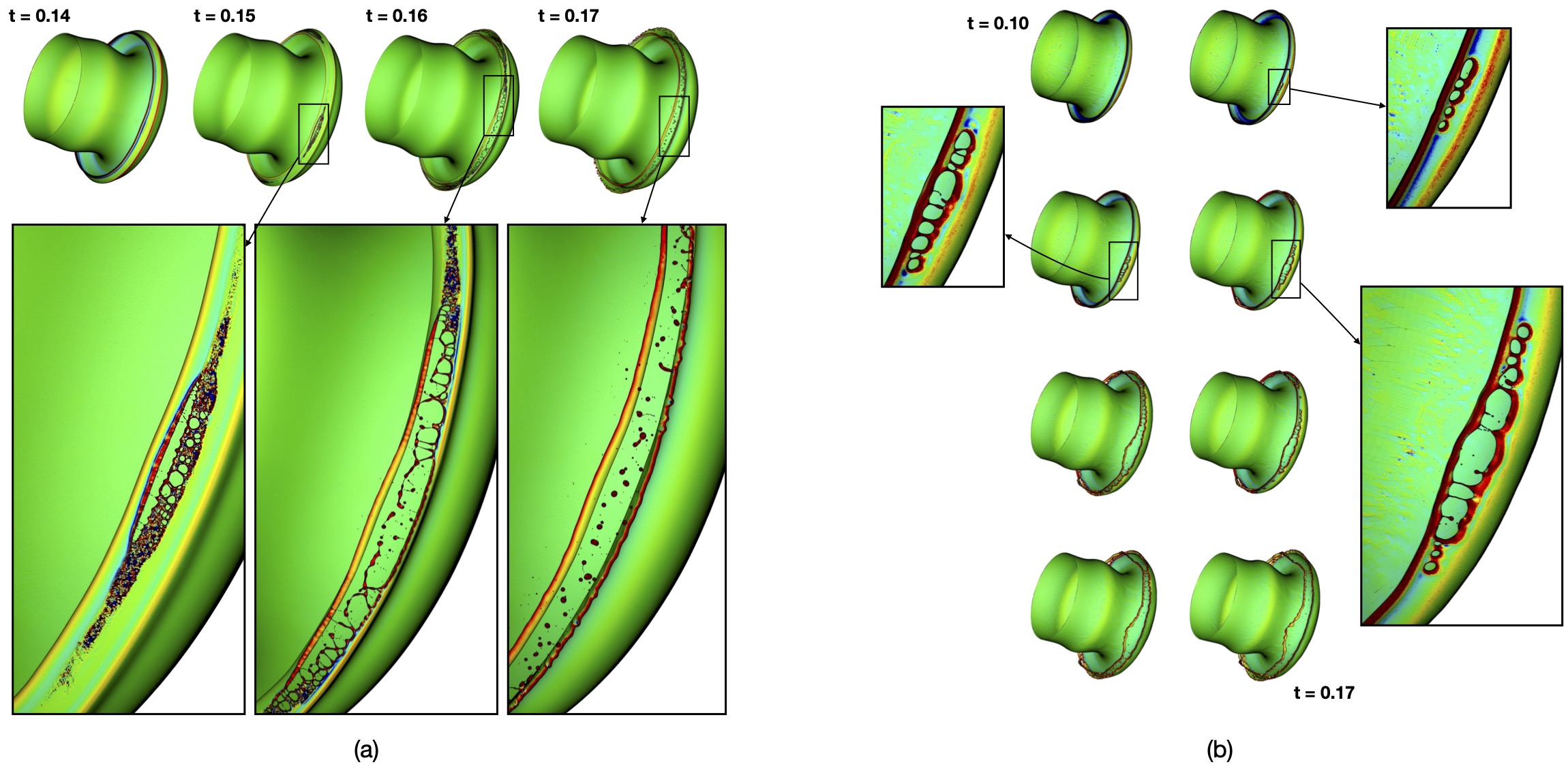}
    \caption{Evolution of the initial mushroom and the first breakup in (a) no manifold death and (b) with manifold death. The interface is colored by the curvature field in both cases. Image (a) shows a detailed view of the first numerical sheet rupture for an $\ell = 14$ simulation. Image (b) corresponds to the simulation with manifold death method applied corresponding to  $(\ell,m) = ({15,13})$. The images shown in (b) are at time $t={0.1, \, 0.11, \, 0.12, \, ... \, , 0.17.}$ The detachment of a circular ligament rim is seen in both cases at $t=0.17$.}
    \label{fig:Initial_Mushroom_No_MD_MD}
\end{figure}

Figure \ref{fig:Initial_Mushroom_No_MD_MD} shows the evolution of the initial mushroom jet and the first sheet rupture. Figure \ref{fig:Initial_Mushroom_No_MD_MD}a shows the no-MD case at ${\ell = 14}$. We can see in the inset at $t=0.15$, numerical sheet rupture happens in the flap connecting the rolled-up mushroom rim with the mushroom head. The inset shows how a spectrum of tiny droplets forms along with a circular ligament rim at $t=0.17$. Figure \ref{fig:Initial_Mushroom_No_MD_MD}b shows the first sheet rupture for the case with manifold death method applied, corresponding to $(\ell,m)$ =$({15,13})$, that the one corresponding to the blue line in Figure \ref{fig:MD_stats_histo_all_LVL_conv}b. One can see the sheet rupture at $t=0.11$ is much more clear now and the curvature oscillations are not present. In this case too we see detachment of the rolled up tip in form a circular ligament at $t=0.17$.
Simple analysis suggests that  sheet thickness near the tip at the moment of rim formation should scale as $D/\We_g$  (see for example \cite{marcotte2019density}). The condition 
$D/\We_g \gg \Delta$ is then necessary for proper resolution of the sheet. This condition is equivalent to
$\We_\Delta \gg 1$  which is verified only for $\ell=15$ and perhaps 14 according to Table \ref{tab:delta}. At lower levels, it is clear that numerical breakup may appear soon behind the rim in no-MD simulations. 

\section{Determination of $d_c$ for manifold death statistics}
\label{appex:d_c_ZOOM_IN}
\begin{figure}
    \centering
    \includegraphics[width=\textwidth]{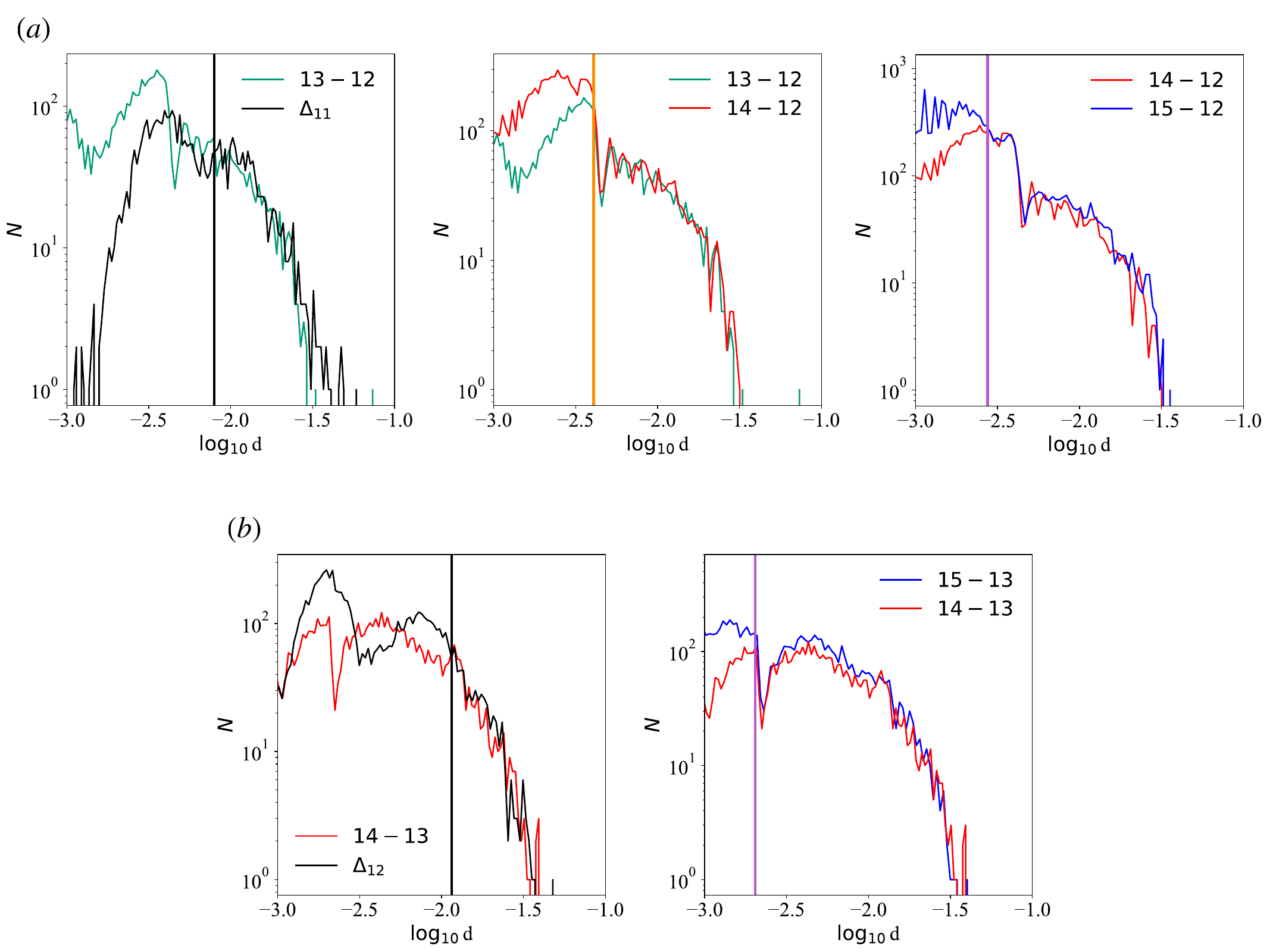}
    \caption{A zoom-in for the converged region corresponding to Figure \ref{fig:MD_stats_histo_all_LVL_conv} showing how $d_c$ (indicated by vertical lines) is obtained. Figures (a) and (b) correspond to a zoom-in for Figure \ref{fig:MD_stats_histo_all_LVL_conv}a and \ref{fig:MD_stats_histo_all_LVL_conv}b respectively. }
    \label{fig:d_c_eyeball}
\end{figure}
The determination of $d_c$, that is the diameter marking the lower bound of the converged region is done by visual inspection of the histogram showing the droplet size distribution. Figure \ref{fig:d_c_eyeball} shows a zoom-in explaining how we determine $d_c$ for each consecutive level. Note that the black curve in the no-MD case is not overlapping in the converged region of the various MD simulations.
This is an important fact since it means that an MD simulation is not ``equivalent'' to a no-MD simulation with a similar ``effective'' critical thickness $h_c$.
Hence it makes more sense to compare the MD plots to each other and treat the black no-MD curves only for the purpose of analyzing the properties of classical non-converging simulations. 
\section{Forensics of the  \cite{Chirco22} study}
\label{appex:forensics_Leo}
\begin{figure}
    \centering
    \includegraphics[width=\textwidth]{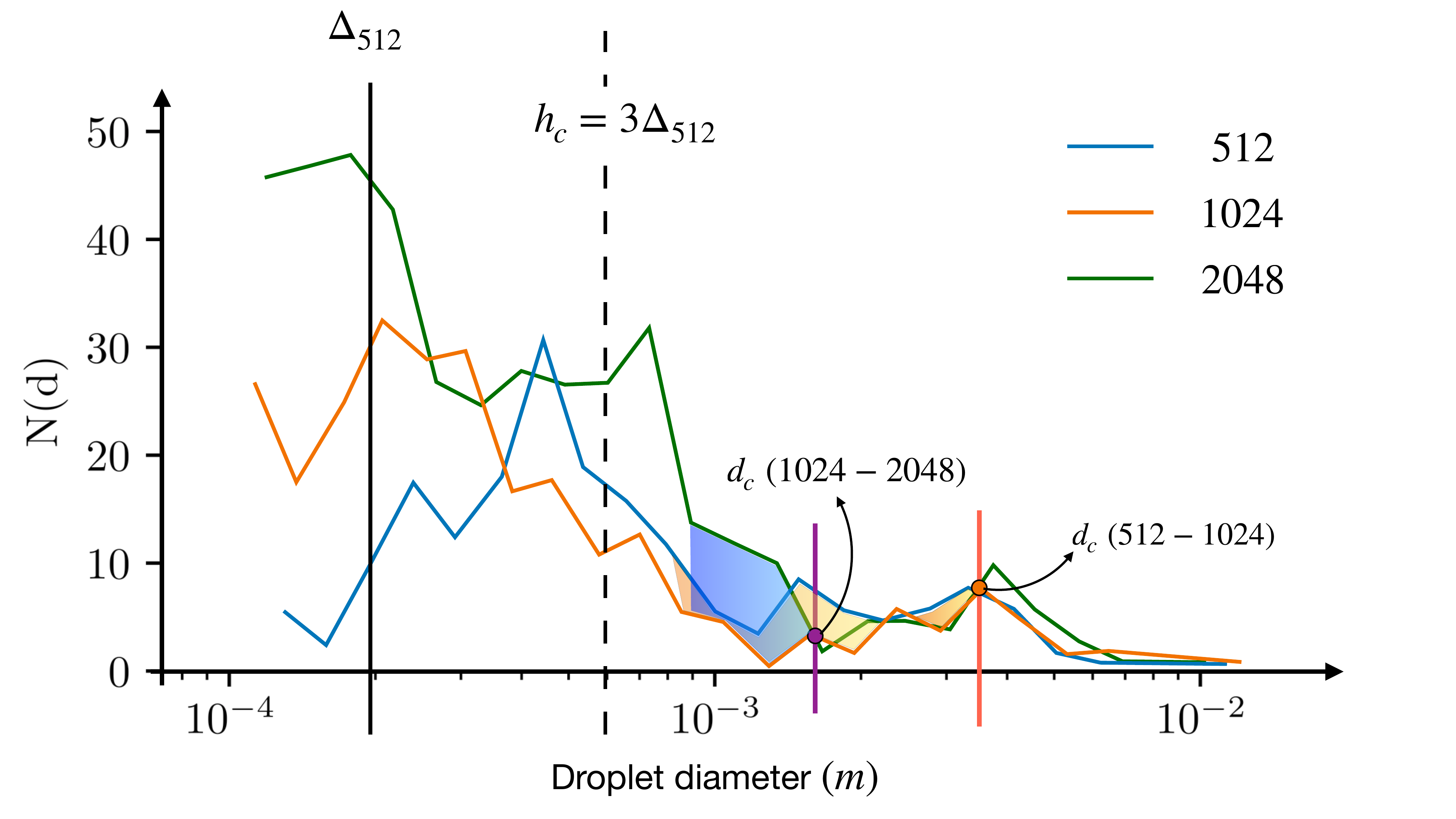}
    \caption{Replotting Figure 12 of \cite{Chirco22} with additional vertical lines showing the critical thickness for manifold death sheet perforation threshold $h_c$, along with two vertical lines for $d_c$. The shaded regions illustrate the departure of the measured distribution from each other below $d_c$ }
    \label{fig:Leo_hc_dc}
\end{figure}
The manifold death method appeared first in the work of  \cite{Chirco22}. 
Here we revisit this previous study with the droplet size distribution comparison method described above. In  \cite{Chirco22}  a region of convergence was identified at $8 \Delta_9$ where $\Delta_9$ was the grid size of the coarsest simulation, also used as the manifold death level. In our study, we have shown that the region of statistical convergence can be reached up to the critical sheet thickness of the manifold death threshold $h_c$, given that the mesh is refined sufficiently. In Figure 12 of \cite{Chirco22}, the authors argue that above $d_c = 8\Delta_{9}$, the simulations seem to converge, where the coarsest mesh had the equivalent of $512^3$ cells ($\ell=9$) and the finest mesh was the equivalent of $2048^3$ cells ($\ell=11$). The case that the authors simulated was a phase inversion case which produces a much smaller number of droplets than our pulsed jet case does. This implies that the number frequencies are affected by a large statistical error. Hence, the conclusions presented in \cite{Chirco22} are qualitative: they demonstrate the rupture of the sheet by manifold death in a clear manner.  To attempt a more quantitative analysis in line with our analysis of the pulsed jet,  we re-plot  Figure 12 of \cite{Chirco22} in our Figure \ref{fig:Leo_hc_dc} . There we mark the MD threshold $h_c = 3\Delta_{9}$ and two values of $d_c$. We see that $d_c$ again moves towards  $h_c$. This movement is in line with our results. Due to computational cost, \cite{Chirco22} stopped at the  $2048^3$ mesh and concluded that the boundary of the converged region lies at $d_c \sim 8 \Delta_{9}$. It is likely that the converged region could be pushed more towards $h_c$ with increased resolution.
\bibliographystyle{jfm}
\bibliography{bibliography,multiphase}

\begin{thebibliography}{57}
\expandafter\ifx\csname natexlab\endcsname\relax\def\natexlab#1{#1}\fi
\def\au#1{#1} \def\ed#1{#1} \def\yr#1{#1}\def\at#1{#1}\def\jt#1{\textit{#1}}
  \def\bt#1{#1}\def\bvol#1{\textbf{#1}} \def\vol#1{#1} \def\pg#1{#1}
  \def\publ#1{#1}\def\arxiv#1{#1}\def\org#1{#1}\def\st#1{\textit{#1}}

\bibitem[Agbaglah(2021)]{agbaglah2021breakup}
{\sc \au{Agbaglah, GG}} \yr{2021}  \at{Breakup of thin liquid sheets through
  hole--hole and hole--rim merging}.  \jt{Journal of Fluid Mechanics}
  \bvol{911},  \pg{A23}.

\bibitem[Anez {\em et~al.\/}(2019)Anez, Ahmed, Hecht, Duret, Reveillon \&
  Demoulin]{Anez2018}
{\sc \au{Anez, J}, \au{Ahmed, A}, \au{Hecht, N}, \au{Duret, B}, \au{Reveillon,
  J} \& \au{Demoulin, FX}} \yr{2019}  \at{Eulerian--lagrangian spray
  atomization model coupled with interface capturing method for diesel
  injectors}.  \jt{International Journal of Multiphase Flow}  \bvol{113},
  \pg{325--342}.

\bibitem[Balachandar {\em et~al.\/}(2020)Balachandar, Zaleski, Soldati, Ahmadi
  \& Bourouiba]{bala2020}
{\sc \au{Balachandar, S.}, \au{Zaleski, S.}, \au{Soldati, A.}, \au{Ahmadi, G.}
  \& \au{Bourouiba, L.}} \yr{2020}  \at{Host-to-host airborne transmission as a
  multiphase flow problem for science-based social distance guidelines}.
  \jt{International Journal of Multiphase Flow}  \bvol{132},  \pg{103439}.

\bibitem[Ben~Rayana {\em et~al.\/}(2006)Ben~Rayana, Cartellier \&
  Hopfinger]{Benrayana06}
{\sc \au{Ben~Rayana, F.}, \au{Cartellier, A.} \& \au{Hopfinger, E.}} \yr{2006}
  Assisted atomization of a liquid layer: investigation of the parameters
  affecting the mean drop size prediction.  \bt{In {\em Proc. ICLASS 2006, Aug.
  27-Sept.1, Kyoto Japan\/}}.  \publ{Academic Publ. and Printings},
  iSBN4-9902774-1-4.

\bibitem[Bianchi {\em et~al.\/}(2005)Bianchi, Pelloni, Toninel, Scardovelli,
  Leboissetier \& Zaleski]{Bianchi05}
{\sc \au{Bianchi, G.~M.}, \au{Pelloni, P.}, \au{Toninel, S.}, \au{Scardovelli,
  R.}, \au{Leboissetier, A.} \& \au{Zaleski, S.}} \yr{2005} A quasi-direct {3D}
  simulation of the atomization of high-speed liquid jets.  \bt{In {\em
  Proceedings of ICES05, 2005 ASME ICE Division Spring Technical
  Conference\/}}. Chicago, Illinois, USA, April 5-7, 2005.

\bibitem[Brujan(2010)]{brujan2010cavitation}
{\sc \au{Brujan, E.}} \yr{2010} {\em Cavitation in Non-Newtonian fluids: with
  biomedical and bioengineering applications\/}.  \publ{Springer Science \&
  Business Media}.

\bibitem[Chesnel {\em et~al.\/}(2011)Chesnel, M{\'e}nard, R{\'e}veillon \&
  D{e}moulin]{chesnel2011subgrid}
{\sc \au{Chesnel, J.}, \au{M{\'e}nard, T.}, \au{R{\'e}veillon, J.} \&
  \au{D{e}moulin, F.-X.}} \yr{2011}  \at{Subgrid analysis of liquid jet
  atomization}.  \jt{Atomization and Sprays}  \bvol{21}~(1).

\bibitem[Chirco {\em et~al.\/}(2022)Chirco, Maarek, Popinet \&
  Zaleski]{Chirco22}
{\sc \au{Chirco, L.}, \au{Maarek, J.}, \au{Popinet, S.} \& \au{Zaleski, S.}}
  \yr{2022}  \at{Manifold death: A volume of fluid implementation of controlled
  topological changes in thin sheets by the signature method}.  \jt{Journal of
  Computational Physics}  \bvol{467},  \pg{111468}.

\bibitem[Chorin(1968)]{chorin68}
{\sc \au{Chorin, A.~J.}} \yr{1968}  \at{Numerical solution of the
  {N}avier--{S}tokes equation}.  \jt{Mathematics of Computing}  \bvol{22},
  \pg{745--762}.

\bibitem[Culick(1960)]{Culick60}
{\sc \au{Culick, F. E.~C.}} \yr{1960}  \at{Comments on a ruptured soap film}.
  \jt{J. Appl. Phys.}  \bvol{31},  \pg{1128--1129}.

\bibitem[DeBar(1974)]{DeBar74}
{\sc \au{DeBar, R.}} \yr{1974}  \bt{Fundamentals of the {KRAKEN} code}. {\em
  Tech. Rep.\/} UCIR-760.  \org{LLNL}.

\bibitem[Debr{\'e}geas {\em et~al.\/}(1998)Debr{\'e}geas, De~Gennes \&
  Brochard-Wyart]{debregeas1998life}
{\sc \au{Debr{\'e}geas, G.~D.}, \au{De~Gennes, P.-G.} \& \au{Brochard-Wyart,
  F.}} \yr{1998}  \at{The life and death of ``bare" viscous bubbles}.
  \jt{Science}  \bvol{279}~(5357),  \pg{1704--1707}.

\bibitem[Duke {\em et~al.\/}(2017)Duke, Kastengren, Matusik, Swantek, Powell,
  Payri, Vaquerizo, Itani, Bruneaux, Grover~Jr {\em
  et~al.\/}]{duke2017internal}
{\sc \au{Duke, D~J}, \au{Kastengren, A~L}, \au{Matusik, K~E}, \au{Swantek,
  A~B}, \au{Powell, C~F}, \au{Payri, R}, \au{Vaquerizo, D}, \au{Itani, L},
  \au{Bruneaux, G}, \au{Grover~Jr, R~O} \& \au{others}} \yr{2017}  \at{Internal
  and near nozzle measurements of engine combustion network “{Spray G}”
  gasoline direct injectors}.  \jt{Experimental Thermal and Fluid Science}
  \bvol{88},  \pg{608--621}.

\bibitem[Fuster {\em et~al.\/}(2009)Fuster, Bagu{\'e}, Boeck, Le~Moyne,
  Leboissetier, Popinet, Ray, Scardovelli \& Zaleski]{fuster2009a}
{\sc \au{Fuster, D.}, \au{Bagu{\'e}, A.}, \au{Boeck, T.}, \au{Le~Moyne, L.},
  \au{Leboissetier, A.}, \au{Popinet, S.}, \au{Ray, P.}, \au{Scardovelli, R.}
  \& \au{Zaleski, S.}} \yr{2009}  \at{{Simulation of primary atomization with
  an octree adaptive mesh refinement and VOF method}}.  \jt{International
  Journal of Multiphase Flow}  \bvol{35}~(6),  \pg{550--565}.

\bibitem[Fuster {\em et~al.\/}(2013)Fuster, Matas, Marty, Popinet, Hoepffner,
  Cartellier \& Zaleski]{Fuster2013}
{\sc \au{Fuster, D.}, \au{Matas, J.~P.}, \au{Marty, S.}, \au{Popinet, S.},
  \au{Hoepffner, J.}, \au{Cartellier, A.} \& \au{Zaleski, S.}} \yr{2013}
  \at{{Instability regimes in the primary breakup region of planar coflowing
  sheets}}.  \jt{Journal of Fluid Mechanics}  \bvol{736},  \pg{150--176}.

\bibitem[Gavrilov(1969)]{gavrilov1969size}
{\sc \au{Gavrilov, L.R.}} \yr{1969}  \at{On the size distribution of gas
  bubbles in water}.  \jt{Sov. Phys. Acoust}  \bvol{15}~(1),  \pg{22--24}.

\bibitem[Gorokhovski \& Herrmann(2008)]{gorokhovski08}
{\sc \au{Gorokhovski, M.} \& \au{Herrmann, M.}} \yr{2008}  \at{{Modeling
  Primary Atomization}}.  \jt{Annual Reviews of Fluid Mechanics}  \pg{pp.
  343--366}.

\bibitem[Herrmann(2010)]{Herrmann10}
{\sc \au{Herrmann, M.}} \yr{2010}  \at{A parallel eulerian interface
  tracking/lagrangian point particle multi-scale coupling procedure}.
  \jt{Journal of Computational Physics}  \bvol{229}~(3),  \pg{745--759}.

\bibitem[Herrmann(2011)]{Herrmann2011}
{\sc \au{Herrmann, M.}} \yr{2011}  \at{On simulating primary atomization using
  the refined level set grid method}.  \jt{Atomization and Sprays}
  \bvol{21}~(4),  \pg{283--301}.

\bibitem[Hirt \& Nichols(1981)]{Hirt1981}
{\sc \au{Hirt, C.W.} \& \au{Nichols, B.D.}} \yr{1981}  \at{{Volume of fluid
  (VOF) method for the dynamics of free boundaries}}.  \jt{Journal of
  Computational Physics}  \bvol{39}~(1),  \pg{201--225}.

\bibitem[Jackiw \& Ashgriz(2022)]{jackiw2022prediction}
{\sc \au{Jackiw, I.~M.} \& \au{Ashgriz, N.}} \yr{2022}  \at{Prediction of the
  droplet size distribution in aerodynamic droplet breakup}.  \jt{Journal of
  Fluid Mechanics}  \bvol{940},  \pg{A17}.

\bibitem[Jarrahbashi \& Sirignano(2014)]{Jarrahbashi2014}
{\sc \au{Jarrahbashi, D.} \& \au{Sirignano, W.~A.}} \yr{2014}  \at{{Invited
  Article: Vorticity dynamics for transient high-pressure liquid injection}}.
  \jt{Physics of Fluids}  \bvol{26}~(10).

\bibitem[Jarrahbashi {\em et~al.\/}(2016)Jarrahbashi, Sirignano, Popov \&
  Hussain]{Jarrahbashi2016}
{\sc \au{Jarrahbashi, D.}, \au{Sirignano, W.~A.}, \au{Popov, P.~P.} \&
  \au{Hussain, F.}} \yr{2016}  \at{{Early spray development at high gas
  density: hole, ligament and bridge formations}}.  \jt{Journal of Fluid
  Mechanics}  \bvol{792},  \pg{186--231}.

\bibitem[Jiang \& Ling(2021)]{jiang2021impact}
{\sc \au{Jiang, D} \& \au{Ling, Y}} \yr{2021}  \at{Impact of inlet gas
  turbulence on the formation, development and breakup of interfacial waves in
  a two-phase mixing layer}.  \jt{Journal of Fluid Mechanics}  \bvol{921},
  \pg{A15}.

\bibitem[Kant {\em et~al.\/}(2023)Kant, Pairetti, Saade, Popinet, Zaleski \&
  Lohse]{kant2023bag}
{\sc \au{Kant, P.}, \au{Pairetti, C.}, \au{Saade, Y.}, \au{Popinet, S.},
  \au{Zaleski, S.} \& \au{Lohse, D.}} \yr{2023}  \at{Bag-mediated film
  atomization in a cough machine}.  \jt{Phys. Rev. Fluids}  \bvol{8}~(7),
  \pg{074802}.

\bibitem[Khanwale {\em et~al.\/}(2022)Khanwale, Saurabh, Ishii, Sundar \&
  Ganapathysubramanian]{khanwale2022breakup}
{\sc \au{Khanwale, M.~A}, \au{Saurabh, K.}, \au{Ishii, M.}, \au{Sundar, H.} \&
  \au{Ganapathysubramanian, B.}} \yr{2022}  \at{Breakup dynamics in primary jet
  atomization using mesh-and interface-refined cahn-hilliard navier-stokes}.
  \jt{arXiv preprint arXiv:2209.13142} .

\bibitem[Kibble(1976)]{kibble1976topology}
{\sc \au{Kibble, T. W.~B.}} \yr{1976}  \at{Topology of cosmic domains and
  strings}.  \jt{Journal of Physics A: Mathematical and General}  \bvol{9}~(8),
   \pg{1387}.

\bibitem[Lasheras \& Hopfinger(2000)]{Lasheras00}
{\sc \au{Lasheras, J.~C.} \& \au{Hopfinger, E.~J.}} \yr{2000}  \at{Liquid jet
  instability and atomization in a coaxial gas stream}.  \jt{Annu. Rev. Fluid
  Mech.}  \bvol{32},  \pg{275--308}.

\bibitem[Lebas {\em et~al.\/}(2009)Lebas, Menard, Beau, Berlemont \&
  Demoulin]{lebas2009}
{\sc \au{Lebas, R.}, \au{Menard, T.}, \au{Beau, P.A.}, \au{Berlemont, A.} \&
  \au{Demoulin, Fran{\c{c}}ois-Xavier}} \yr{2009}  \at{{Numerical simulation of
  primary break-up and atomization: DNS and modelling study.}}
  \jt{International Journal of Multiphase Flow}  \bvol{35}~(3),  \pg{247--260}.

\bibitem[Ling {\em et~al.\/}(2017{\natexlab{{\em a\/}}})Ling, Fuster, Zaleski
  \& Tryggvason]{ling17}
{\sc \au{Ling, Y.}, \au{Fuster, D.}, \au{Zaleski, S.} \& \au{Tryggvason, G.}}
  \yr{2017{\natexlab{{\em a\/}}}}  \at{{Spray formation in a quasiplanar
  gas--liquid mixing layer at moderate density ratios: A numerical closeup}}.
  \jt{Phys. Rev. Fluids}  \bvol{2}~(1),  \pg{014005}.

\bibitem[Ling {\em et~al.\/}(2017{\natexlab{{\em b\/}}})Ling, Fuster, Zaleski
  \& Tryggvason]{Ling2017}
{\sc \au{Ling, Yue}, \au{Fuster, Daniel}, \au{Zaleski, St{\'{e}}phane} \&
  \au{Tryggvason, Gr{\'{e}}tar}} \yr{2017{\natexlab{{\em b\/}}}}  \at{{Spray
  formation in a quasiplanar gas-liquid mixing layer at moderate density
  ratios: A numerical closeup}}.  \jt{Physical Review Fluids}  \bvol{2}~(1),
  \pg{014005}.

\bibitem[Lohse \& Villermaux(2020)]{lohse2020double}
{\sc \au{Lohse, D.} \& \au{Villermaux, E.}} \yr{2020}  \at{Double threshold
  behavior for breakup of liquid sheets}.  \jt{Proceedings of the National
  Academy of Sciences}  \bvol{117}~(32),  \pg{18912--18914}.

\bibitem[Marcotte \& Zaleski(2019)]{marcotte2019density}
{\sc \au{Marcotte, F.} \& \au{Zaleski, S.}} \yr{2019}  \at{Density contrast
  matters for drop fragmentation thresholds at low {O}hnesorge number}.
  \jt{Phys. Rev. Fluids}  \bvol{4}~(10),  \pg{103604}.

\bibitem[M{\'e}nard {\em et~al.\/}(2007)M{\'e}nard, Tanguy \&
  Berlemont]{menard07}
{\sc \au{M{\'e}nard, T.}, \au{Tanguy, S.} \& \au{Berlemont, A.}} \yr{2007}
  \at{{Coupling level set/VOF/ghost fluid methods: Validation and application
  to 3D simulation of the primary break-up of a liquid jet}}.  \jt{Int. J. of
  Multiphase Flow}  \bvol{33}~(5),  \pg{510--524}.

\bibitem[Opfer {\em et~al.\/}(2014)Opfer, Roisman, Venzmer, Klostermann \&
  Tropea]{opfer14}
{\sc \au{Opfer, L.}, \au{Roisman, I.V.}, \au{Venzmer, J}, \au{Klostermann, M.}
  \& \au{Tropea, C.}} \yr{2014}  \at{Droplet-air collision dynamics: Evolution
  of the film thickness}.  \jt{Phys. Rev. E}  \bvol{89}~(1),  \pg{013023}.

\bibitem[Pairetti {\em et~al.\/}(2021)Pairetti, Villiers \&
  Zaleski]{pairetti2021numerical}
{\sc \au{Pairetti, Cesar}, \au{Villiers, Rapha{\"e}l} \& \au{Zaleski,
  St{\'e}phane}} \yr{2021} A numerical cough machine. Accepted for publication
  in Computer and Fluids.

\bibitem[Pairetti {\em et~al.\/}(2020)Pairetti, Damian, Nigro, Popinet \&
  Zaleski]{pairetti2020mesh}
{\sc \au{Pairetti, C.~I.}, \au{Damian, S.~M.}, \au{Nigro, N.~M}, \au{Popinet,
  S.} \& \au{Zaleski, S.}} \yr{2020}  \at{Mesh resolution effects on primary
  atomization simulations}.  \jt{Atomization and Sprays}  \bvol{30}~(12).

\bibitem[Popinet(2003)]{popinet2003}
{\sc \au{Popinet, S.}} \yr{2003}  \at{Gerris: a tree-based adaptive solver for
  the incompressible euler equations in complex geometries}.  \jt{J. Comput.
  Phys.}  \bvol{190}~(2),  \pg{572--600}.

\bibitem[Popinet(2009)]{popinet2009}
{\sc \au{Popinet, S.}} \yr{2009}  \at{{An accurate adaptive solver for
  surface-tension-driven interfacial flows}}.  \jt{J. Comput. Phys.}
  \bvol{228}~(16),  \pg{5838--5866}.

\bibitem[Popinet(2018)]{Popinet2018}
{\sc \au{Popinet, S.}} \yr{2018}  \at{{Numerical Models of Surface Tension}}.
  \jt{Annu. Rev. Fluid Mech.}  \bvol{50}~(1),  \pg{49--75},  \arxiv{arXiv:
  1507.05135}.

\bibitem[Salvador {\em et~al.\/}(2018)Salvador, Ruiz, Crialesi-Esposito \&
  Blanquer]{salvador2018analysis}
{\sc \au{Salvador, F.~J.}, \au{Ruiz, S.}, \au{Crialesi-Esposito, M.} \&
  \au{Blanquer, I.}} \yr{2018}  \at{Analysis on the effects of turbulent inflow
  conditions on spray primary atomization in the near-field by direct numerical
  simulation}.  \jt{Int. J. of Multiphase Flow}  \bvol{102},  \pg{49--63}.

\bibitem[Saurabh {\em et~al.\/}(2023)Saurabh, Ishii, Khanwale, Sundar \&
  Ganapathysubramanian]{saurabh2023scalable}
{\sc \au{Saurabh, K.}, \au{Ishii, M.}, \au{Khanwale, M.~A}, \au{Sundar, H.} \&
  \au{Ganapathysubramanian, B.}} \yr{2023} Scalable adaptive algorithms for
  next-generation multiphase flow simulations.  \bt{In {\em 2023 IEEE
  International Parallel and Distributed Processing Symposium (IPDPS)\/}},
  \pg{pp. 590--601}. IEEE.

\bibitem[Savva \& Bush(2009)]{savva2009viscous}
{\sc \au{Savva, N.} \& \au{Bush, J. W.~M.}} \yr{2009}  \at{Viscous sheet
  retraction}.  \jt{J. Fluid Mech.}  \bvol{626},  \pg{211--240}.

\bibitem[Shima \& Sakai(1987)]{shima1987equation}
{\sc \au{Shima, A.} \& \au{Sakai, I.}} \yr{1987}  \at{On the equation for the
  size distribution of bubble nuclei in liquids (report 2)}.  \jt{Rep. Inst.
  High Speed Mech., Tohoku Univ., Ser B}  \bvol{54},  \pg{51}.

\bibitem[Shinjo \& Umemura(2010)]{shinjo2010simulation}
{\sc \au{Shinjo, J} \& \au{Umemura, A}} \yr{2010}  \at{Simulation of liquid jet
  primary breakup: Dynamics of ligament and droplet formation}.
  \jt{International Journal of Multiphase Flow}  \bvol{36}~(7),  \pg{513--532}.

\bibitem[Song \& Tryggvason(1999)]{song99}
{\sc \au{Song, M.} \& \au{Tryggvason, G.}} \yr{1999}  \at{The formation of
  thick borders on an initially stationary fluid sheet}.  \jt{Phys. Fluids}
  \bvol{11}~(9),  \pg{2487--93}.

\bibitem[Tang {\em et~al.\/}(2023)Tang, Adcock \&
  Mostert]{Tang_Adcock_Mostert_2023}
{\sc \au{Tang, K.}, \au{Adcock, T.A.A.} \& \au{Mostert, W.}} \yr{2023}  \at{Bag
  film breakup of droplets in uniform airflows}.  \jt{Journal of Fluid
  Mechanics}  \bvol{970},  \pg{A9}.

\bibitem[Taylor(1959)]{Taylor59c}
{\sc \au{Taylor, G.~I.}} \yr{1959}  \at{The dynamics of thin sheets of fluid
  {III}. {D}isintegration of fluid sheets}.  \jt{Proc. Roy. Soc. London A}
  \bvol{253},  \pg{313--321}.

\bibitem[Tolfts {\em et~al.\/}(2023)Tolfts, Deplus \&
  Machicoane]{tolfts2023statistics}
{\sc \au{Tolfts, Oliver}, \au{Deplus, Guillaume} \& \au{Machicoane,
  Nathana{\"e}l}} \yr{2023}  \at{Statistics and dynamics of a liquid jet under
  fragmentation by a gas jet}.  \jt{Physical Review Fluids}  \bvol{8}~(4),
  \pg{044304}.

\bibitem[Torregrosa {\em et~al.\/}(2020)Torregrosa, Payri, Salvador \&
  Crialesi-Esposito]{torregrosa2020study}
{\sc \au{Torregrosa, A.~J.}, \au{Payri, R.}, \au{Salvador, F.~J.} \&
  \au{Crialesi-Esposito, M.}} \yr{2020}  \at{Study of turbulence in atomizing
  liquid jets}.  \jt{International Journal of Multiphase Flow}  \bvol{129},
  \pg{103328}.

\bibitem[Vernay {\em et~al.\/}(2017)Vernay, Ramos, W{\"u}rger \&
  Ligoure]{Vernay:2017bz}
{\sc \au{Vernay, C.}, \au{Ramos, L.}, \au{W{\"u}rger, A.} \& \au{Ligoure, C.}}
  \yr{2017}  \at{{Playing with Emulsion Formulation to Control the Perforation
  of a Freely Expanding Liquid Sheet}}.  \jt{Langmuir}  \bvol{33}~(14),
  \pg{3458--3467}.

\bibitem[Villermaux(2020)]{villermaux2020fragmentation}
{\sc \au{Villermaux, E.}} \yr{2020}  \at{Fragmentation versus cohesion}.
  \jt{J. Fluid Mech.}  \bvol{898}.

\bibitem[Villermaux \& Bossa(2009)]{Villermaux:2009ccz}
{\sc \au{Villermaux, E} \& \au{Bossa, B}} \yr{2009}  \at{{Single-drop
  fragmentation determines size distribution of raindrops}}.  \jt{Nature
  Physics} .

\bibitem[Wang \& Bourouiba(2018)]{wang2018unsteady}
{\sc \au{Wang, Y} \& \au{Bourouiba, L}} \yr{2018}  \at{Unsteady sheet
  fragmentation: droplet sizes and speeds}.  \jt{J. Fluid Mech.}  \bvol{848},
  \pg{946--967}.

\bibitem[Weymouth \& Yue(2010)]{Weymouth:2010hy}
{\sc \au{Weymouth, G.~D.} \& \au{Yue, D. K.~P.}} \yr{2010}  \at{{Conservative
  Volume-of-Fluid method for free-surface simulations on Cartesian-grids}}.
  \jt{J. Comput. Phys.}  \bvol{229}~(8),  \pg{2853--2865}.

\bibitem[Zhang {\em et~al.\/}(2020)Zhang, Popinet \& Ling]{zhang2020modeling}
{\sc \au{Zhang, B.}, \au{Popinet, S.} \& \au{Ling, Y.}} \yr{2020}  \at{Modeling
  and detailed numerical simulation of the primary breakup of a gasoline
  surrogate jet under non-evaporative operating conditions}.  \jt{Int. J. of
  Multiphase Flow}  \bvol{130},  \pg{103362}.

\bibitem[Zurek(1985)]{zurek1985cosmological}
{\sc \au{Zurek, W.~H.}} \yr{1985}  \at{Cosmological experiments in superfluid
  helium?}  \jt{Nature}  \bvol{317}~(6037),  \pg{505--508}.

\end{thebibliography}
\end{document}